\documentclass[galaxies,article,submit,moreauthors,pdftex]{mdpi} 
\usepackage{lmodern}
\usepackage{subcaption}
\usepackage{wasysym}
\firstpage{1} 
\pubvolume{xx}
\issuenum{1}
\articlenumber{5}
\pubyear{2019}
\copyrightyear{2019}
\history{Received: date; Accepted: date; Published: date}

\Title{An In-depth Investigation of Faraday Depth Spectrum Using Synthetic Observations
of Turbulent MHD Simulations}

\Author{Aritra Basu $^{1}$*, Andrew Fletcher $^2$, S.~A.~Mao $^{3}$, Blakesley Burkhart $^{4,5}$,
Rainer Beck~$^3$, Dominic Schnitzeler $^6$}

\AuthorNames{Aritra Basu, S. A. Mao, Andrew Fletcher, Blakesley Burkhart, Rainer Beck and Dominic Schnitzeler}

\address{%
$^{1}$ \quad Fakult\"at f\"ur Physik, Universit\"at Bielefeld, Postfach 100131, 33501 Bielefeld, Germany; aritra@physik.uni-bielefeld.de\\
$^{2}$ \quad School of Mathematics, Statistics and Physics, Newcastle University, Newcastle-upon-Tyne, NE13 7RU, United Kingdom\\
$^{3}$ \quad Max-Planck-Institut f\"ur Radioastronomie, Auf dem H\"ugel 69, 53121 Bonn, Germany\\
$^{4}$ \quad Center for Computational Astrophysics, Flatiron Institute, 162 Fifth Avenue, New York, NY 10010, USA \\
$^{5}$ \quad Department of Physics and Astronomy, Rutgers, The State University of New Jersey, 136 Frelinghuysen Rd, Piscataway, NJ 08854, USA \\
$^{6}$ \quad ASML, De Run 6501, 5504 DR, Veldhoven, The Netherlands}

\corres{Correspondence: aritra@physik.uni-bielefeld.de}

\abstract{In this paper we present a detailed analysis of the Faraday depth
(FD) spectrum and its clean components obtained through the application of the
commonly used technique of Faraday rotation measure synthesis to analyze
spectro-polarimetric data. In order to directly compare the Faraday depth
spectrum with physical properties of a magneto-ionic medium, we generated
synthetic broad-bandwidth spectro-polarimetric observations from
magnetohydrodynamic (MHD) simulations of a transonic, isothermal, compressible
turbulent medium. We find that correlated magnetic field structures give rise
to a combination of spiky, localized peaks at certain FD values, and broad
structures in the FD spectrum. Although the majority of these spiky FD
structures appear narrow, giving an impression of a Faraday thin medium, we
show that they arise from strong synchrotron emissivity at that FD. Strong
emissivity at a FD can arise because of both strong spatially-local polarized
synchrotron emissivity at a FD or accumulation of weaker emissions along the
distance through a medium that have Faraday depths within half the width of the
rotation measure spread function. Such a complex Faraday depth spectrum is a
natural consequence of MHD turbulence when the lines of sight pass through a
few turbulent cells. This therefore complicates the convention of attributing
narrow FD peaks to presence of a Faraday rotating medium along the line of
sight. Our work shows that it is difficult to extract the FD along a line of
sight from the Faraday depth spectrum using standard methods for a turbulent
medium in which synchrotron emission and Faraday rotation occur
simultaneously.}


\begin{document}


\section{Introduction}

The advent of broad-band receivers on all major radio telescopes in the last
decade have opened up new avenues for investigating the properties of
synchrotron-emitting relativistic plasma in astrophysical objects. Previously,
one had to observe the same source at multiple, widely-spaced frequencies to
understand its polarization behaviour. Broad-band radio spectro-polarimetric
measurements of the Stokes $Q$ and $U$ parameters makes it possible to measure
the variation of the polarized synchrotron emission over a wide contiguous
frequency range which can provide crucial insights into the properties of the
magneto-ionic medium for a large number of sources in a much reduced amount of
observing time. Broad-band spectro-polarimetry has played a crucial role in
unveiling the properties of magnetic fields in nearby galaxies \citep{mao15,
damas16, stein19}, in high redshift galaxies \citep{kim16, mao17}, in active
galactic nuclei \citep[AGN;][]{sulli12, sulli17, paset18} and the intergalactic
medium \citep{sulli19}. In order to physically interpret such data, the
technique of Faraday rotation measure (RM) synthesis \citep{burn66, brent05}
and direct fitting of the Stokes $Q$ and $U$ spectra of a polarized source with
models of the magneto-ionic media, known as Stokes $Q,U$ fitting
\citep{farnsw11, sulli12, ander15} have been developed. It is often not
straightforward to interpret the results from these techniques and connect them
to the physical properties of the magnetized plasmas being investigated
\citep{ander16}. 

Currently, several large-scale spectro-polarimetric campaigns are underway,
mostly in the 1 to 5~GHz frequency range. On one hand, dedicated surveys with
interferometers are being conducted to study the broad-band polarization
properties of millions of extragalactic sources. For example, the Karl G.
Jansky Very Large Array Sky Survey \citep[VLASS;][]{myers14, lacy19}, the
Polarization Sky Survey of the Universe's Magnetism \citep[POSSUM;][]{gaens10},
the QU Observations at Cm wavelength with Km baselines using
ATCA\footnote{\url{https://research.csiro.au/quocka/}} (QUOCKA), the MeerKAT
International GHz Tiered Extragalactic Exploration (MIGHTEE) Survey
\citep{jarvi16}, and the recent LOFAR Two-meter Sky Survey
\citep[LoTSS;][]{eck18a, eck19} and S-PASS/ATCA \citep{schni19}. Many more
broad-band polarization surveys are planned in the future with the upcoming
Square Kilometre Array (SKA).

On the other hand, to study the diffuse Galactic magneto-ionic medium,
broad-band, large sky-area surveys below $\sim15$~GHz with single dish
telescopes have been undertaken, e.g., the Global Magneto-Ionic Medium Survey
\citep[GMIMS;][]{wolle09}, the GALFA Continuum Transit Survey
\citep[GALFACTS;][]{taylor13}, the S-band Polarization All Sky Survey
\citep[S-PASS;][]{carre19}, survey with the SKA-MPG prototype telescope
\citep{basu19}, the C-Band All Sky Survey \citep[C-BASS;][]{jones18} and the
Q-U-I JOint TEnerife \citep[QUIJOTE;][]{quijote15a, quijote15b}.

With these broad-band polarization surveys, pressing astrophysical problems,
such as, black hole accretion and its connection to AGN jet launching
mechanism, cosmic evolution of magnetic fields in galaxies, structure and
strength of magnetic fields in the interstellar, intra-cluster and
intergalactic medium will be investigated \citep[see e.g.][]{AASKA14}. In
addition to addressing astrophysical questions, these surveys will also
contribute to the solution of fundamental cosmological questions via sensitive
measurements of the Galactic diffuse synchrotron emission. This emission
contaminates cosmological signals from the early Universe, such as, the cosmic
microwave background radiation, the cosmic dawn and the epoch of reionization.

Quantities related to the plasma properties of a magneto-ionic medium that can
be derived from observations are: the intrinsic fractional polarization and
angle of the linearly polarized synchrotron emission, the Faraday depth (FD)
and its dispersion. The intrinsic fractional polarization is determined by the
ratio of turbulent to ordered magnetic field strengths in the plane of the sky
averaged over the telescope beam \citep{sokol98}. The intrinsic angle of the
linearly polarized emission gives us information about the orientation of the
ordered magnetic fields in the plane of the sky. The Faraday depth gives us
information on the strength and direction of the average magnetic field
parallel to the line of sight and its spatial variation can help us to
distinguish between coherent and anisotropic random magnetic fields
\citep{jaffe10}. The dispersion of FD depends upon the properties of turbulent
magnetic fields parallel to the line of sight. Hence, robust measurements of
these quantities can provide insights into the 3-dimensional properties of
magneto-ionic media in astrophysical sources. To measure them, RM synthesis and
Stokes $Q,U$ fitting techniques are applied to broad-band spectro-polarimetric
observations. Therefore, it is imperative to investigate, in detail, the scope
of application of these data analysis techniques and their limitations.
Understanding the efficacy and efficiency of these tools are of paramount
importance for the success of the above mentioned surveys.

In this paper we will focus on the technique of RM synthesis when it is applied
to infer properties of diffuse medium, e.g., the Galactic interstellar medium
(ISM). Starting with magnetohydrodynamic (MHD) simulations of isothermal,
transonic, compressible turbulent plasma, similar to that observed in the
Galactic ISM \citep{gaens11, koley19}, we use ray-tracing to simulate
broad-band spectro-polarimetric observations. Then, we apply RM synthesis to
test what we can learn about the medium. The major questions we will
investigate are: (1)~Is there a difference in the nature of the Faraday depth
spectrum obtained by applying RM synthesis to a medium with a realistic model
of turbulence from MHD simulations {\it versus} the commonly used model of
turbulence as a Gaussian random field? (2)~What is the origin of complexity in
the Faraday depth spectrum? (3)~Can RM synthesis recover the Faraday depth
of a diffuse medium which is simultaneously Faraday rotating and emitting
polarized synchrotron radiation?

This paper is organized as follows. In Section~\ref{sec:techniques} we discuss
in brief the techniques of RM synthesis and Stokes $Q,U$ fitting.  We present
in brief the salient features of the software package, {\tt COSMIC}, for
generating synthetic broad-band spectro-polarimetric data from MHD simulations
in Section~\ref{sec:cosmic}. In Section~\ref{sec:benchmark} we test the
numerical performance of {\tt COSMIC} using simulated media for which the
broad-band polarization behaviour is known analytically. We present details of
the MHD simulations in Section~\ref{sec:mhd} and synthetic polarization
observations performed by applying {\tt COSMIC} to the simulations are
discussed in Section~\ref{sec:syntheticObs}. In Section~\ref{sec:FDanalysis} we
present a detailed analysis of the results obtained from RM synthesis and
compare them to the intrinsic properties of the medium, and we summarize our
findings in Section~\ref{sec:conclusion}.

\section{Common spectro-polarimetric data analysis techniques}
\label{sec:techniques}

RM synthesis and Stokes $Q,U$ fitting are two commonly used techniques that are
used to extract information from broad-band observations of polarized emission.
The parameters of interest are the intrinsic fractional polarization ($p_{\rm
int}$), intrinsic orientation of the polarization angle ($\theta_0$), Faraday
depth (FD) and the intrinsic dispersion of FD ($\sigma_{\rm FD}$). In order to
gain physical insights into an astrophysical system, robust measurement of
these quantities are essential.

$\bullet$ \textit{Stokes parameter ($Q,U$) fitting: }
The technique of Stokes $Q,U$ fitting is a \textit{parametric} fitting of the
Stokes $Q$ and $U$ parameters' wavelength ($\lambda$) dependent variation using
models of a turbulent magneto-ionic media analytically derived, for example, in
\citet{burn66}, \citet{tribb91}, \citet{sokol98} and \citet{rosse08}, and
therefore requires assumptions on the nature of the medium being investigated.
In Stokes $Q,U$ fitting, $p_{\rm int}$, $\theta_0$, FD and $\sigma_{\rm FD}$
are directly fitted for as model parameters. However, fitting is limited to a
set of source models that might oversimplify the physics inside real radio
sources and their environments. Moreover, often the broad-band Stokes $Q$ and
$U$ data cannot be fitted by a single model and therefore linear combinations
of models are used.  Since increasing the number of models improves the quality
of the fit by increasing the number of free parameters \citep[principle of
parsimony;][]{ander07book}, statistical measures, such as, the Bayesian
inference criterion and/or Akaike information criterion are used to limit the
number of models or to choose between degenerate fits \citep{sulli12, schni18}.
However, Stokes $Q,U$ fitting has the advantage of also fitting for the
spectral index of the polarized flux density \citep[see][]{schni18}, which is
not possible in RM synthesis and is one of the origins of complexity in the
Faraday depth spectrum \citep[e.g.,][]{schni18, schni19}.

$\bullet$  \textit {Rotation measure synthesis: }
RM synthesis is a Fourier transform-like operation in which it is assumed that
the linearly polarized radio source can be described as a sum of emitters at
their respective Faraday depths, and is the Fourier transform of the frequency
spectrum of the complex polarization\footnote{Complex polarization, $P$, is
defined as $P = Q + i\,U$, where, $Q$ and $U$ are the Stokes parameters.}
\citep{burn66, brent05}. Therefore, RM synthesis makes only a weak assumption
about the physical properties of the magneto-ionic medium and is a
\textit{non-parametric} approach. Like any Fourier transform performed over a
finite space (in this case a finite frequency coverage), the Faraday depth
spectrum (variation of fractional polarization $p$ or polarized intensity $PI$
as a function of FD) obtained from RM synthesis is convolved with a complicated
response function determined by the frequency coverage of the observations. The
response function is known as the {\it rotation measure spread function} (RMSF)
and has sidelobes due to sharp cut-offs and gaps in the frequency coverage.
Artefacts produced by the RMSF sidelobes can be deconvolved by applying the
technique of RM clean \citep{heald09}. In the case when Faraday rotation
originates at multiple Faraday depths or Faraday depth varies contiguously
through a volume, RM clean models a source as discrete $\delta$-function
emitters in Faraday depth space known as \textit{clean components}. The
full-width at half-maximum (FWHM) of the RMSF is determined by the bandwidth in
$\lambda^2$ of the observations and it determines how well the emitters can be
resolved in the Faraday depth space. The channel width of observations
determines the maximum observable FD and the highest frequency end determines
the sensitivity to the largest scale structure in the Faraday depth space
\citep[see][for details]{brent05}. This means that the results depend heavily
on the frequency coverage used, and interpreting the RM cleaned Faraday depth
spectra or the clean components can be a challenging task.

Another challenge of RM synthesis is the way $p_{\rm int}$, $\theta_0$, FD and
$\sigma_{\rm FD}$ are determined from the Faraday depth spectrum. This comes
down to a matter of choice for an individual investigator. In order to extract
information on the physical quantities of magneto-ionic media, attempts have
recently been made to describe the Faraday depth spectrum using parametric
functions. In such an approach, the Faraday depth spectrum is modelled as a
$\delta$-function for a purely Faraday rotating medium, as a top-hat function
for a simultaneously Faraday rotating and synchrotron emitting medium
containing regular magnetic fields and constant densities of thermal and
relativistic electrons, and as super-Gaussian function for a medium which
contains both turbulent and regular magnetic fields \citep{ander16, idegu17,
idegu18}. Recently, \citet{eck18b} proposed a model-free description of
mapping the Faraday depth spectrum to polarized emission as a function of
distance. The parametric descriptions, for both Faraday depth spectra and
Stokes $Q,U$ fitting, implicitly assume a Gaussian random distribution of the
components of the turbulent magnetic field. In contrast, the turbulent magnetic
field and free electron distribution in the diffuse ISM, the focus of this
paper, are expected to have spatially correlated structures, and are often
non-Gaussian \citep{armst95, burkh09, haver08, holli17, makar18}.  Further,
observations of extragalactic sources and the diffuse Galactic emission have
revealed that FD spectra often show complicated structures \citep{ander16,
eck17, wolle19} referred to as {\it Faraday complexity}. A part of the
complexity can be introduced by the spectral index of the polarized emission,
and hence RM synthesis is typically performed on fractional polarized
parameters. Dedicated efforts, both mathematical and computational, are
required to incorporate information on spectral index and Faraday
depolarization into RM synthesis.

Because of these caveats, it is necessary to investigate the results of RM
synthesis using MHD turbulence simulations which provide a more realistic
magnetic field and Faraday depth distribution that are not described by simple
Gaussian statistics.

\section{{\tt COSMIC}: from physical quantities to Stokes parameters and
Faraday rotation} \label{sec:cosmic}

We have developed an end-to-end, fully parallelized, Python based software
package --- Computerized Observations of Simulated MHD Inferred Cubes ({\tt
COSMIC}) --- to generate synthetic data cubes of Stokes parameters as a
function of frequency for further analysis. RM synthesis and RM clean are
performed by integrating the {\tt pyrmsynth}
package\footnote{\url{https://github.com/mrbell/pyrmsynth}} into {\tt COSMIC}.
The code requires 3-dimensional (3-D) spatial cubes of the three magnetic field
components (in units of $\mu$G) and the distribution of neutral or ionized gas
density (in units of cm$^{-3}$) computed from MHD simulations as inputs, in a
Cartesian coordinate system. Depending on the type of MHD simulations, other
optional inputs can be provided, such as, the 3-D spatial distributions of
temperature and the number density or energy spectrum of cosmic ray electrons
(CREs). The default coordinate system is chosen such that the line of sight
(LOS) is along the $z$-axis and, $x$- and $y$-axes are in the plane of the sky.
However, {\tt COSMIC} allows the user to choose the LOS axis
perpendicular to any of the six faces of the cube.

For MHD simulations which do not contain cosmic rays, to compute the total
synchrotron emission and the Stokes $Q$ and $U$ parameters of the linearly
polarized synchrotron emission, a user can choose from several options to
determine the number density of CREs ($n_{\rm CRE}$) and their energy spectrum.
Similarly, to compute the Faraday depth, the number density of free electrons
($n_{\rm e}$) is estimated by choosing a suitable ionization model depending on
the type of the simulations and other ancillary data computed in the
simulations, such as, the gas temperature. Using these, {\tt COSMIC} computes
the 2-D distribution of the total synchrotron intensity and the Stokes $Q$ and
$U$ parameters on the plane of the sky across a frequency range specified by
the user. Details of how observables are computed numerically and the different
user-specified options are presented in Appendix~\ref{sec:cosmic_app}.

\section{Benchmarking {\tt COSMIC} with analytic models of magneto-ionic media}
\label{sec:benchmark}

To test the accuracy of the numerical calculations performed by {\tt COSMIC},
we first compared outputs from it with simple models of magneto-ionic media
whose frequency-dependent polarization behaviour have analytic solutions. We
simulated two types of radio sources which are simultaneously synchrotron
emitting and Faraday rotating: (1) a uniform slab containing a regular magnetic
field and constant densities of free electrons and CREs \citep{burn66,
sokol98}; (2) a volume that contains both regular and turbulent magnetic fields
in which the turbulent fields have a Gaussian random distribution, known as the
internal Faraday dispersion (IFD) model \citep{sokol98}. 

\begin{table} \centering
 \caption{Setup parameters used to generate synthetic observations to test {\tt
COSMIC} with uniform slab and internal Faraday dispersion (IFD) models.}
  \begin{tabular}{@{}lcc}
 \hline
Parameter  & Uniform slab   & IFD \\
 \hline
\multicolumn{1}{@{}l}{Regular field strengths} & 
\multicolumn{2}{@{}c@{}}{$\langle B_x \rangle = 4\,\mu$G, $\langle B_y \rangle = 7\,\mu$G, $\langle B_z \rangle = 5\,\mu$G} \\
Random field strengths: & & \\
$\sigma_x$ ($\mu$G) & 0 & 10 \\
$\sigma_y$ ($\mu$G) & 0 & 10 \\
$\sigma_z$ ($\mu$G) & 0 & 10 \\
$n_{\rm e}$ (cm$^{-3}$) & 0.05 & 0.05 \\
 \hline
 & & \\
\multicolumn{1}{@{}l}{Box size} & 
\multicolumn{2}{c}{$512\times512\times 512\,{\rm pc^3}$} \\
\multicolumn{1}{@{}l}{Mesh size} & 
\multicolumn{2}{c}{$1\times1\times 1\,{\rm pc^3}$} \\
\multicolumn{1}{@{}l}{Spectral index} & 
\multicolumn{2}{c}{$\alpha = -0.8$}\\
\multicolumn{1}{@{}l}{Spectral curvature} & 
\multicolumn{2}{c}{None}\\
\multicolumn{1}{@{}l}{Frequency range} & 
\multicolumn{2}{c}{$\nu_{\rm min}=0.5$\,GHz, $\nu_{\rm max}=6$\,GHz}\\
\multicolumn{1}{@{}l}{Number of channels} & 
\multicolumn{2}{c}{$n_{\rm chan}=500$}\\
\hline
\end{tabular}
\label{tab:pars}
\end{table}

\begin{figure}[t]
\centering
\begin{tabular}{cc}
{\mbox{\includegraphics[width=7cm]{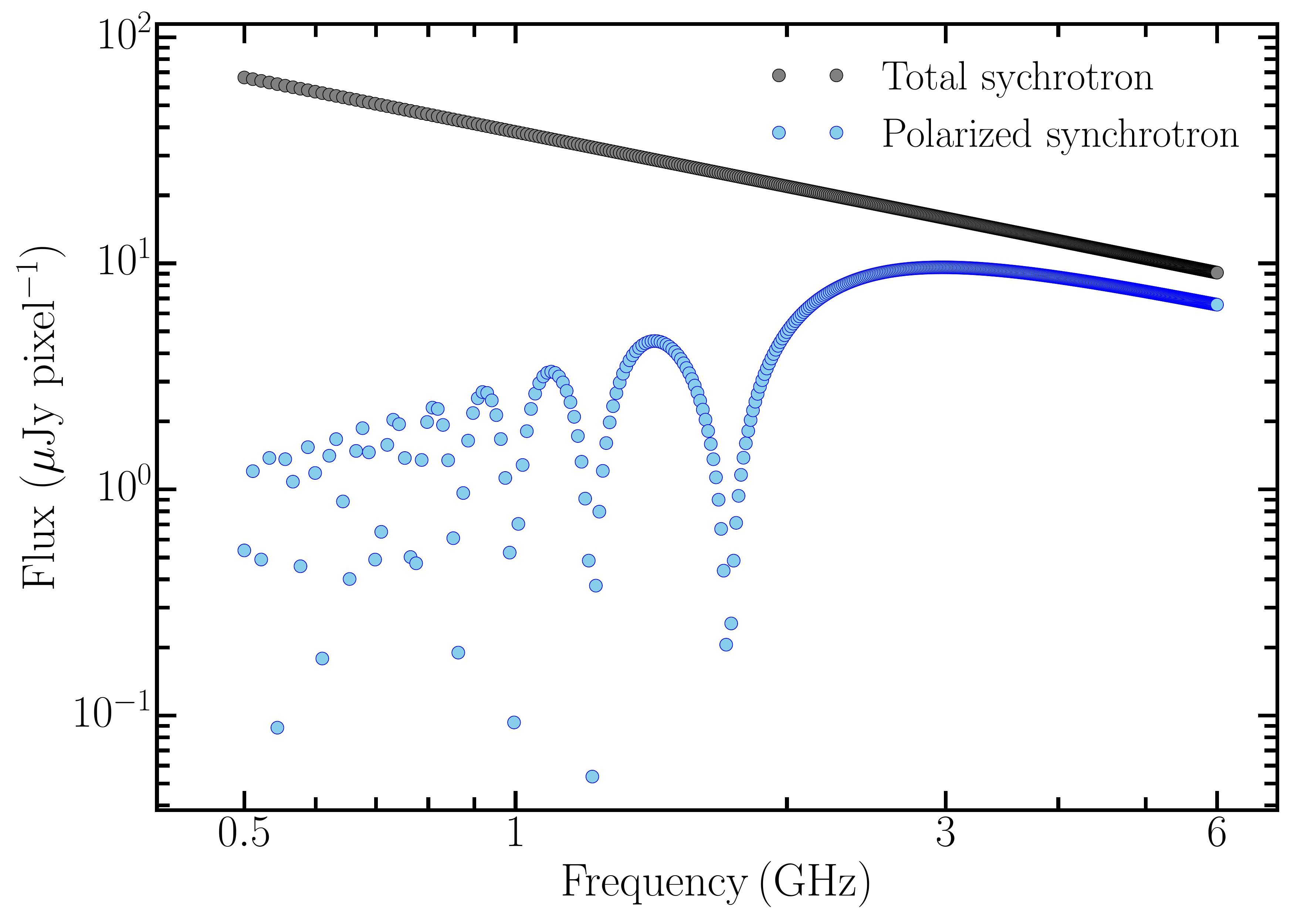}}}&
{\mbox{\includegraphics[width=7cm]{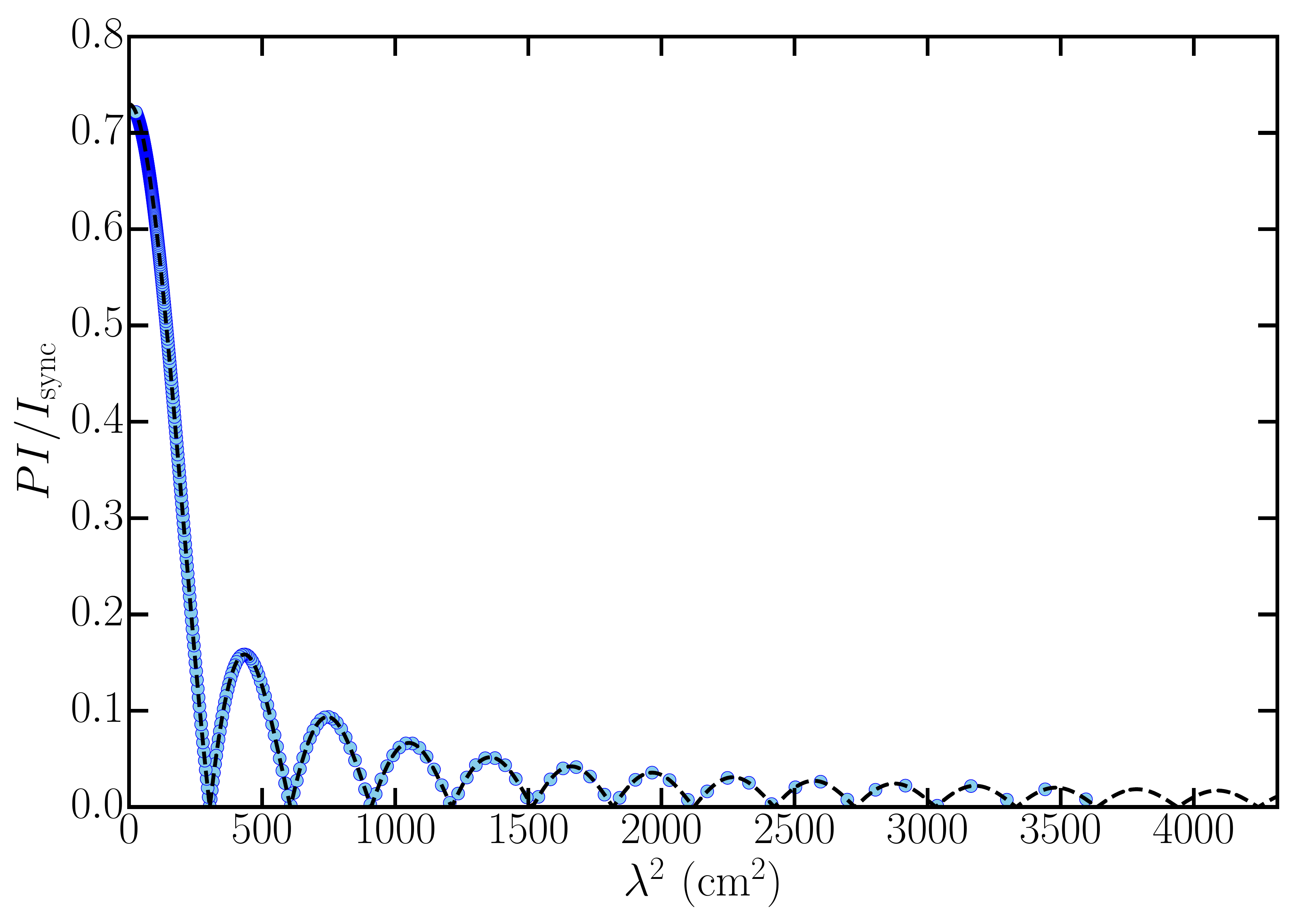}}}\\
{\mbox{\includegraphics[width=7cm]{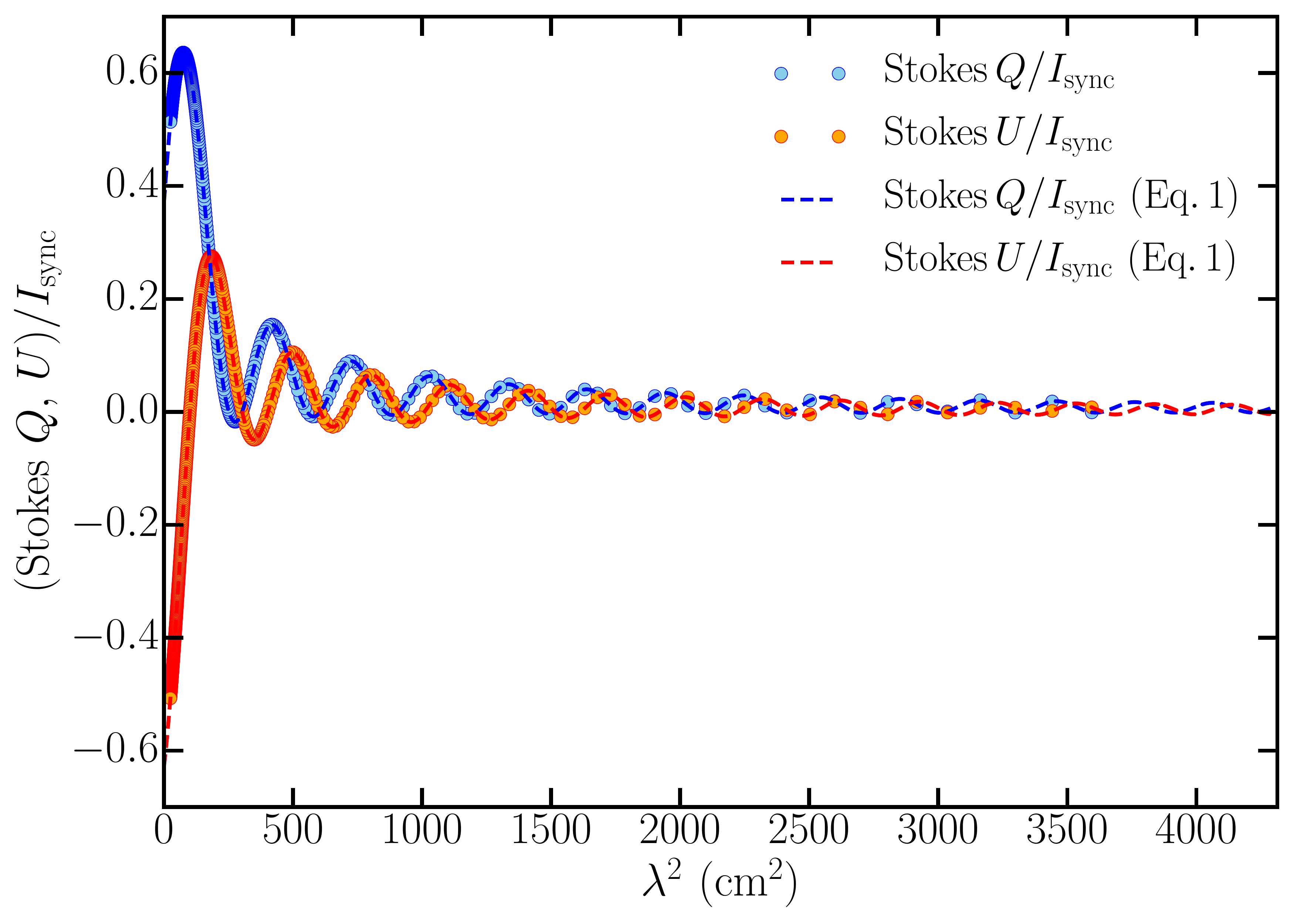}}}&
{\mbox{\includegraphics[width=7cm]{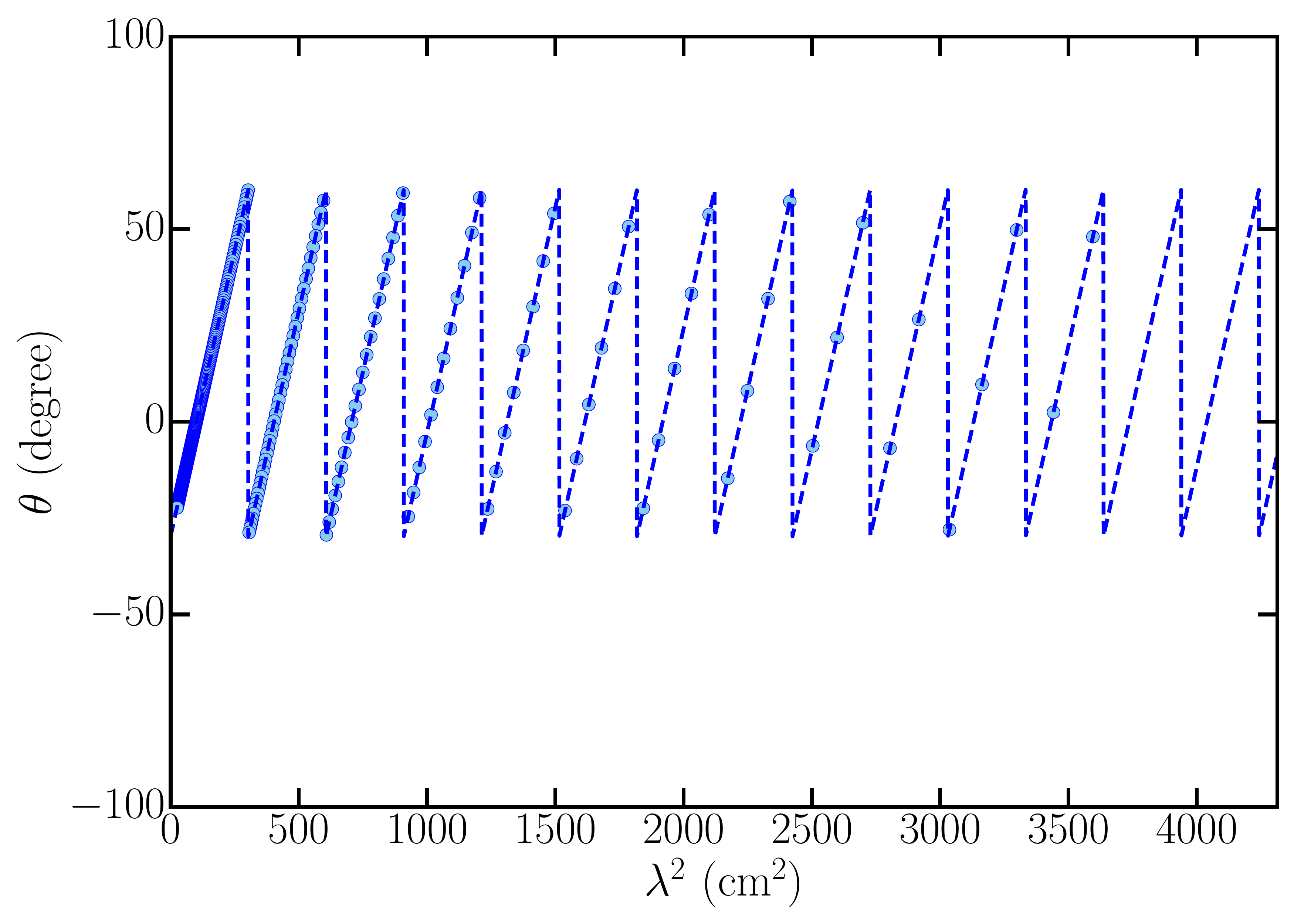}}}\\
\end{tabular}
\caption{Synthetic spectra of Stokes parameters of synchrotron emission
generated by {\tt COSMIC} for a uniform slab. Numerically computed quantities
are shown as the data points and the analytical functions for the linearly
polarized quantities are shown as dashed lines. {\it Top left}: Spectrum of the
total synchrotron flux density (grey points) and the linearly polarized flux
density (blue points). {\it Top right}: Variation of the factional polarization
with $\lambda^2$. {\it Bottom left}: Variation of fractional Stokes $Q$ and
$U$ parameters as a function of $\lambda^2$. {\it Bottom right}: Variation of
the angle of the plane of linear polarization ($\theta$) with $\lambda^2$.}
\label{fig:burn_slab}
\end{figure}

\subsection{Uniform slab model} \label{sec:burn}

A uniform slab, also commonly referred to as a `Burn slab' after
\citet{burn66}, is a synchrotron emitting medium in which the strengths of the
three magnetic field components and the densities of CRE and thermal electrons
are all spatially constant. We generated such a medium on a $512 \times 512
\times 512$~pixel$^3$ mesh and assumed the mesh points to be separated by 1 pc.
Therefore, the physical volume of the simulated uniform medium is $512 \times
512 \times 512$~pc$^3$. The values of the physical quantities used to simulate
a uniform slab are listed in Table~\ref{tab:pars}. We have used CRE density
such that the total synchrotron flux density of the medium is 10~Jy at 1~GHz
(see Section~\ref{sec:sync_em}). We assumed a power-law energy spectrum of the
CREs, so that, the synchrotron emission also follow a power-law frequency
spectrum with spectral index $\alpha = -0.8$ ($I_\nu \propto \nu^\alpha$, where
$I_\nu$ is the intensity at frequency $\nu$).

For such a medium, the complex fractional polarization varies with wavelength
$\lambda$ as \citep{burn66, sokol98},
\begin{equation}
p(\lambda) = p_{\rm int}\,\dfrac{\sin {\rm FD}\,\lambda^2}{{\rm FD}\,\lambda^2}\,
{\rm e}^{2\,i\,\left(\theta_0 + \frac{1}{2}\,{\rm FD}\,\lambda^2 \right)}.
\label{eq:burn}
\end{equation}
Here, $p_{\rm int}$ is the intrinsic fractional polarization of the medium,
after accounting for any frequency-independent beam depolarization of the
synchrotron emission that might be present, for example due to unresolved
turbulent magnetic fields. As there is no beam or random magnetic field in this
model, $p_{\rm int}$ is the same as the maximum fractional polarization $p_{\rm
max}$ given in Eq.~\eqref{eq:pmax}. The expected values of $p_{\rm int}$, FD
and $\theta_0$ are listed in Table~\ref{tab:vals}.

Spectra of the total and linearly polarized synchrotron flux densities, $I_{\rm
sync}$ and $PI$, respectively, obtained from {\tt COSMIC} for a single LOS
through the domain are shown in the top-left panel in Fig.~\ref{fig:burn_slab}.
The top-right and bottom-left panels in Fig.~\ref{fig:burn_slab} show the
fractional Stokes parameters, i.e., $PI/I_{\rm sync}$, and $Q/I_{\rm sync}$ and
$U/I_{\rm sync}$, respectively, as a function of $\lambda^2$, computed from the
data using Eqs.~\eqref{eq:sync_em}, \eqref{eq:compQ}, \eqref{eq:compU} and
\eqref{eq:compPI}. The bottom-right panel show the variation of the
polarization angle, $\theta_0$, with $\lambda^2$. The dashed lines in
Fig.~\ref{fig:burn_slab} show the analytical function given by
Eq.~\eqref{eq:burn}. To check the robustness of our numerical calculations, the
parameters of the analytical model, namely, $p_{\rm int}$, FD and $\theta_0$,
were calculated directly from the {\tt COSMIC} output. Note that the fractional
polarization and the angle of polarization are given by the amplitude and phase
of the complex polarization, and are dependent on the wavelength (e.g.,
Eq.~\ref{eq:burn}). Therefore, $p_{\rm int}$ and $\theta_0$ can either be
determined at $\lambda = 0$, which is unphysical, or when FD and/or
$\sigma_{\rm FD}$ are $0~\rm rad\,m^{-2}$. We therefore computed $p_{\rm int}$
and $\theta_0$ by setting FD in each pixel of the 3-D cube to zero and then
performing the synthetic observations. In Table~\ref{tab:vals}, we present the
expected values of $p_{\rm int}$, FD and $\theta_0$, and compare them with
those computed from the cubes. All the values are in excellent agreement with
the theoretical values.

\begin{table} \centering
 \caption{Expected values of physical parameters for the setup parameters used
in Table~\ref{tab:pars} and values obtained using {\tt COSMIC}.}
  \begin{tabular}{@{}lcccccc}
 \hline
\multicolumn{1}{l}{Parameter}  & 
\multicolumn{1}{c}{~~~~~~~~}&
\multicolumn{2}{c}{Uniform slab}   & 
\multicolumn{1}{c}{~~~~~~~~}&
\multicolumn{2}{c}{IFD}   \\
	  & & Expected & Obtained & & Expected & Obtained \\
 \hline
	  $p_{\rm int}$ & & 0.73 & 0.73 & & 0.176 & 0.174 \\
	  FD ($\rm rad\,m^{-2}$) & & $103.94$  & $103.94$ & & 103.94 & 103.63 \\
	  $\sigma_{\rm FD}$ ($\rm rad\,m^{-2}$) & & 0.0  & 0.0 & & 9.20 & 9.18 \\
	  $\theta_0$ ($^\circ$) & & $150.26$ & $150.25$ & & 150.26 & 150.25\\
 \hline
\end{tabular}
\label{tab:vals}
\end{table}

\subsection{Internal dispersion model} \label{sec:IFD}

The internal Faraday dispersion (IFD) model describes a medium that is
simultaneously synchrotron emitting and Faraday rotating in the presence of
both regular and random magnetic fields, wherein the random field is isotropic,
has Gaussian statistics and is delta-correlated (i.e. there are no correlated
structures in the field). In other words, the correlation length is same as
the length of the mesh point separation. We generated a volume of the same
dimension as in Section~\ref{sec:burn}, but in this case, the magnetic field
strengths of each component in each 1 pc$^3$ pixel were drawn at random from a
Gaussian distribution. The mean value of the Gaussian represents the regular
field strength along the corresponding direction and the standard deviation is
a measure of the strength of the turbulent fields. Values of the different
parameters used to generate the IFD volume are listed in Table~\ref{tab:pars}.

\begin{figure}[t]
\centering
\begin{tabular}{cc}
{\mbox{\includegraphics[width=7.0cm]{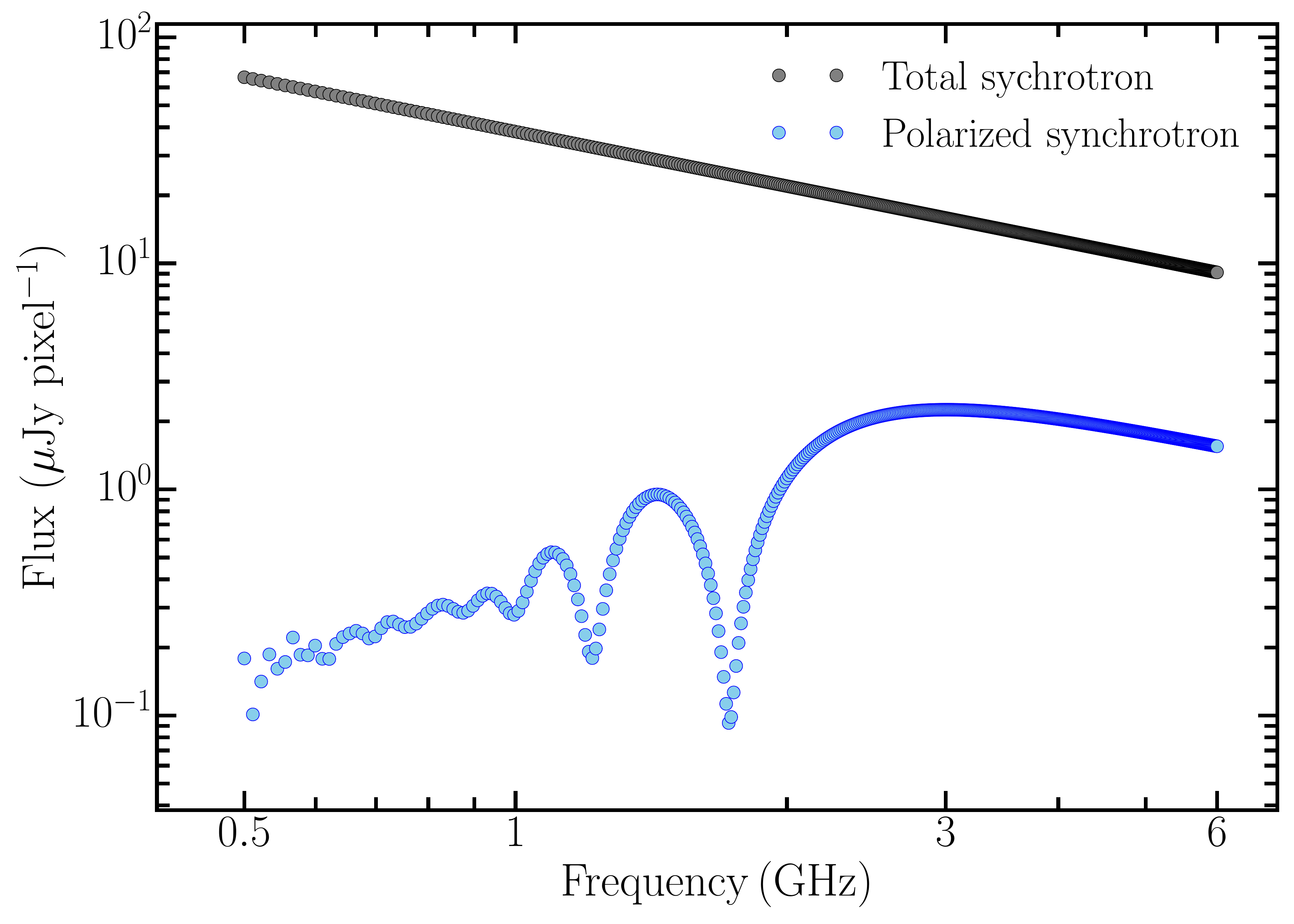}}}&
{\mbox{\includegraphics[width=7.0cm]{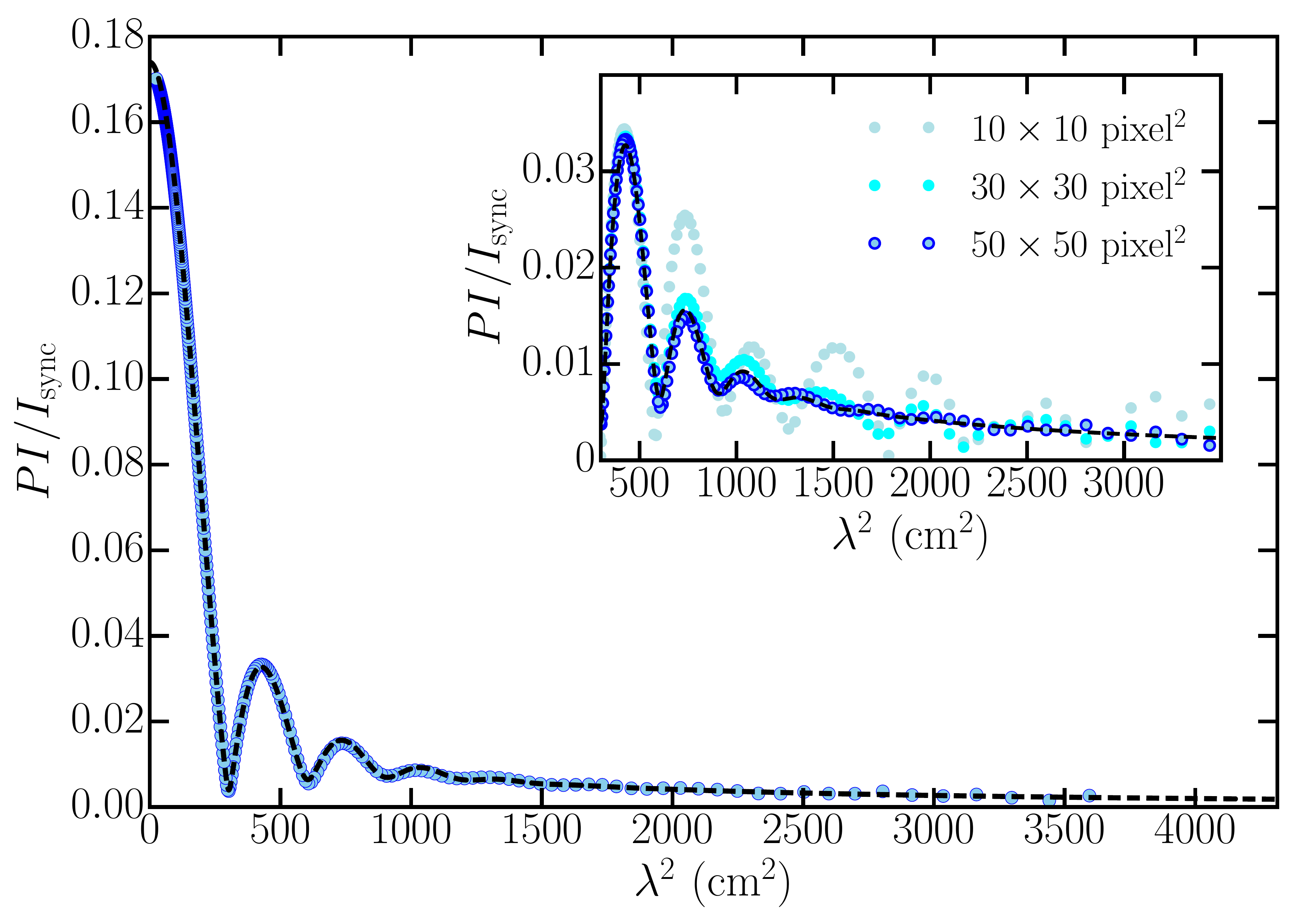}}}\\
{\mbox{\includegraphics[width=7.0cm]{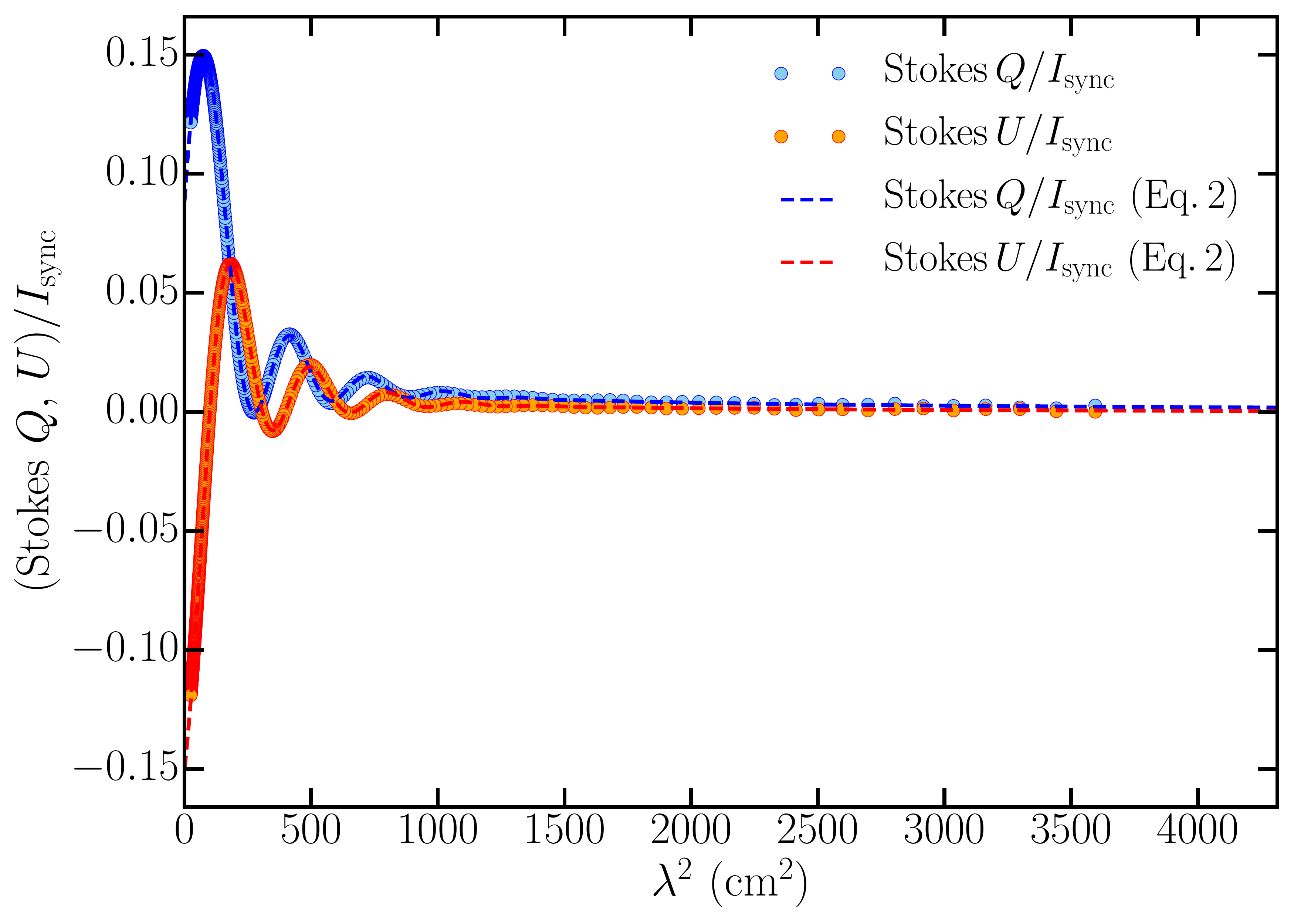}}}&
{\mbox{\includegraphics[width=7.0cm]{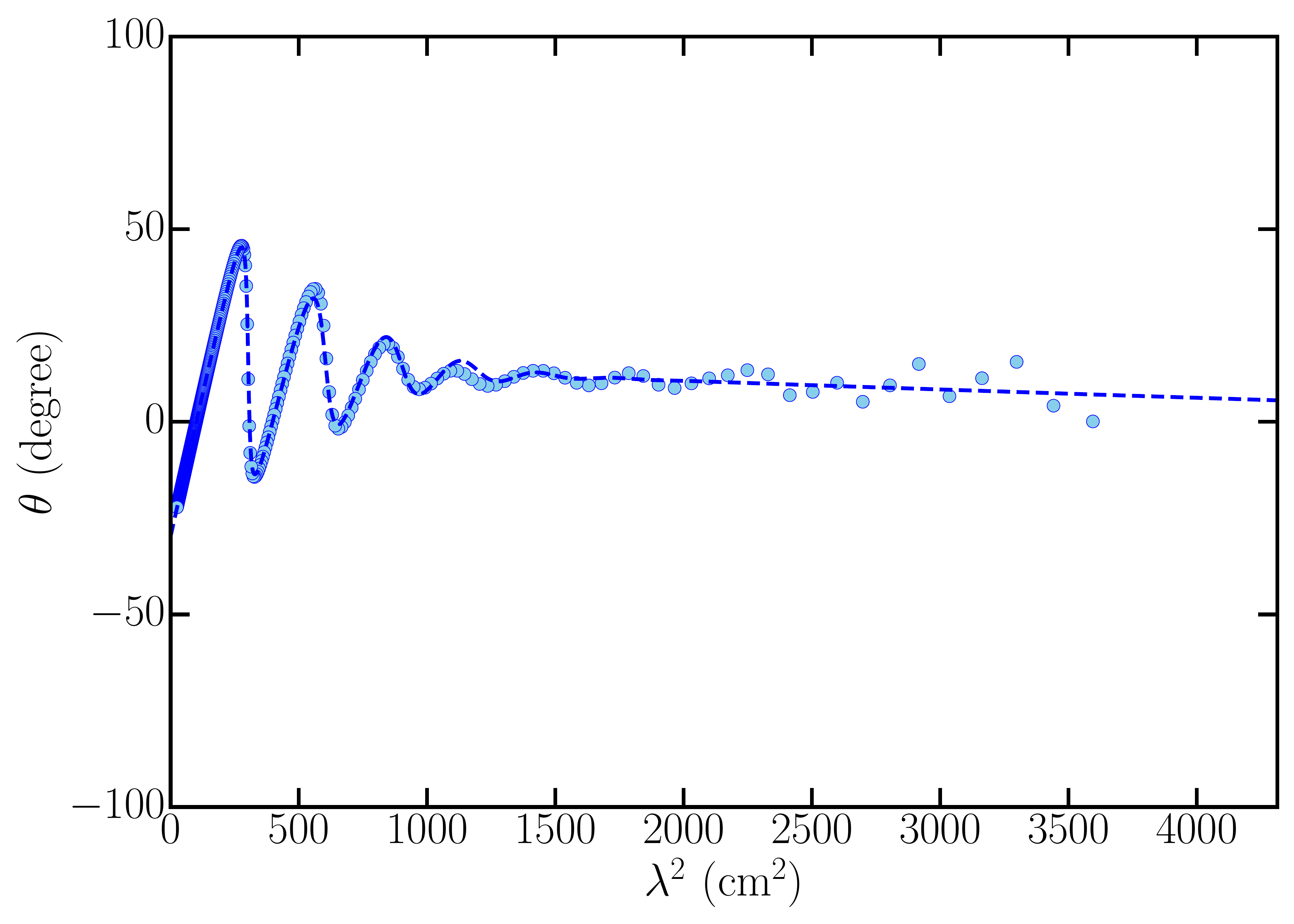}}}\\
\end{tabular}
\caption{Synthetic spectra of Stokes parameters of synchrotron emission
generated by {\tt COSMIC} for an internal Faraday dispersion model. Different
shades of data points in the inset of the top right panel show the fractional
polarization computed by averaging over different areas. All other panels, data
points and lines have the same meaning as in Fig.~\ref{fig:burn_slab}. In the
main plots only the results for averaging over $50\times 50~\rm pixel^2$ are
shown.}
\label{fig:IFD}
\end{figure}

The complex fractional polarization in such a medium varies with
$\lambda$ as \citep{sokol98},
\begin{equation} 
p(\lambda) = p_{\rm int}\,{\rm e}^{2i\theta_0}\,\left(\dfrac{1 \, - \,
\exp[-(2\,\sigma_{\rm FD}^2\,\lambda^4 - 2\,i\,{\rm FD}\,\lambda^2)]}
{2\,\sigma_{\rm FD}^2\,\lambda^4 - 2\,i\,{\rm FD}\,\lambda^2} \right).  
\label{eq:int_disp} 
\end{equation}
Here, $\sigma_{\rm FD}$ is the intrinsic dispersion of FD and is given by
\citep{beck07}, 
\begin{equation}
\sigma_{\rm FD} = 0.812\,\langle n_{\rm e}\rangle \, \sigma_{\rm \|}\,\sqrt{\dfrac{l_0\,L}{f_{\rm V}}},
\label{eq:sigmaRM}
\end{equation}
where $\sigma_{\rm \|}$ is the strength of turbulent magnetic fields along the
line of sight, $l_0$ is the correlation length of the product $n_{\rm
e}\,\sigma_\|$, $L = 512$~pc is the path-length through the magneto-ionic
medium, and $f_{\rm V}$ is the volume filling factor of $n_{\rm e}$. In this
example, since $n_{\rm e}$ is constant, $f_{\rm V}=1$, and $l_0 = 1$~pc as the
random field is delta-correlated.

The intrinsic fractional linear polarization ($p_{\rm int}$) of the synchrotron
emission originating from superposition of regular and isotropic random
magnetic fields, and constant number density of CREs, is given by
\citep{sokol98}, 
\begin{equation}
p_{\rm int} = p_{\rm max}\,\dfrac{\langle B_\perp\rangle^2}{\langle B_\perp\rangle^2 + \sigma_x^2 + \sigma_y^2}.
\label{eq:polfrac}
\end{equation}

For our choice of parameters in Table~\ref{tab:pars}, $p_{\rm int}$,
$\theta_0$, FD and $\sigma_{\rm FD}$ are theoretically expected to be 0.176,
150.26$^\circ$, $103.94~\rm rad\,m^{-2}$ and $9.2~\rm rad\,m^{-2}$,
respectively. In Table~\ref{tab:vals}, we compare these values with those
directly obtained from {\tt COSMIC} output using the same method as described
at the end of the previous section. The values agree well with each other. In
Fig.~\ref{fig:IFD}, we show the variation of polarization fraction with
$\lambda$ computed from synthetic observations of the simulated volume and they
agree well with the analytic function given in Eq.~\eqref{eq:int_disp}. 

We should point out that, in this case, we determined the frequency spectra of
the polarization parameters shown in Fig.~\ref{fig:IFD} by averaging over $50
\times 50$ pixels$^2$ in the $I$, $Q$ and $U$ images. This is done to ensure
that there are a sufficient number of pixels to capture the Gaussian statistics
generated by standard routines for generating random numbers in Python. For
smaller averaging areas, significant deviations of the synthetic data from the
analytic function is observed, especially towards longer wavelengths. A
comparison of the variation of fractional polarization with $\lambda^2$
obtained from synthetic observations by averaging over different areas is shown
in the inset of the top right-hand panel of Fig.~\ref{fig:IFD}. At shorter
wavelengths, all synthetic observations agrees well with the analytic function.

\section{Applying {\tt COSMIC} to MHD simulations of a turbulent medium}
\label{sec:mhd}

The magnetic fields and free electrons in the ISM have more complicated
distributions than can be described by simple delta-correlated Gaussian
statistics, for example they can have spatial correlations in their structure
\citep[see e.g.,][]{haver08, mao15, makar18}. Therefore, investigating the
broad-band properties of the linearly polarized synchrotron emission from
realistic MHD simulations is required, especially with several on-going and
upcoming polarization surveys of the Milky Way's diffuse emission.

Here we apply {\tt COSMIC} to MHD simulations of isothermal, compressible and
transonic turbulence in the ISM from Burkhart {\it et al.} \cite{burkh09,
burkh13}. Polarization observations of the Galactic plane and Galactic H{\sc i}
21~cm observations of the warm gas have confirmed the transonic nature of
turbulence in the ISM \citep{gaens11, herron2016, koley19}. The simulation set
up is similar to that of past works \citep{kowal07, burkh09, bialy17}. We refer
to these works for the details of the numerical set-up and we only provide a
short overview here. The simulation used here is a 3-D turbulent box of
isothermal compressible MHD turbulence with a resolution $512^3$, and each mesh
point is separated by 1~pc. The code is a third-order accurate essentially
nonoscillatory scheme which solves the ideal MHD equations in a periodic box
with purely solenoidal driving of the flow at a scale $2.5$ times smaller than
the domain size, i.e., $\sim200$~pc. The simulation has two control parameters:
the sonic Mach number ($M_{\rm s} = v/c_{\rm s}$, where $v$ is the flow
velocity and $c_{\rm s}$ is the sound speed) and Alfv\'enic Mach number
($M_{\rm A} = v/v_{\rm A}$, where $v_{\rm A}$ is the Alfv\'en speed). In these
simulations, we used $M_{\rm s} \approx 2.0$ and $M_ {\rm A} \approx 0.7$. The
magnetic field consists of a regular background field ($B_{\rm ext}$) and a
turbulent field ($b$), i.e., $B = B_{\rm ext} + b$ with the magnetic field
initialized along a single preferred direction. In the choice of our coordinate
system, the regular field is along the $x$-direction and has a strength of
$10\,\mu$G. This regular field will not contribute to Faraday depth as the LOS
is along the $z$-direction.

\begin{figure}
\centering
\begin{tabular}{c}
{\mbox{\includegraphics[width=6.5cm]{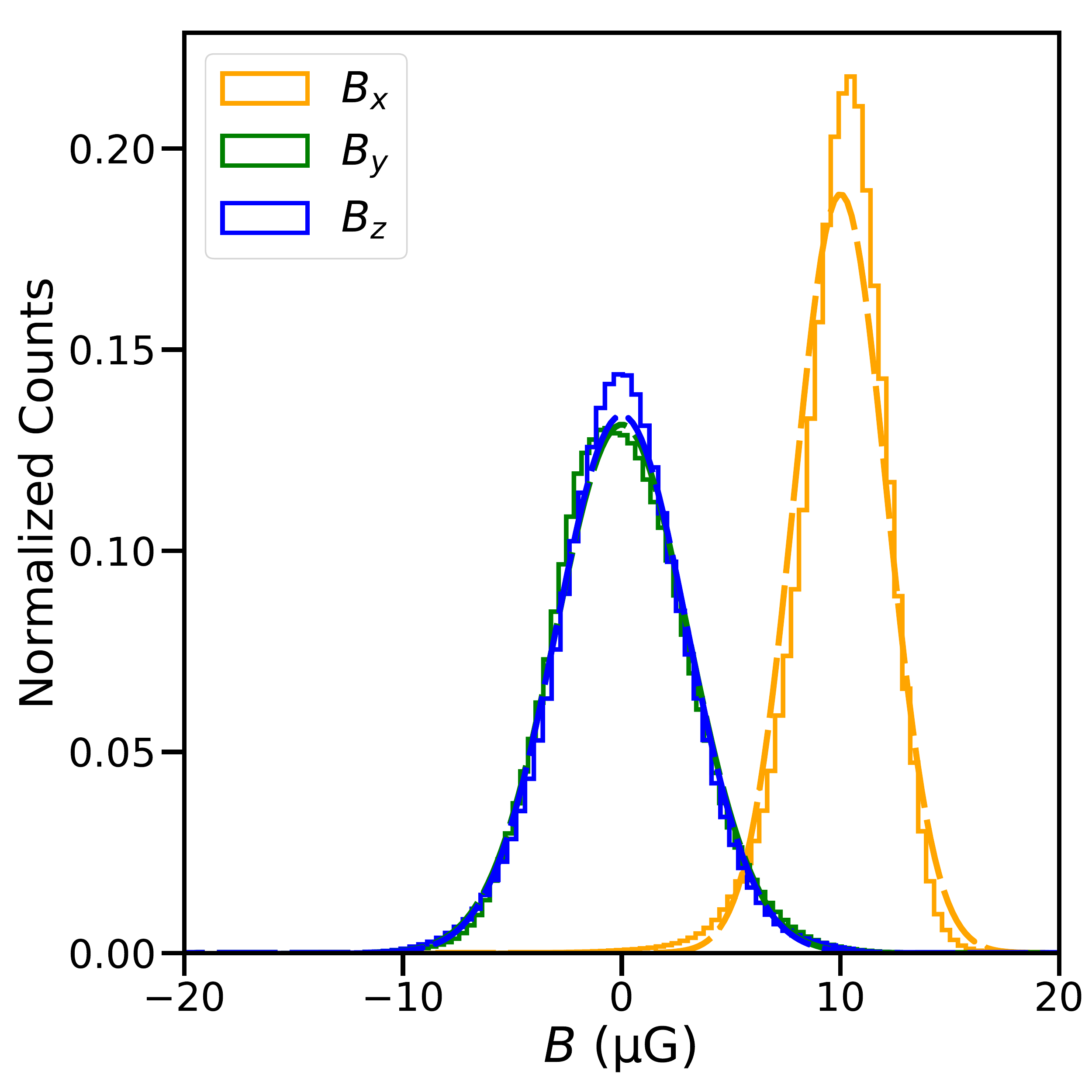}}}\\
\end{tabular}
\caption{Distribution of strengths of the three components of the magnetic
field from the isothermal, compressible turbulence simulations used for our
analysis are shown as the histograms. The dashed lines show Gaussian
distributions with mean and standard deviation computed from the corresponding
field component.}
\label{fig:bcomp}
\end{figure}

The distribution of the strengths of the three magnetic field components over
the entire simulation volume is shown in Fig.~\ref{fig:bcomp}. The
corresponding dashed lines are for equivalent Gaussian distributions computed
from the mean and standard deviation of the corresponding field component. The
distribution of $B_x$ shows deviation from a Gaussian distribution, while that
for $B_y$ and $B_z$ agrees well with Gaussian distributions. We must stress,
although the distributions of the magnetic field components closely resembles a
Gaussian distribution, spatially they are correlated on the driving scale of
turbulence in these simulations, unlike the delta-correlated Gaussian fields
described in Section~\ref{sec:IFD}. For details regarding their structural and
statistical properties we refer interested readers to Burkhart {\it et al.}
\citep{burkh09, burkh13}. To summarize, physical specifications of the
simulations are as follows: (1) Physical size of the simulation volume is $512
\times 512 \times 512$~pc$^3$. (2) Mesh resolution of the simulation is $1
\times1 \times1$~pc$^3$. (3) The mean magnetic field strengths are: $\langle
B_x \rangle=10\,\mu$G and $\langle B_y \rangle = \langle B_z \rangle \approx
0\,\mu$G; and the three components have dispersions $\sigma_x = 2\,\mu$G,
$\sigma_y = 3\,\mu$G and $\sigma_z = 3\,\mu$G. 
Thus, the ratio of regular to turbulent field strengths in this
simulation is $\sim2$. We would like to emphasize that this regular field,
local to the simulated volume, would only contribute to the polarized intensity
of the synthetic observations, unlike those observed in external galaxies.
Polarization measurements in external galaxies are performed with comparatively
lower spatial resolution. Therefore, they are are mostly sensitive to the
ordered component of the magnetic field within the beam, or to the large-scale
regular fields, and therefore in external galaxies, the ordered fields are
found to be about three times weaker than the turbulent fields
\citep{fletc10}.

\begin{figure}
\centering
\begin{tabular}{cc}
{\mbox{\includegraphics[height=6.5cm]{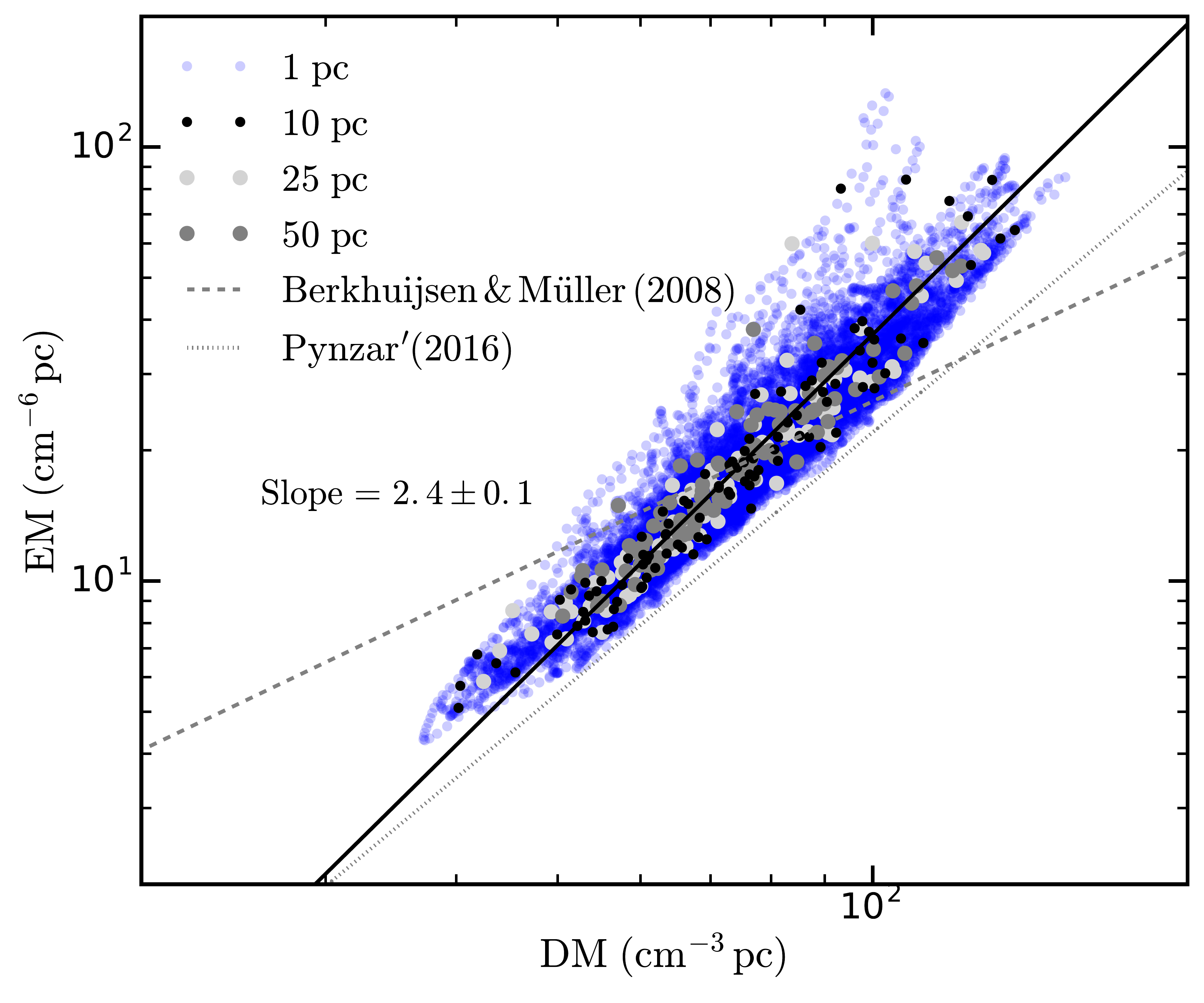}}} &
{\mbox{\includegraphics[width=6.5cm]{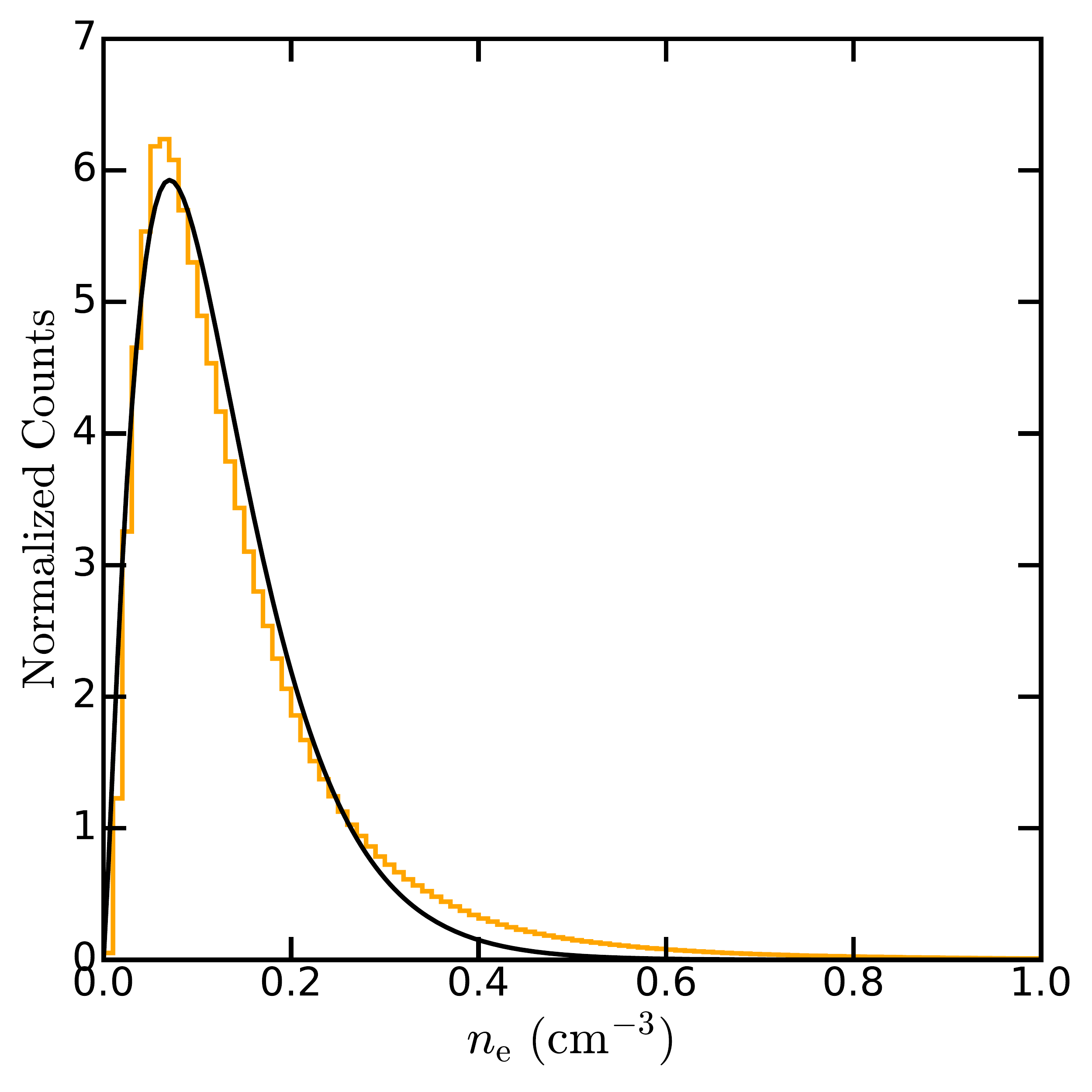}}}\\
\end{tabular}
\caption{{\it Left}: Variation of emission measure (EM) as a function of
dispersion measure (DM) for an ionization fraction of 0.5 of the simulation
volume. The different shades for the symbols represents different averaging
scales. The black line shows the best-fit, and the grey dashed and dotted lines
are from \citet{berkh08} and \citet{pynzar16}, respectively. {\it Right}:
Distribution of free electron density $n_{\rm e}$ in the simulation volume. The
solid black line shows the best-fit gamma function representation of the
distribution.}
\label{fig:em_dm}
\end{figure}

As the MHD simulation used in this work is isothermal and in thermal
equilibrium, the ionization fraction, $f_{\rm ion}$, of neutral hydrogen is
assumed to be constant throughout the volume and can be chosen to compute the
free electron density $n_{\rm e}$ (see Appendix~\ref{sec:FaradayRotation}).
Increasing $f_{\rm ion}$ increases $n_{\rm e}$ and thereby the amount of
Faraday rotation ($\rm FD_{cell}$) in each cell of the MHD cube, which
increases the amount of Faraday depolarization through the entire LOS. To find
physically motivated value for $f_{\rm ion}$, we produced maps of emission
measure (${\rm EM} = \int n_{\rm e}^2\,{\rm d}l$) and dispersion measure (${\rm
DM} = \int n_{\rm e}\,{\rm d}l$) for various values of $f_{\rm ion}$ and
compared EM vs. DM by averaging over different areas \citep{schni08phd}. In the
left-hand panel of Fig.~\ref{fig:em_dm} we show the variation of EM as a
function of DM. The slope of the best-fit line (shown as the black line) is
found to be $2.4\pm0.1$ indicating that the simulation used here is strongly
clumpy with low $f_{\rm V}$. We varied $f_{\rm ion}$ such that the amplitude
roughly matches with observations of the Galactic ISM. For example,
\citet{berkh08} found a slope of 1.15 for the EM vs. DM relation for the
diffuse ionized gas around the Solar neighbourhood with mean $f_{\rm V}=0.08$.
In denser and clumpier ISM with $f_{\rm V} < 0.06$, the slope of the EM vs. DM
relation was estimated to be 2 by \citet{pynzar16}.

For the simulations used here, the amplitude of the EM vs. DM relation agrees
well with that observed in our Galaxy for $f_{\rm ion} = 0.5$ (see
Fig.~\ref{fig:em_dm}). In the right-hand panel of Fig.~\ref{fig:em_dm}, we show
the distribution of $n_{\rm e}$ in the simulation volume for $f_{\rm ion} =
0.5$. Here, the median $n_{\rm e}$ is $0.11~{\rm cm^{-3}}$ and it has a maximum
value of $3.5~{\rm cm^{-3}}$. Such values of $n_{\rm e}$ are typically observed
within Galactic latitudes $\pm 15^\circ$ \citep{ne2001II, pynzar16}. The
distribution of $n_{\rm e}$ is well approximated by a Gamma distribution (solid
black line) with shape parameter $\approx 2.25$ and inverse scale parameter
$\approx0.06$, i.e., an expectation value for $n_{\rm e} \approx 0.13~\rm
cm^{-3}$.

For the values of $n_{\rm e}$ and $B_\|$ in these MHD simulations, $<0.3\%$ of
the $1~{\rm pc^3}$ mesh points have $|{\rm FD_{cell}}| > 3~{\rm rad\,m^{-2}}$
and $<1\%$ mesh points have $|{\rm FD_{cell}}| > 2~{\rm rad\,m^{-2}}$. That
means, within the cell, $<0.3\%$ and $<1\%$ of the mesh points undergo Faraday
depolarization of $>20\%$ and $>10\%$, respectively, at frequencies below
$\sim0.5$~GHz. Calculations of the polarization parameters in {\tt COSMIC}
assumes that the Faraday depolarization in a single mesh point is negligible
(see Appendices~\ref{sec:FaradayRotation}, \ref{sec:polInt} and
\ref{sec:limitations}). Therefore, for these simulations {\tt COSMIC} can be
safely applied to frequencies above 0.5~GHz.

The following options were used to compute the synthetic spectra of $I_{\rm
sync}$, $PI$, Stokes $Q$ and $U$ in the rest of this paper. (1)~As the
simulations do not contain cosmic rays and the typical CRE propagation lengths
are expected to be comparable to or larger than the size of the simulation box
(see Appendix~\ref{sec:sync_em}), we assume that the CREs are uniformly
distributed throughout the 3-D volume of the simulation. (2)~The CREs follow a
power-law energy spectra with the same energy index throughout the volume.
Thus, the synchrotron intensity spectral index ($\alpha$) is constant, both
spatially and with frequency, with $\alpha = -0.8$ defined as $I_{\rm
sync}\propto \nu^{\alpha}$. (3)~$n_{\rm CRE}$ is normalized such that the total
synchrotron flux density at 1~GHz integrated over the entire volume is 10~Jy
(see Appendix~\ref{sec:sync_em}). Note that the amplitude of the frequency
spectrum of intrinsic emissivities of the total synchrotron emission, and
Stokes $Q$ and $U$ parameters can be scaled depending on the unknown $n_{\rm
CRE}$ without affecting the relative frequency variation. (4)~The ionization
fraction is constant throughout the volume with $f_{\rm ion} = 0.5$. 

In the following analyses, we have not added any systematic effects arising
from combining data observed using multiple telescope receivers into one large
bandwidth or telescope noise. Also, the synthetic observations are not sampled
using $u - v$ coverage of an interferometer and therefore mimicks observations
performed using a single dish radio telescope. This is because we want to
investigate what can be learnt about a medium from RM synthesis under ideal
conditions.

\section{Synthetic observations of MHD simulations} \label{sec:syntheticObs}

We compute synthetic spectra of the total and linearly polarized quantities
that describe the synchrotron emission of the volume in the frequency range 0.5
to 6~GHz divided into 500 frequency channels. The choice of lowest frequency is
determined by the need to minimize Faraday depolarization within each 3-D mesh
point (see Section~\ref{sec:mhd}) and the highest frequency is chosen such that
the synthetic observations are sensitive to broad structures in the Faraday
depth space and roughly corresponds to the high frequency end of on-going
spectro-polarimetric surveys of the diffuse Galactic synchrotron emission
\citep[e.g.][]{jones18}.

\begin{figure*}[t]
\centering
\begin{tabular}{cc}
{\mbox{\includegraphics[width=6.5cm]{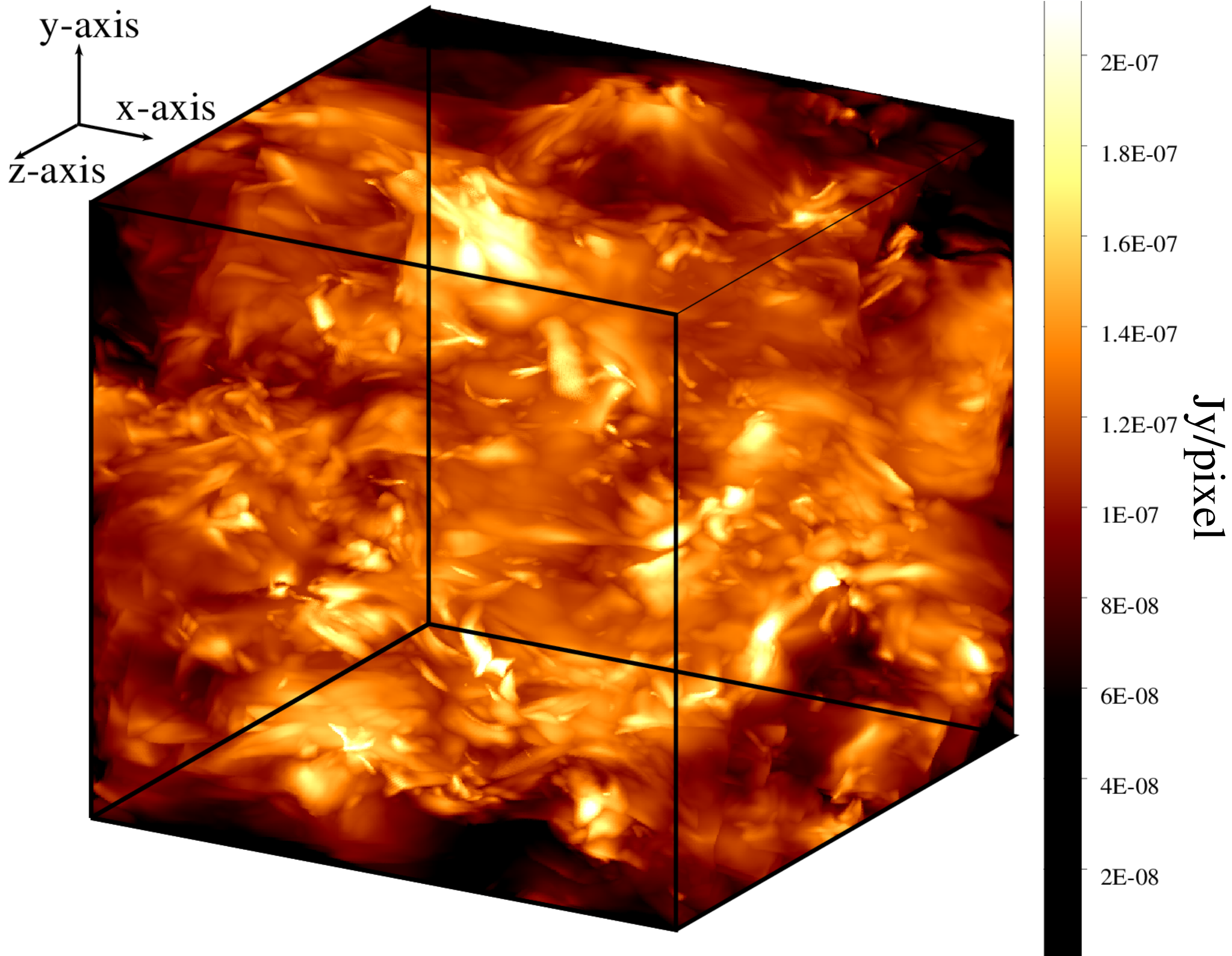}}} &
{\mbox{\includegraphics[width=7.0cm, trim=0cm 0cm 0cm 0.1cm, clip]{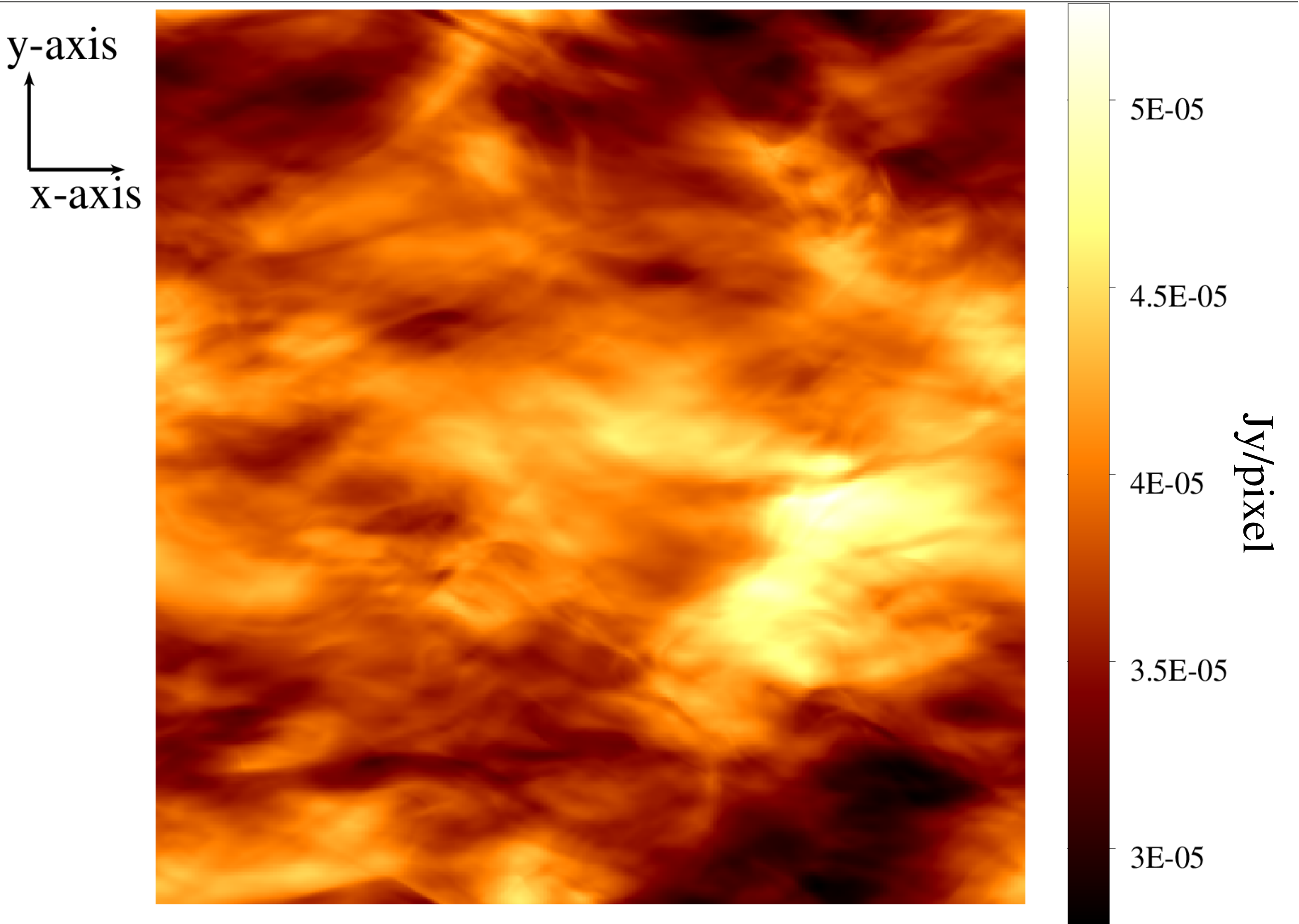}}} \\
\end{tabular}
\caption{{\it Left}: 3-D synchrotron emissivity per pixel at 1 GHz. Synchrotron
emissivity was computed assuming a constant density of CREs throughout the 3-D
volume. {\it Right}: 2-D synchrotron intensity at 1 GHz integrated along
$z$-axis.}
\label{fig:sync}
\end{figure*}

\begin{figure*}
\begin{centering}
\begin{tabular}{ccc}
{\mbox{\includegraphics[width=5.0cm, trim=4.5cm 15cm 5cm 1cm, clip]{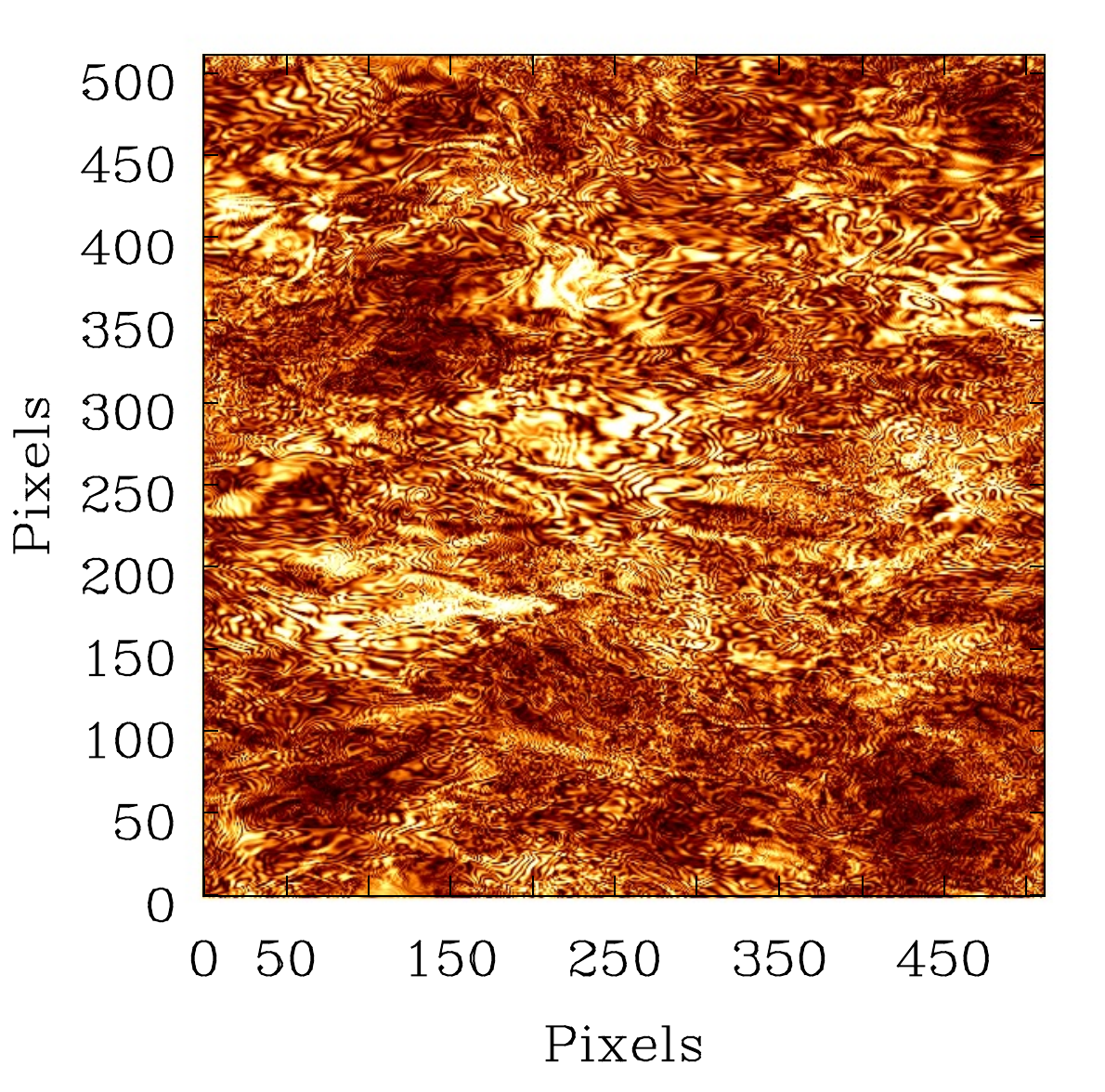}}} &
{\mbox{\includegraphics[width=5.0cm, trim=4.5cm 15cm 5cm 1cm, clip]{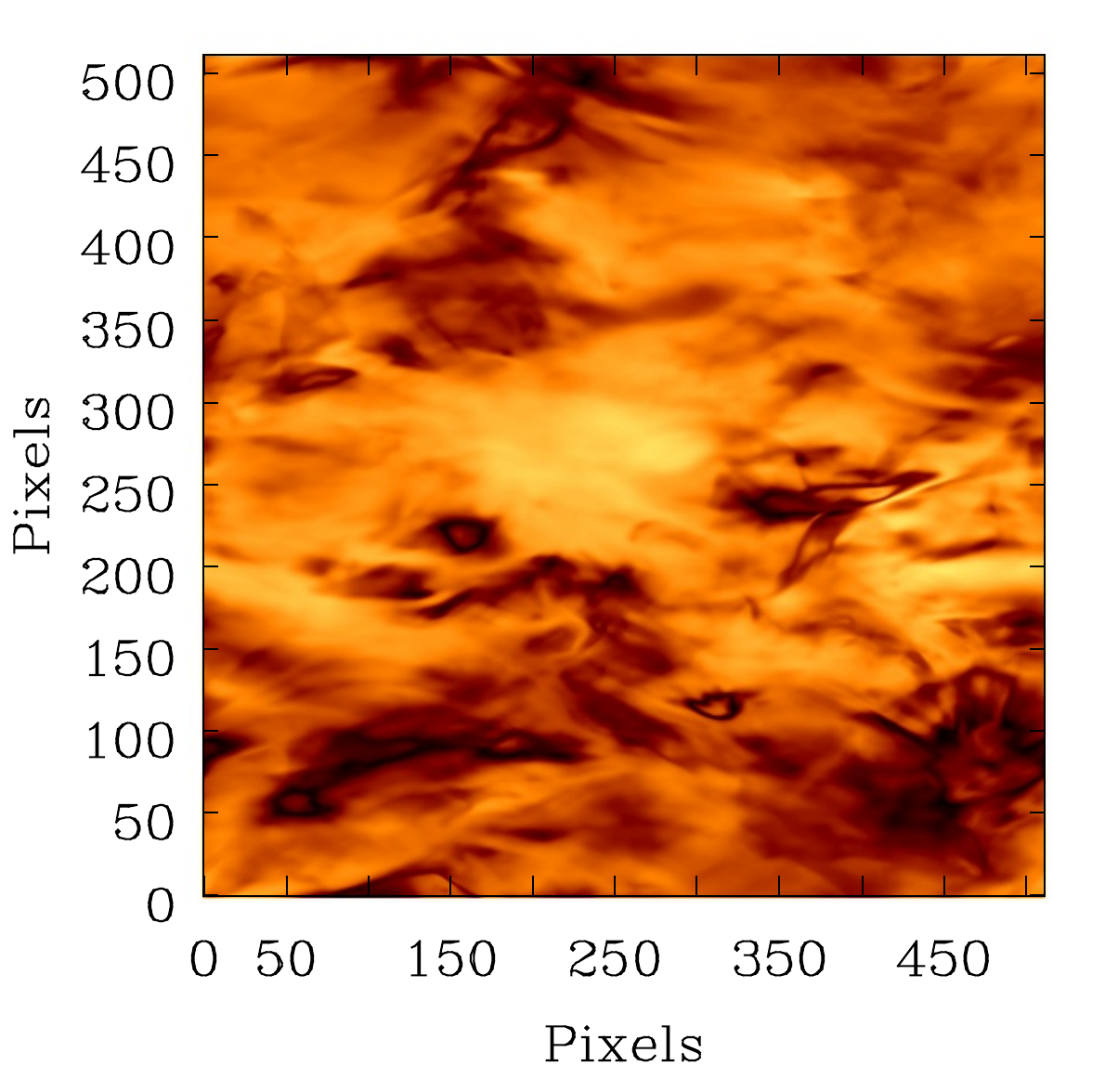}}} &
{\mbox{\includegraphics[height=4.8cm, width=6.3cm]{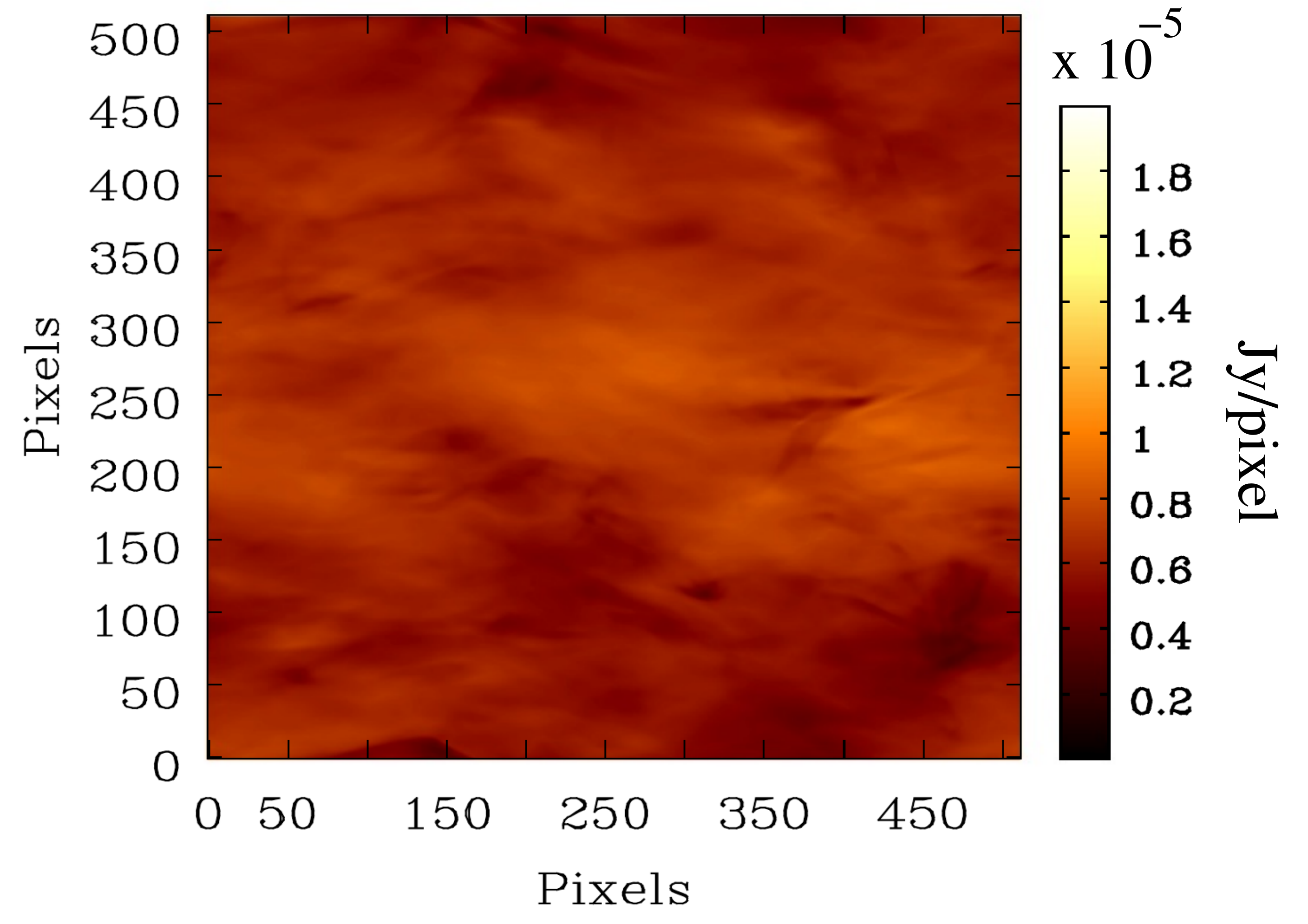}}} \\
\end{tabular}
\end{centering}
\caption{Linearly polarized intensities at 0.5 GHz (left), 1.5 GHz (middle) and
5 GHz (right) including the effects of Faraday depolarization.}
\label{fig:polI_mhd}
\end{figure*}

In the left-hand panel of Fig.~\ref{fig:sync}, we show the 3-D total
synchrotron emissivity of the simulated volume at 1 GHz. The right-hand panel
shows the projected 2-D map of the total synchrotron intensity ($I_{\rm sync}$)
obtained by integrating the 3-D cube in the left-hand panel along the $z$-axis,
our LOS axis.  Since we have assumed a constant density of CREs, all of the
structure in this image arises due to the variations of synchrotron emissivity
caused by fluctuations of the magnetic field component in the plane of the sky,
both along and across the LOS.

In Fig.~\ref{fig:polI_mhd}, we show the polarized intensities at 0.5, 1.5 and 5
GHz. Due to frequency dependent Faraday depolarization, the polarized emission
develops small-scale structures at lower frequencies. This is due to
fluctuations in $B_\|$ along and across the lines of sight.  For frequencies
below $\sim3$~GHz, where most of the polarized surveys of Galactic diffuse
emission have been performed \citep[e.g.,][]{duncan97, wolle06, testo08,
wolle19}, the polarized emission shows some `canal-like' small scale structures
\citep[e.g.][]{shuku03}. Due to severe Faraday depolarization at 0.5~GHz, the
polarized emission show structures on scales of a few pixels. The maps of
Stokes $Q$ and $U$ parameters also show a similar trend in their structural
properties with frequency. This underlines the challenge of combining
interferometric observations of diffuse emission at different frequencies. The
smooth, diffuse structure in Stokes $Q$ and $U$ at the highest frequencies will
be filtered out from the data. This does not happen at lower frequencies
because different sightlines undergo different amounts of Faraday
depolarization, leading to rapid variations on small scales. This issue cannot
be overcome even if frequency-scaled interferometer arrays are used to match
the $u-v$ coverage at different frequencies.

\begin{figure*}[t]
\centering
\begin{tabular}{cc}
{\mbox{\includegraphics[width=7.0cm]{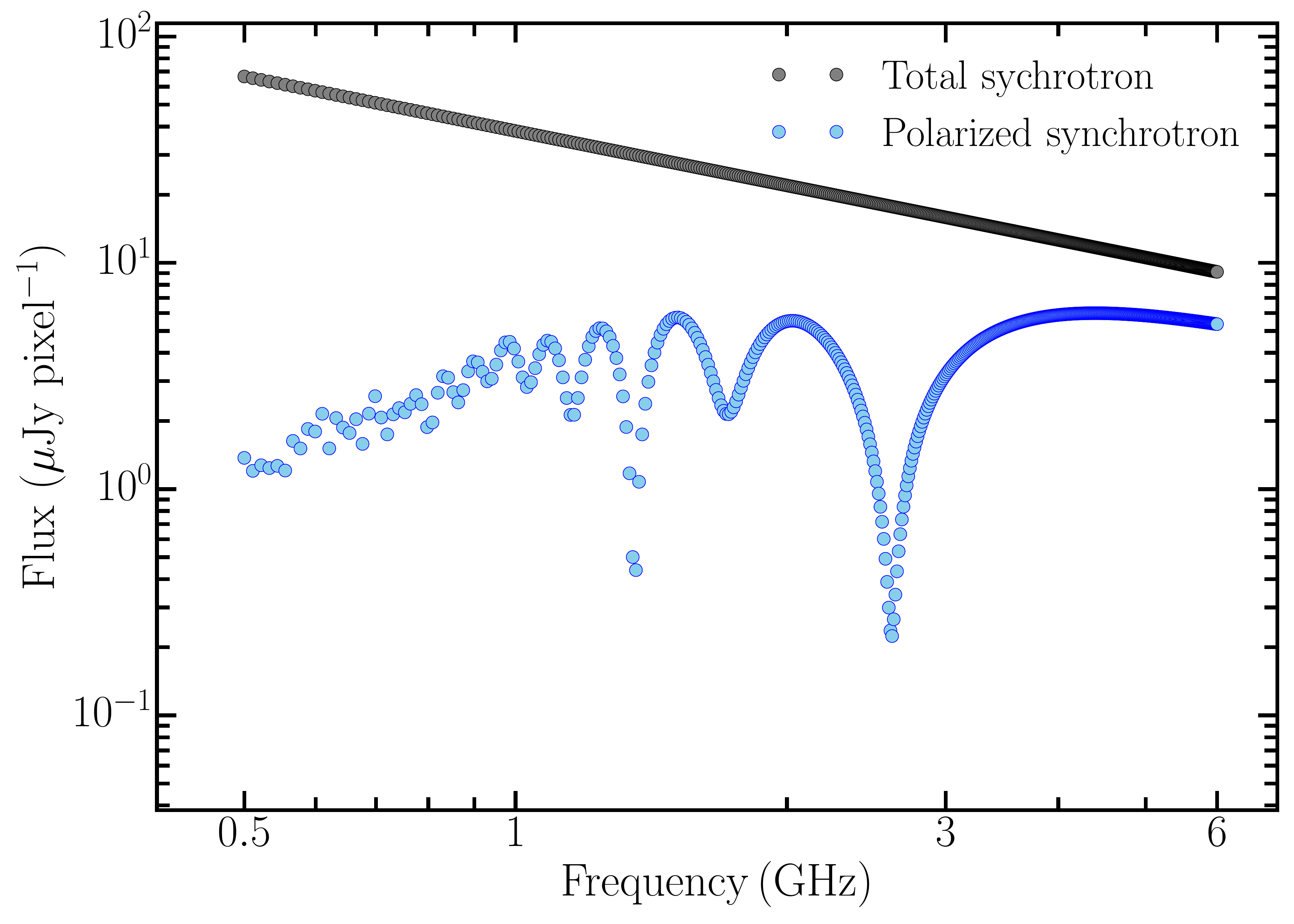}}} &
{\mbox{\includegraphics[width=7.0cm]{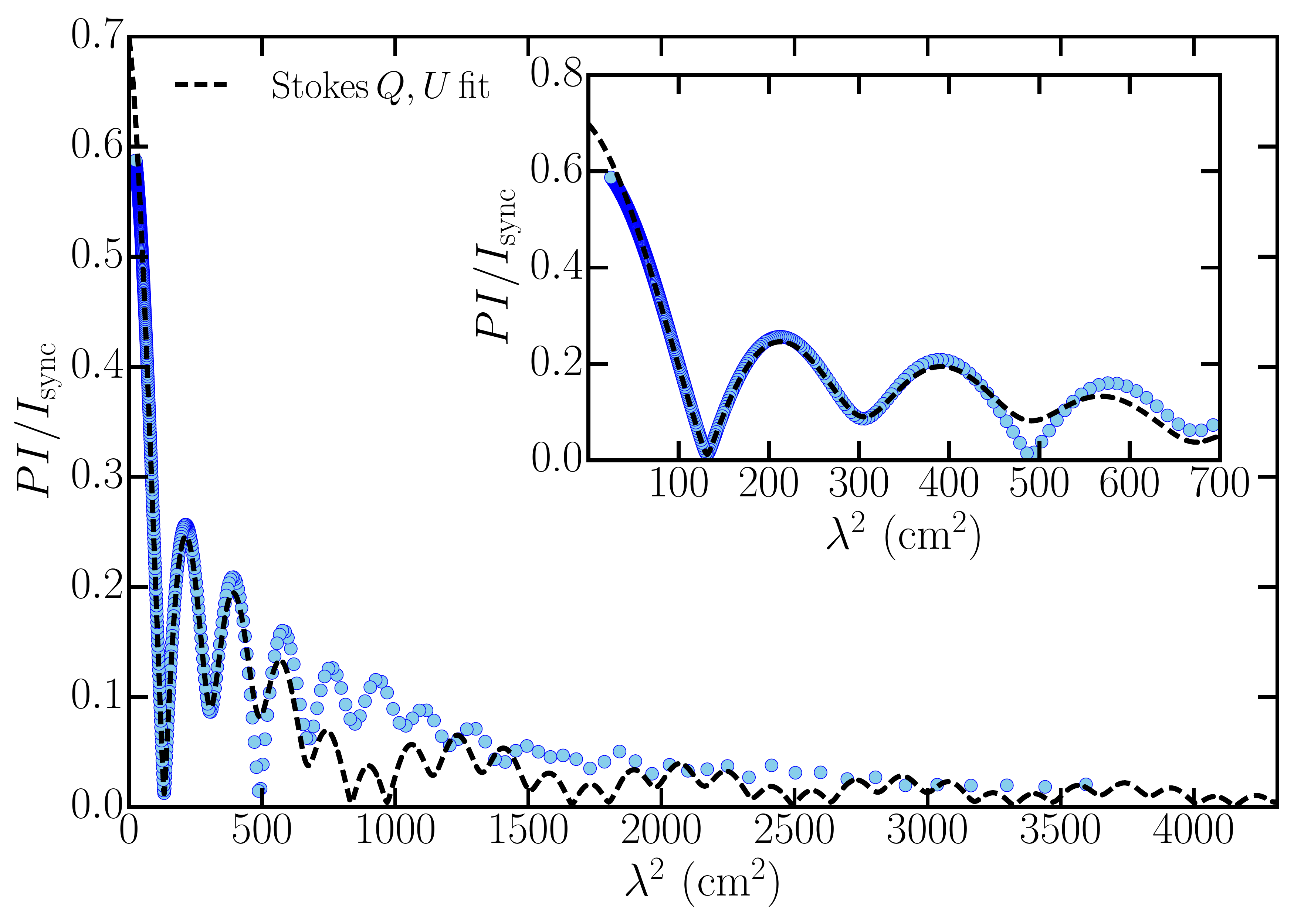}}} \\
{\mbox{\includegraphics[width=7.0cm]{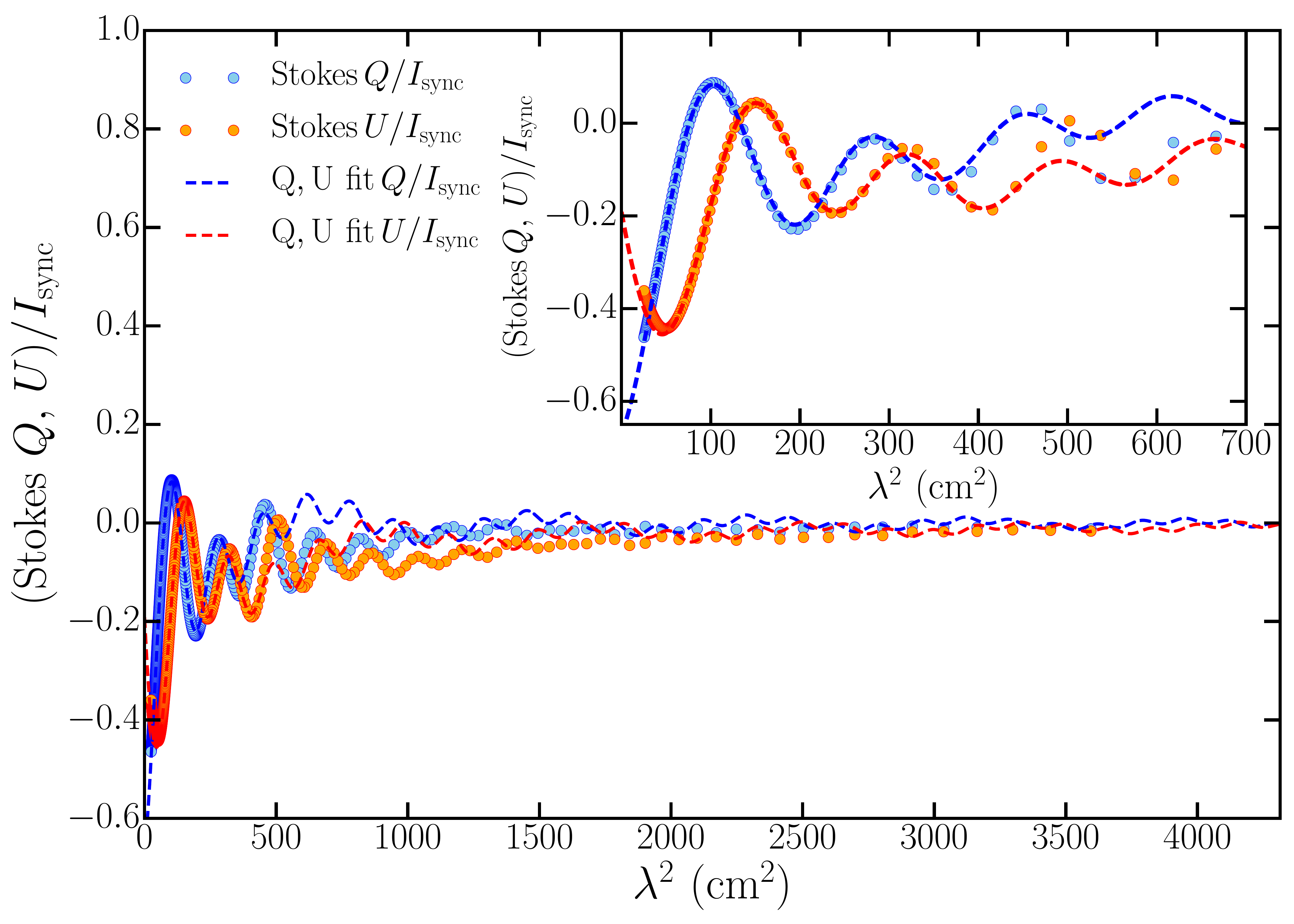}}} &
{\mbox{\includegraphics[width=7.0cm]{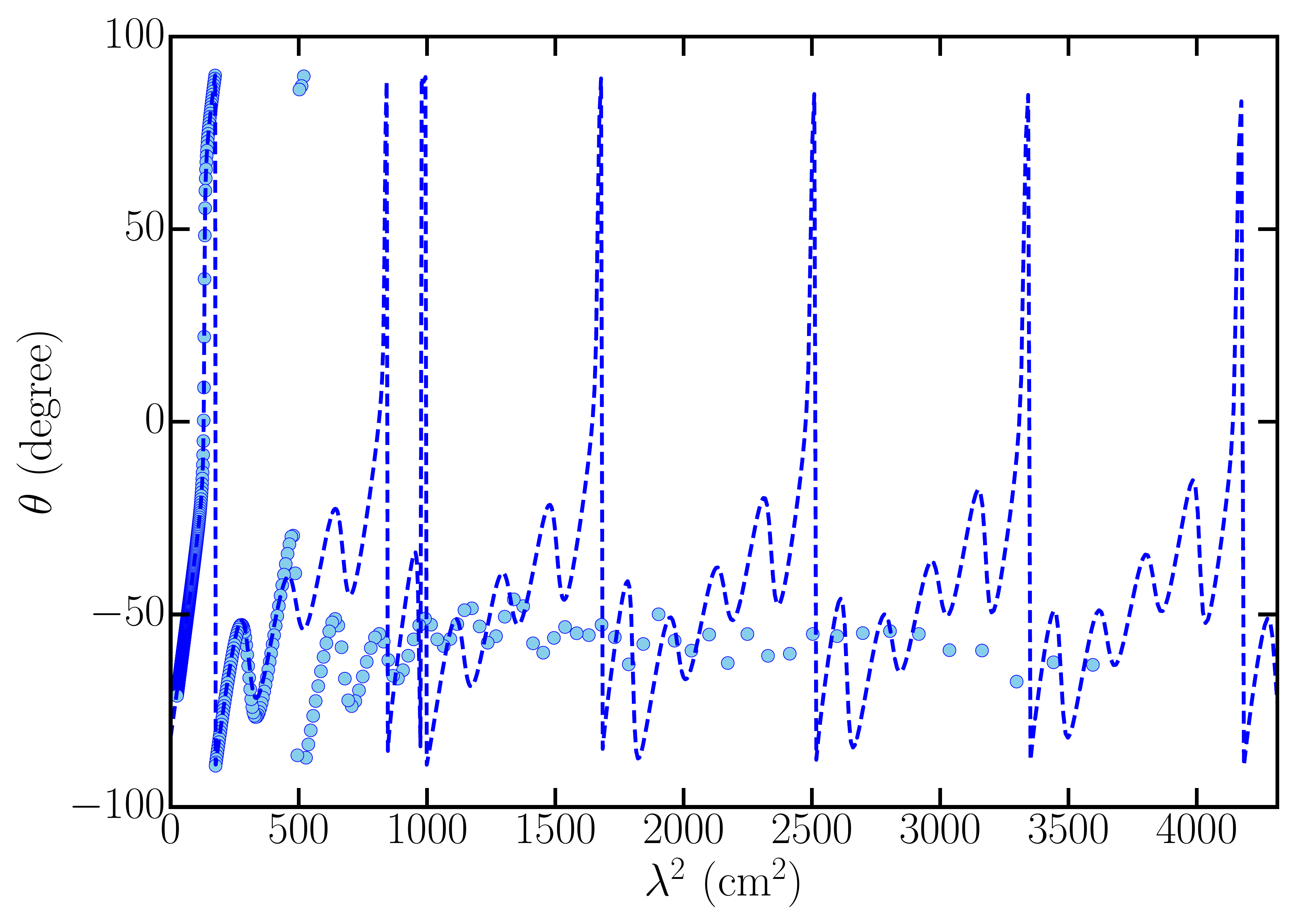}}} \\
\end{tabular}
\caption{Synthetic spectra of Stokes parameters of synchrotron emission
generated by {\tt COSMIC} from MHD simulations. Here we have used $\alpha =
-0.8$ and assumed optically thin synchrotron emission. Quantities computed from
{\tt COSMIC} are shown as the data points. {\it Top left}: Spectrum of the
total synchrotron intensity is shown as the grey points and the linearly
polarized intensity is shown as the blue points. {\it Top right}:
Variation of the factional polarization with $\lambda^2$. {\it Bottom left}:
Variation of Stokes $Q/I_{\rm sync}$ and $U/I_{\rm sync}$ parameters as a
function of $\lambda^2$. {\it Bottom right}: Variation of the angle of the
plane of linear polarization ($\theta$) with $\lambda^2$. All the dashed lines
represents the best-fit obtained from Stokes $Q,U$ fitting. The insets shows
the wavelength regime below $700$~cm$^2$ within which successful fits were
obtained (see text for details).}
\label{fig:mhd_spectrum}
\end{figure*}

\subsection{Broad-band spectra of Stokes parameters} \label{sec:qufit}

Fig.~\ref{fig:mhd_spectrum} shows the synthetic spectra of the total
synchrotron emission and parameters of linearly polarized synchrotron emission
between 0.5 and 6 GHz. All the quantities are computed by averaging over a
randomly chosen $30\times30$ pixel$^2$ region corresponding to a spatial scale
of $30\times30$ pc$^2$. The polarized emission at lower frequencies ($< 1$~GHz)
show strong frequency-dependent variations. This implies that at low
frequencies, the small-scale polarized structures seen in the left-hand panel
of Fig.~\ref{fig:mhd_spectrum} are also expected to vary strongly with slight
changes in frequency.

Although Stokes $Q,U$ fitting is not the main focus of this work, we
none-the-less present in brief the result of fitting the synthetic data shown
in Fig.~\ref{fig:mhd_spectrum}. Stokes $Q,U$ fitting was performed using analytical
functions for several types of depolarization models described in
\citet{sokol98} and \citet{sulli17} and their linear combinations. In our
Stokes $Q,U$ fitting routine, we allow combination of a maximum of three
depolarization models and use the corrected Akaike information criteria
\citep{hurvi89, cavanaugh97} to choose the best-fit model. Because of the
strong variation of the Stokes $Q$ and $U$ parameters in the longer wavelength
regime (in this case roughly $>700$~cm$^2$ corresponding to frequencies
$<1.1$~GHz), the combination of three depolarization models is not sufficient
to fit the Stokes $Q$ and $U$ parameter spectra for the entire wavelength range
considered here. However, for wavelengths below $700$~cm$^2$, fits converged
successfully (inset in Fig.~\ref{fig:mhd_spectrum}). For the synthetic data in
Fig.~\ref{fig:mhd_spectrum}, the best-fit was obtained with a linear combination of
three internal Faraday depolarization components (Eq.~\ref{eq:int_disp}). In
fact, none of the regions we have investigated in these simulations could be
fitted by a single depolarization model, in contrast to the simulated volumes
in Section~\ref{sec:benchmark}. This implies that spatially correlated
distributions of the magnetic fields and/or thermal electron densities, as in
these MHD simulations, could possibly give rise to multiple polarized
components (also see Section~\ref{sec:single_pix_los}) that have different
wavelength-dependent depolarization behaviours.

The failure of Stokes $Q,U$ fitting in being able to fit the synthetic data
below $\sim1$~GHz with three or less components brings to light an important
aspect about the technique. To our knowledge, most of the Stokes $Q,U$ fitting
in the literature has been applied to data above $\sim1$~GHz, or to high
redshift sources for which the rest-frame emission originated close to 1~GHz or
higher, and up to three polarization components has been sufficient to fit
those data.  It will be interesting to investigate the performance of Stokes
$Q,U$ fitting when broad-band spectro-polarimetric data is acquired below 1~GHz
for a large number of sources with future surveys. A detailed investigation of
the results of Stokes $Q,U$ fitting, the generic properties of the method in
connection to different types of MHD simulations and the optimum number of
depolarization models required to extract maximum physical insights into a
diffuse magneto-ionic medium is beyond the scope of this paper, and will be
discussed elsewhere.

\subsection{Reconstructed Faraday depth map} \label{sec:fd_map}

\begin{figure*}
\centering
\begin{tabular}{cc}
{\mbox{\includegraphics[height=6.0cm, trim=0cm 0cm 8cm 0cm, clip]{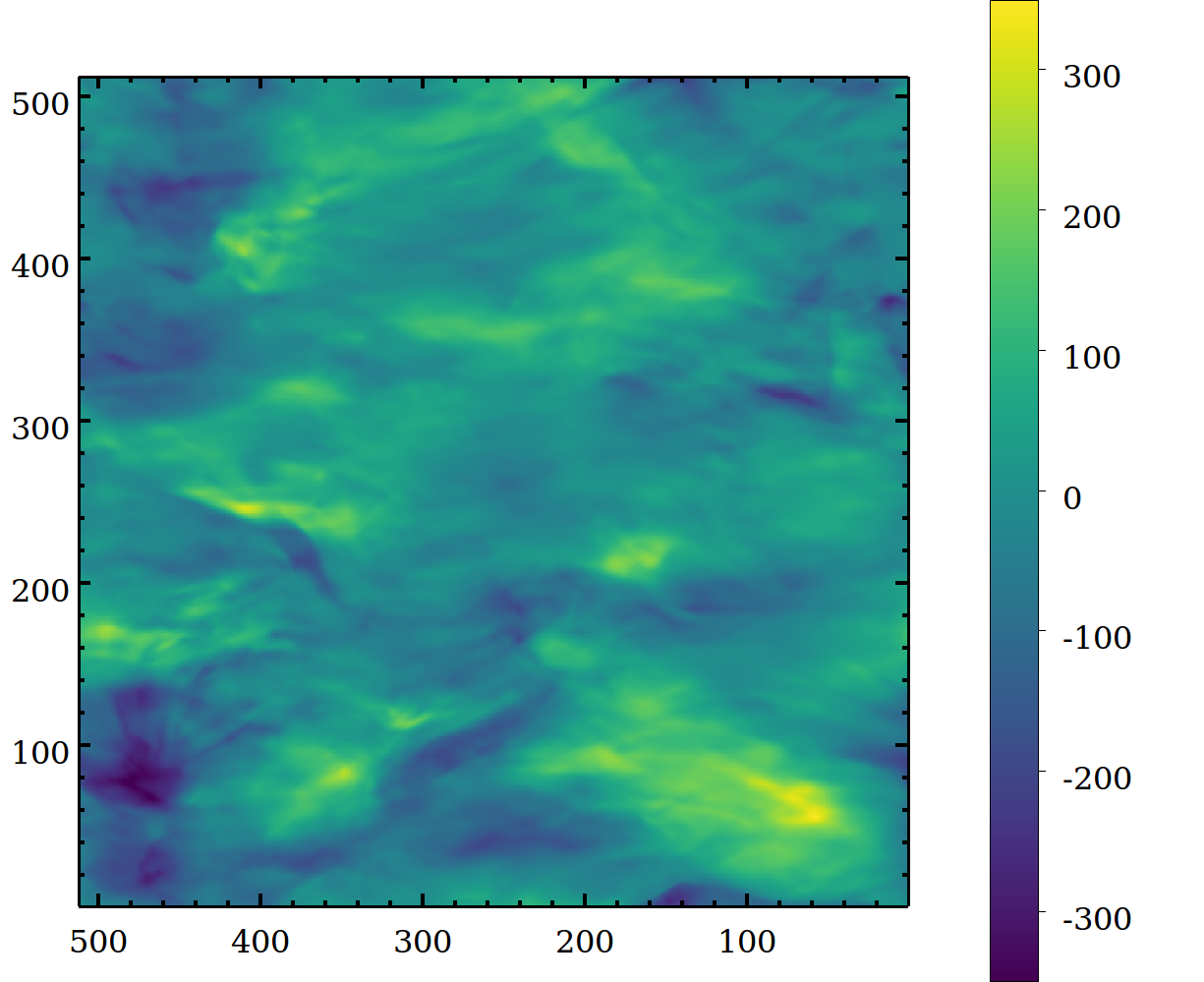}}} &
{\mbox{\includegraphics[width=8.0cm, height=6.0cm]{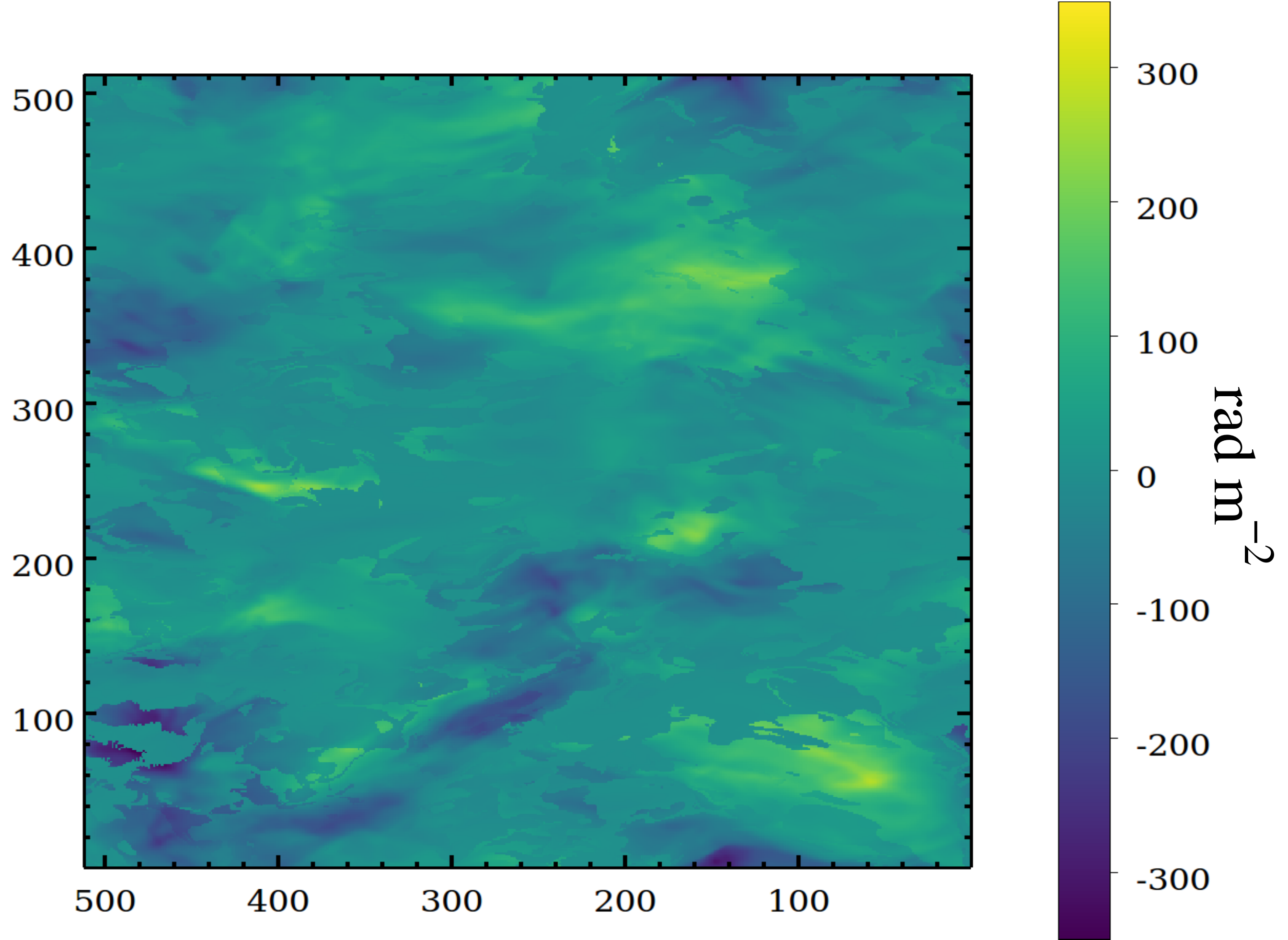}}} \\
\end{tabular}
\caption{Comparison of the Faraday depth map computed from the MHD simulations
(left) with the Faraday depth map reconstructed from RM synthesis applied to
synthetic observations in the frequency range 0.5 to 6 GHz (right).}
\label{fig:rm_compare}
\end{figure*}

To construct the Faraday depth map from the synthetic Stokes $Q$ and $U$
parameters, we applied the technique of RM synthesis. In order to avoid
complications arising from the effects of spectral index \citep{schni18,
schni19}, all analysis pertaining to RM synthesis were performed on fractional
polarization quantities, i.e., $q = Q/I_{\rm sync}$, $u = U/I_{\rm sync}$ and
$p = PI/I_{\rm sync}$. For the frequency setup used here, the RMSF has a FWHM
of $10~\rm rad\,m^{-2}$ and is sensitive to extended Faraday depth structures
up to $\sim1250~\rm rad\,m^{-2}$ \citep[see][]{brent05}. This is sufficient
to perform high Faraday depth resolution investigation of these synthetic
observations without being affected by missing large-scale Faraday depth
structures.

The Faraday depth spectrum was computed for the search range $-3000$ to $+3000$
$\rm rad\,m^{-2}$ with a step size of $0.5~\rm rad\,m^{-2}$.  Since we have not
added any noise to the synthetic data, RM clean was performed by setting loop
gain to 0.02 and with 1000 cleaning iterations. This allows cleaning down to
the minimum fractional emissivity of the 3-D mesh points in the simulated
volume. Synthetic observations are processed using the {\tt pyrmsynth}
implementation of RM synthesis to produce a Faraday depth spectrum which is
then deconvolved using RM clean. In Fig.~\ref{fig:rm_compare}, we show the
expected Faraday depth map from the MHD simulations (${\rm FD_{MHD}} =
0.812\,\int\,n_{\rm e}\,B_\|\,{\rm d}l$, computed using
Eq.~\eqref{eq:losFD}) on the left-hand side and the reconstructed Faraday
depth map computed using RM synthesis ($\rm FD_{RM\,synthesis}$) on the
right-hand side. We have determined $\rm FD_{RM\,synthesis}$ in a pixel as the
location of the peak of the Faraday depth spectrum computed by fitting a
parabola. The spatial features in the reconstructed FD map broadly matches with
the ${\rm FD_{MHD}}$ map, but the magnitudes differ significantly. Also, some
regions of the reconstructed FD map show sharp jumps across neighbouring
pixels. In the next section we will discuss in detail the complexity of
determining the Faraday depth from the Faraday depth spectrum.

\begin{figure}
\centering
\begin{tabular}{cc}
{\mbox{\includegraphics[width=6.5cm]{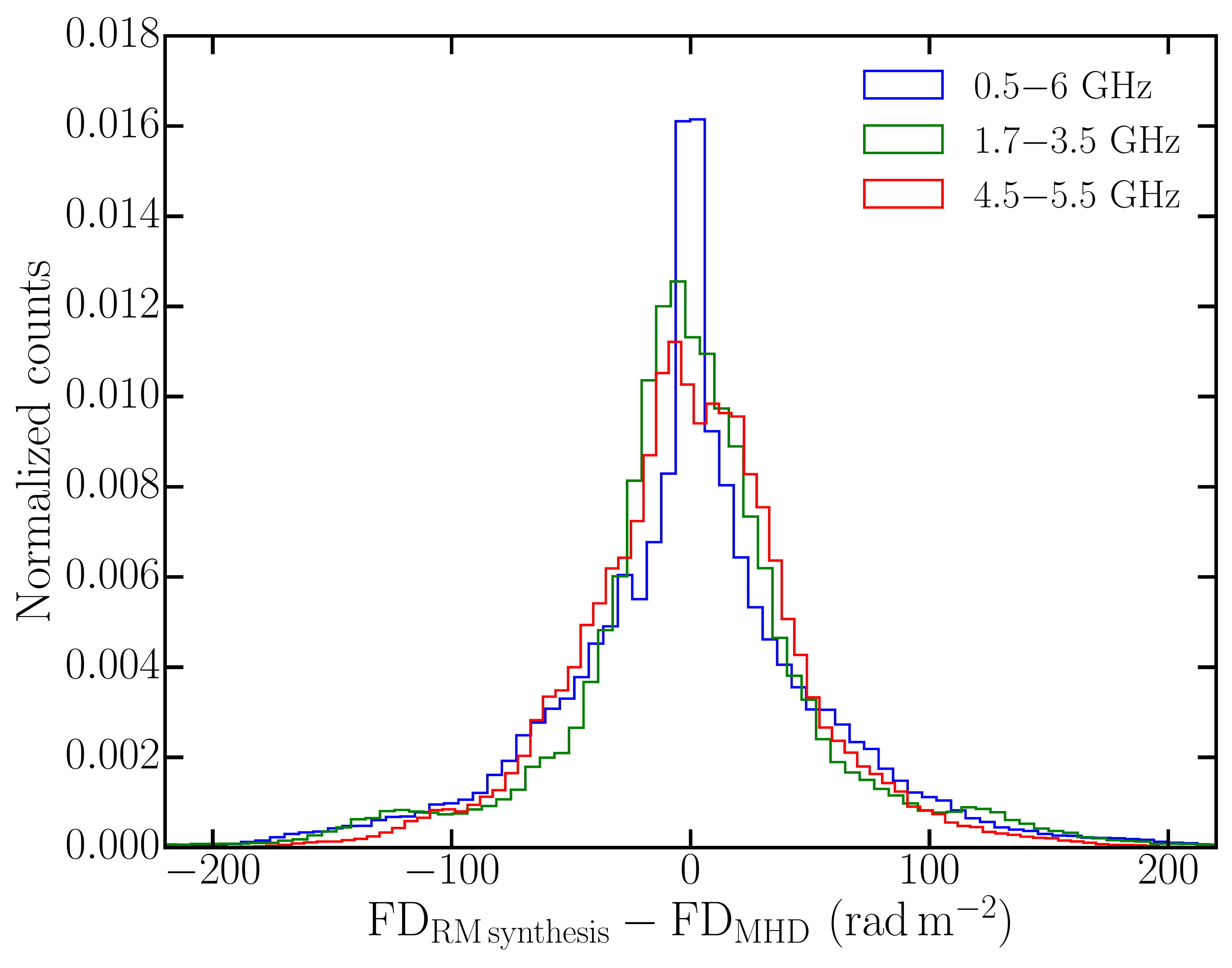}}}&
{\mbox{\includegraphics[width=6.5cm]{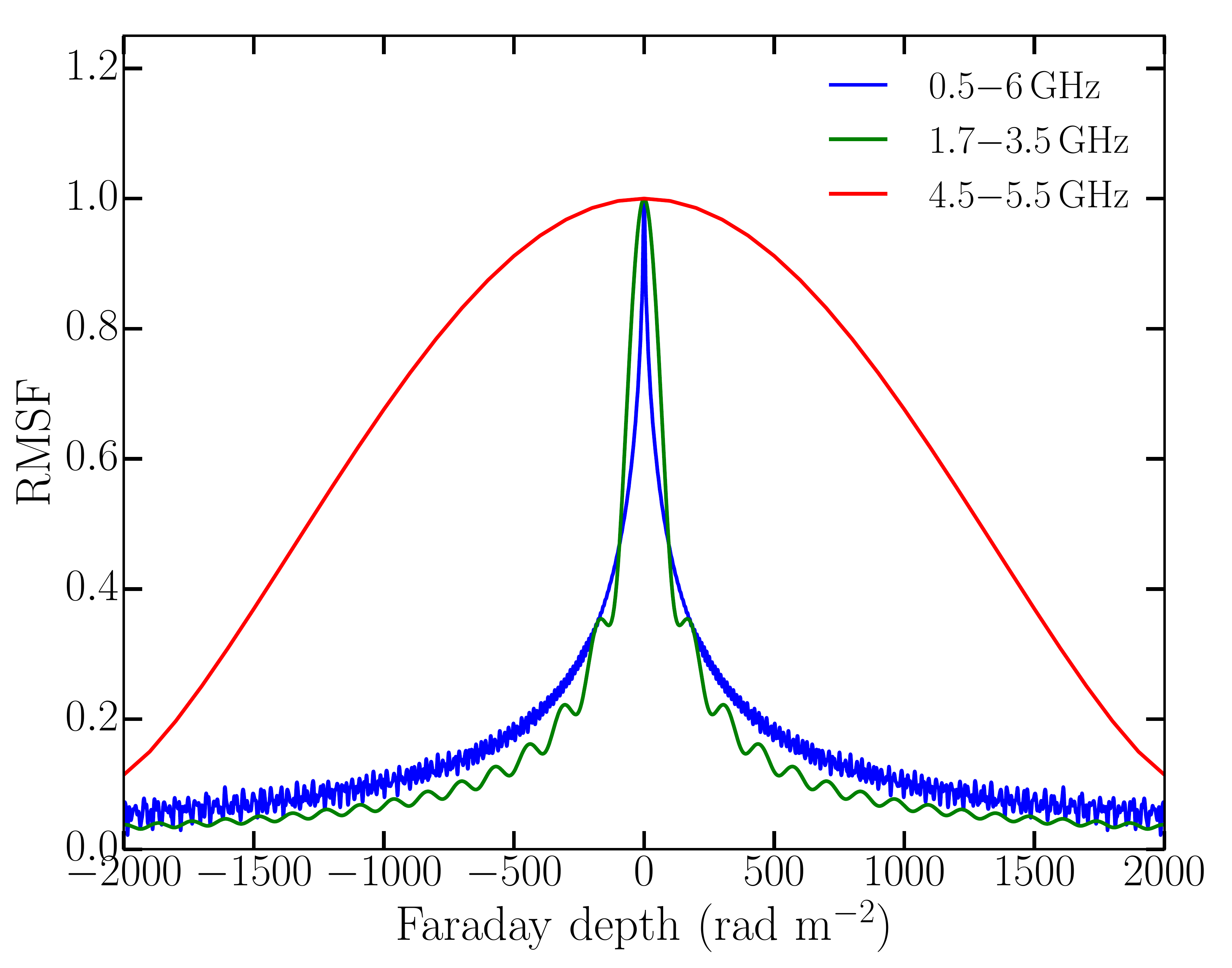}}}\\
\end{tabular}
\caption{{\it Left}: Distribution of the difference between FD estimated by
applying RM synthesis ($\rm FD_{RM\,synthesis}$) and the FD obtained
from the MHD simulation ($\rm FD_{\rm MHD}$) computed using
Eq.~\eqref{eq:losFD}. The different colours for the histograms are for
different frequency coverages. {\it Right}: RMSF for the corresponding
frequency coverages shown in the left-hand panel.}
\label{fig:rm_dist_compare}
\end{figure}

In the left-hand panel of Fig.~\ref{fig:rm_dist_compare}, we show the
distribution of the difference $\rm FD_{RM\,synthesis} - FD_{\rm MHD}$. The
FWHM of the RMSF for the synthetic observations of only $10~\rm rad\,m^{-2}$,
which in combination with the signal-to-noise ratio, gives an estimate of the
error of determined FD. Since, we have not added any noise to the synthetic
data, the estimated $\rm FD_{RM\,synthesis}$ is expected to have vanishing
error. However, $\rm FD_{RM\,synthesis}$ is significantly off with respect to
the expected $\rm FD_{\rm MHD}$ by up to $\pm200~\rm rad\,m^{-2}$. This is
mainly because of the way FD is determined from Faraday depth spectra that
are highly complex as discussed in the next section. To assess possible
systematic effects arising from the RMSF and/or sensitivity to large Faraday
depth structures, we also computed $\rm FD_{RM\,synthesis}$ for different
frequency coverages. The different histograms in Fig.~\ref{fig:rm_dist_compare}
(left-hand panel) show the distribution of $\rm FD_{RM\,synthesis} - FD_{\rm
MHD}$ for the different frequency coverages. The respective RMSFs are shown in
the right-hand panel of Fig.~\ref{fig:rm_dist_compare}. For none of the
frequency ranges is FD recovered with sufficient accuracy.

\section{Comparing Faraday depth spectrum obtained by RM synthesis with the
intrinsic spectrum of a turbulent medium} \label{sec:FDanalysis}

In this section we examine Faraday depth spectrum of individual lines of sight
through both the simple benchmark data cubes of Section~\ref{sec:benchmark} and
the MHD turbulence data cube described in Section~\ref{sec:mhd}, using two
different methods. First, the {\tt COSMIC} code is used to generate Stokes $Q$
and $U$ values using the observational setup described in
Section~\ref{sec:syntheticObs} and Faraday depth spectra were computed in the
same way described in Section~\ref{sec:fd_map}. Second, we calculate the
polarized emissivity and the local Faraday depth at each position along the
same lines of sight to generate a model of the intrinsic Faraday depth spectrum
for that LOS.

We wish to know how close the two Faraday depth spectra are for a realistic
model of a turbulent, synchrotron emitting medium. In particular, does a
spectrum obtained by RM synthesis give a true representation of the emissivity
distribution by Faraday depth that the medium actually produces? If the answer
to this is no, then what is the connection between the Faraday depth spectrum
and the physical state of the emitting region? Whilst the answers to these
questions will probably not be surprising, with hindsight, to someone familiar
with the method, we will gain valuable insights into the nature of the problem
from this investigation.

\subsection{Faraday depth spectra of analytical models}

We first illustrate the process using the two simple models for the
distribution of emissivity and Faraday depth described in Section
\ref{sec:benchmark}: a uniform slab and a internal Faraday dispersion volume
with delta-correlated Gaussian random fields. Artefacts introduced by the RM
synthesis and RM clean algorithms will be clearly visible, the data will
provide a useful reference point for comparisons later, and we shall begin to
see the difficulties in discriminating between very different sources using the
technique.  

\begin{figure*}
\centering
\begin{tabular}{cc}
{\mbox{\includegraphics[height=7cm]{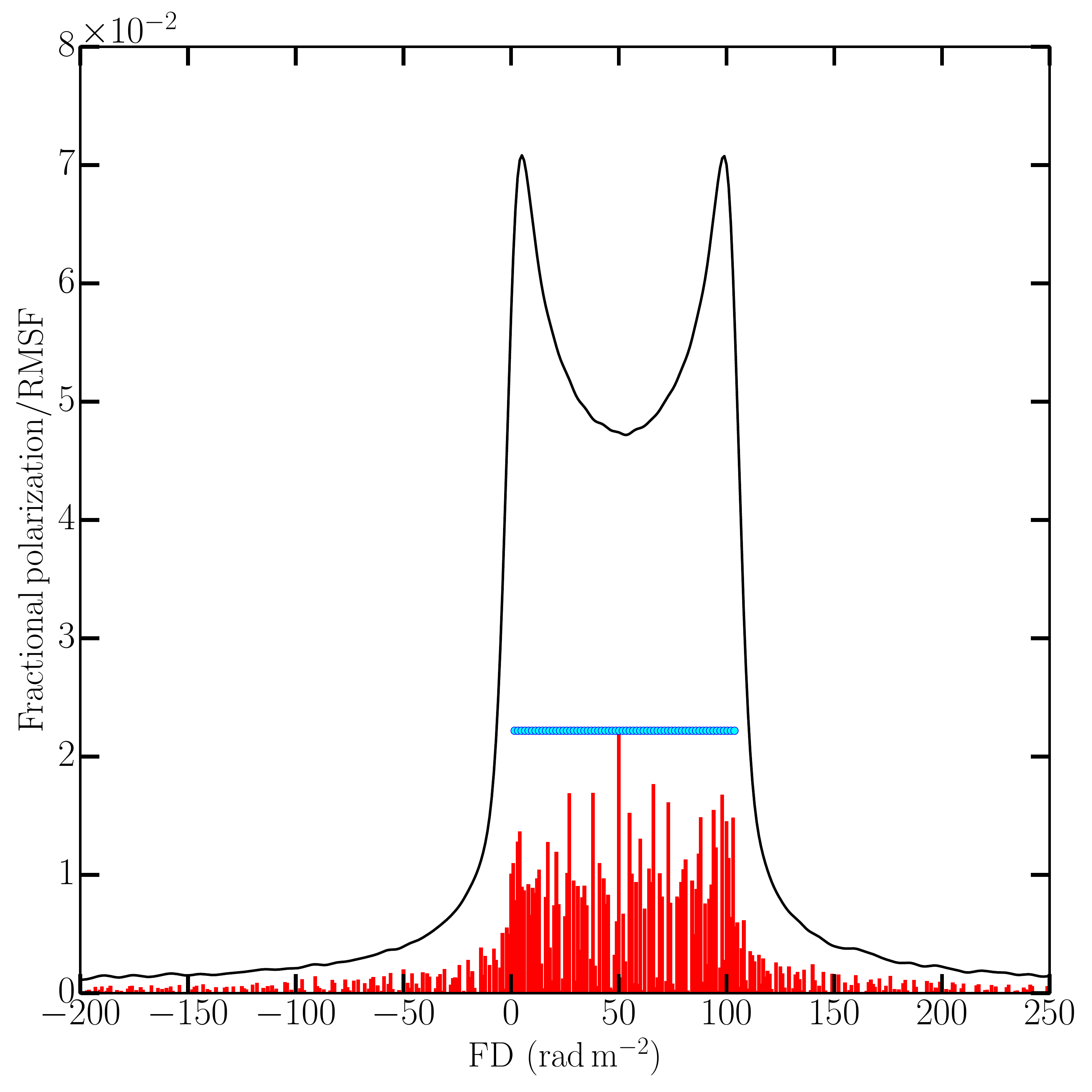}}} &
{\mbox{\includegraphics[height=7cm]{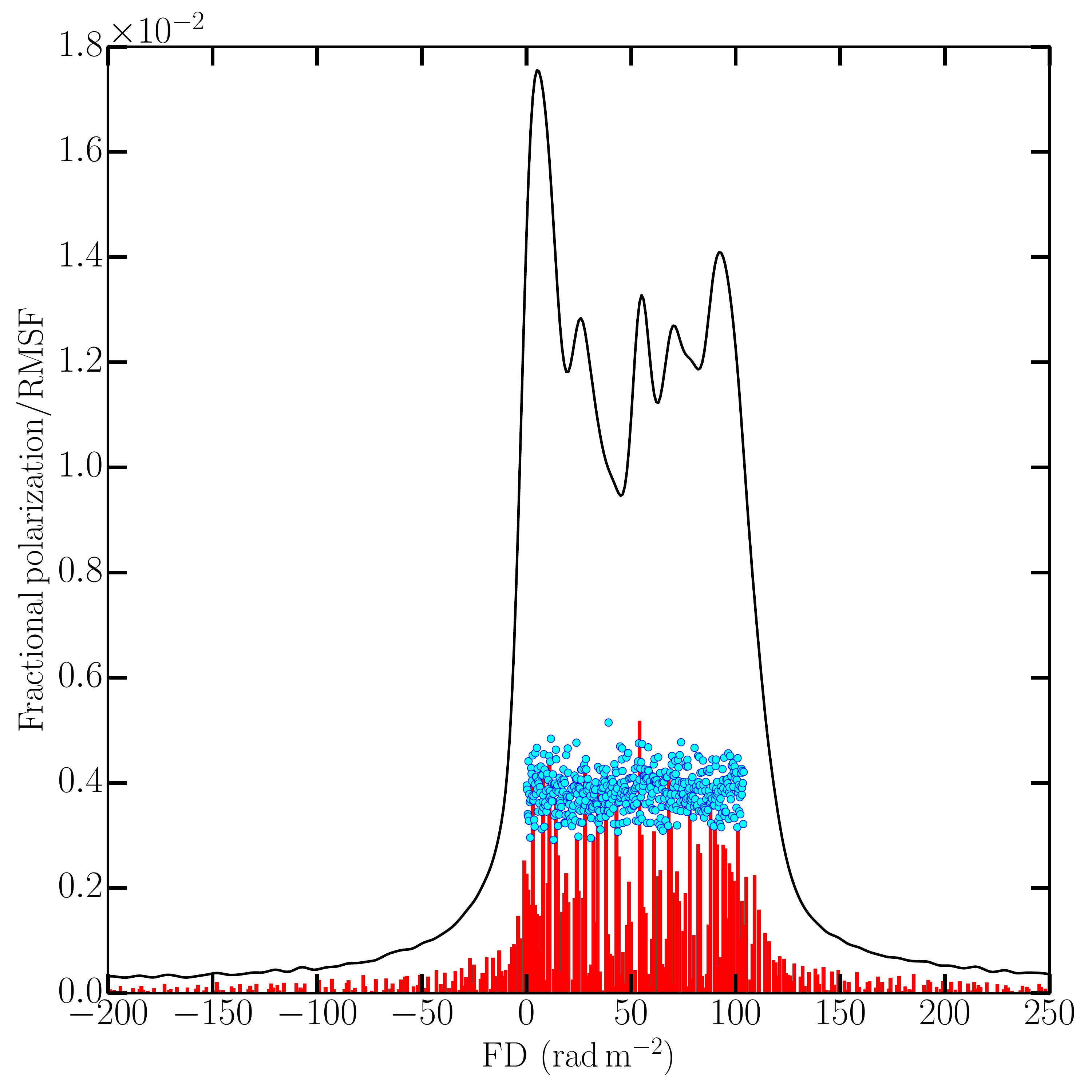}}} \\
\end{tabular}
\caption{Faraday depth spectra for the uniform slab (left) and internal Faraday
dispersion (IFD) models (right). The red lines are the clean components. The blue
dots shows the relative intrinsic synchrotron emissivities at respective FD 
computed from the simulated volume described in Section~\ref{sec:benchmark} and
represents a model of the intrinsic Faraday depth spectrum. For the IFD
model, we have averaged $10\times 10~{\rm pixel}^2$ pixels.}
\label{fig:rm_analytic}
\end{figure*}

In Fig.~\ref{fig:rm_analytic}, for both the uniform slab and Gaussian random
field models, we show three different versions of the Faraday depth spectra
along a single LOS: the continuous lines show the RM cleaned Faraday depth
spectrum; the vertical red lines show the position and amplitude of individual
clean components in this spectrum selected by the RM clean algorithm; the blue
dots show the polarized emission at each Faraday depth computed directly from
the simulations and serves as a model Faraday depth spectrum. At
$\lambda = 0$, the last representation is equivalent to equation 2 in
\citet{eck18b}.

For the uniform slab (Fig.~\ref{fig:rm_analytic}, left-hand panel), the Faraday
depth spectrum for a single LOS shows the double-horn feature that
arises due to sharp edges in the magnetic field distribution and finite
frequency coverage for performing RM synthesis \citep[see e.g.,][for detailed
discussions]{frick11, beck12, schni15b}. One of the peaks is close to 0 $\rm
rad\,m^{-2}$, while the other peak is located at $102.1~\rm rad\,m^{-2}$,
consistent with the Faraday depth through the entire layer of $103.94~\rm
rad\,m^{-2}$. The Faraday depth spectrum for the IFD model
(Fig.~\ref{fig:rm_analytic}, right-hand panel) also shows peaks in the spectrum
from the front of the layer at $0~\rm rad\,m^{-2}$ and the rear at $~100~\rm
rad\,m^{-2}$, but three additional maxima are detected by RM synthesis.  It is
worth noting that all 512 mesh points along this LOS have a random magnetic
field component which is not correlated with the magnetic field of its
neighbours (the random field is said to be delta-correlated). These random
fluctuations will produce point-to-point, uncorrelated, variation in both the
polarized emissivity and the local Faraday depth, as seen for the blue points
in Fig.~\ref{fig:rm_analytic} (right-hand panel). The reason why there are a
few, not 512 peaks, in the spectrum is because these local variations are
blended together by both the resolution of the RMSF (see right-hand panel of
Fig.~\ref{fig:rm_dist_compare}) and because emission from different positions
along the LOS may occur at the same Faraday depth. In principle, the former
effect can be removed by deconvolution using RM clean but the latter effect is
\emph{irreversible}. It is important to note, a single LOS through an
IFD medium does not contain sufficient statistical information on the
delta-correlated random fields, therefore in Fig.~\ref{fig:rm_analytic}
(right-hand panel) we have averaged over $10\times 10~\rm pixel^2$. For a
comparison with single LOS results presented later as closely as possible, we
have chosen averaging over $10\times 10~\rm pixel^2$ instead of larger area as
was done in Section~\ref{sec:benchmark}.

The blue points in Fig.~\ref{fig:rm_analytic} show the intrinsic Faraday depth
spectra of the models, which should be compared to the red bars showing the
components selected by RM clean. The only noticeable difference between the
distribution of clean components, which are roughly constant in each model, is
the lower amplitudes for the random field model which is a consequence of lower
intrinsic fractional polarization (Eq.~\ref{eq:polfrac}). The Faraday depth
spectra allow for an easy identification of the uniform slab, but the clean
components do not. In both models the amplitudes of the clean components appear
to be systematically lower than the intrinsic emissivities because of
normalization: the peak intrinsic emissivity is scaled to be the same magnitude
as the peak clean component, as described in Section~\ref{sec:single_pix_fd}.

\subsection{The origin of complexity in the Faraday depth spectrum of a
turbulent medium} \label{sec:single_pix_los}

\begin{figure}[!t]
\centering
\begin{tabular}{ccc}
{\mbox{\includegraphics[height=4.0cm]{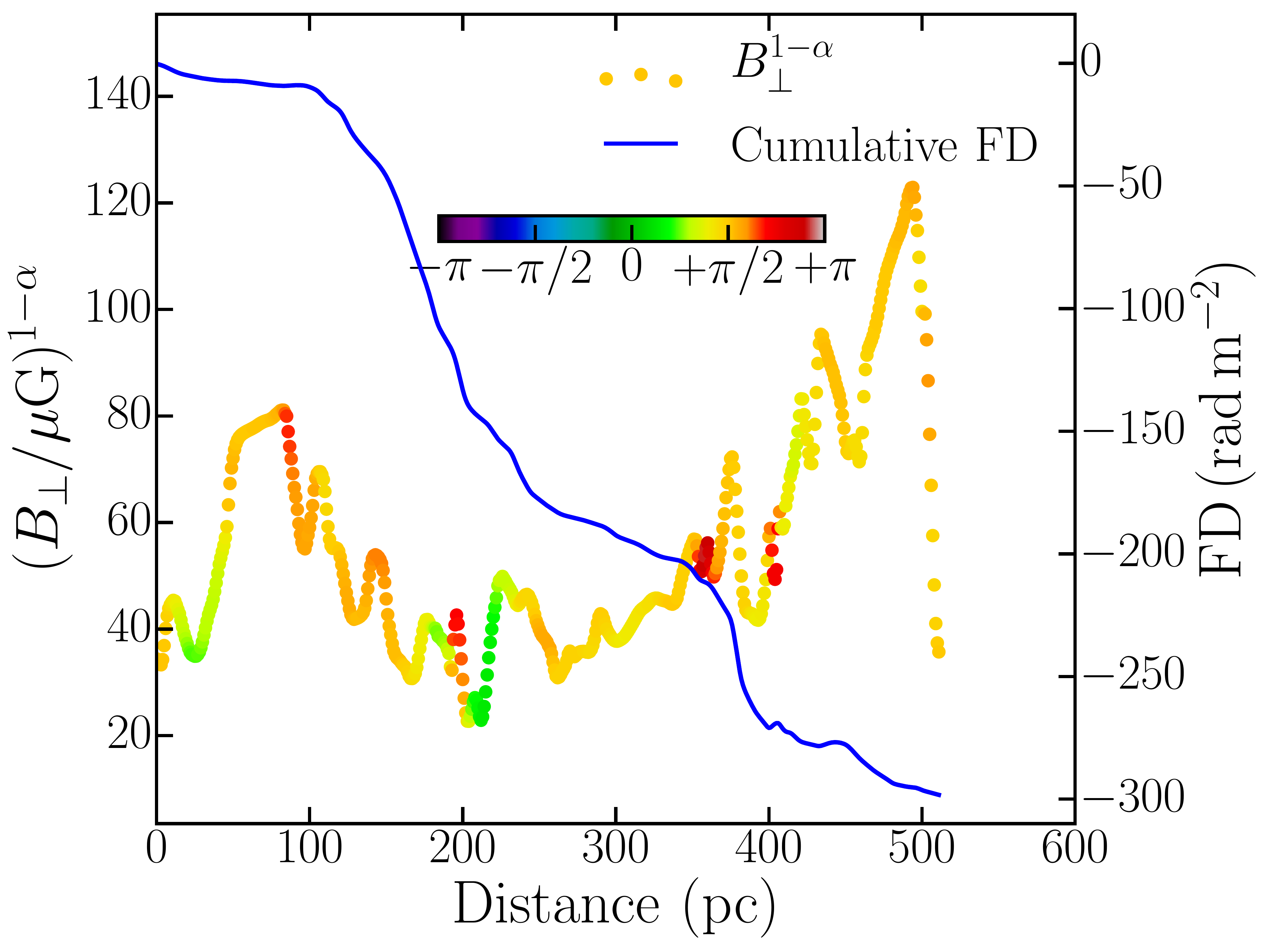}}} &
{\mbox{\includegraphics[height=4.2cm]{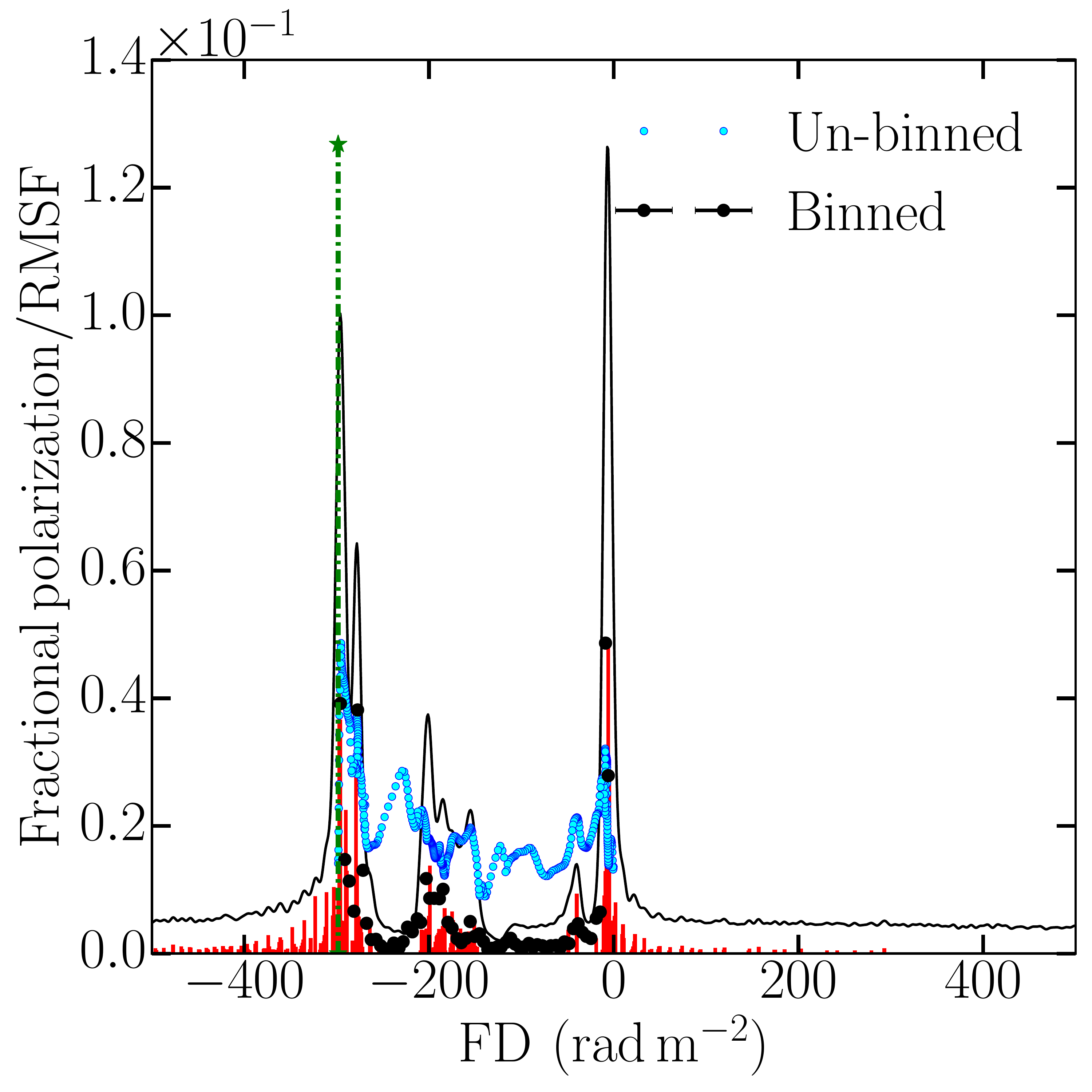}}} &
{\mbox{\includegraphics[height=4.2cm]{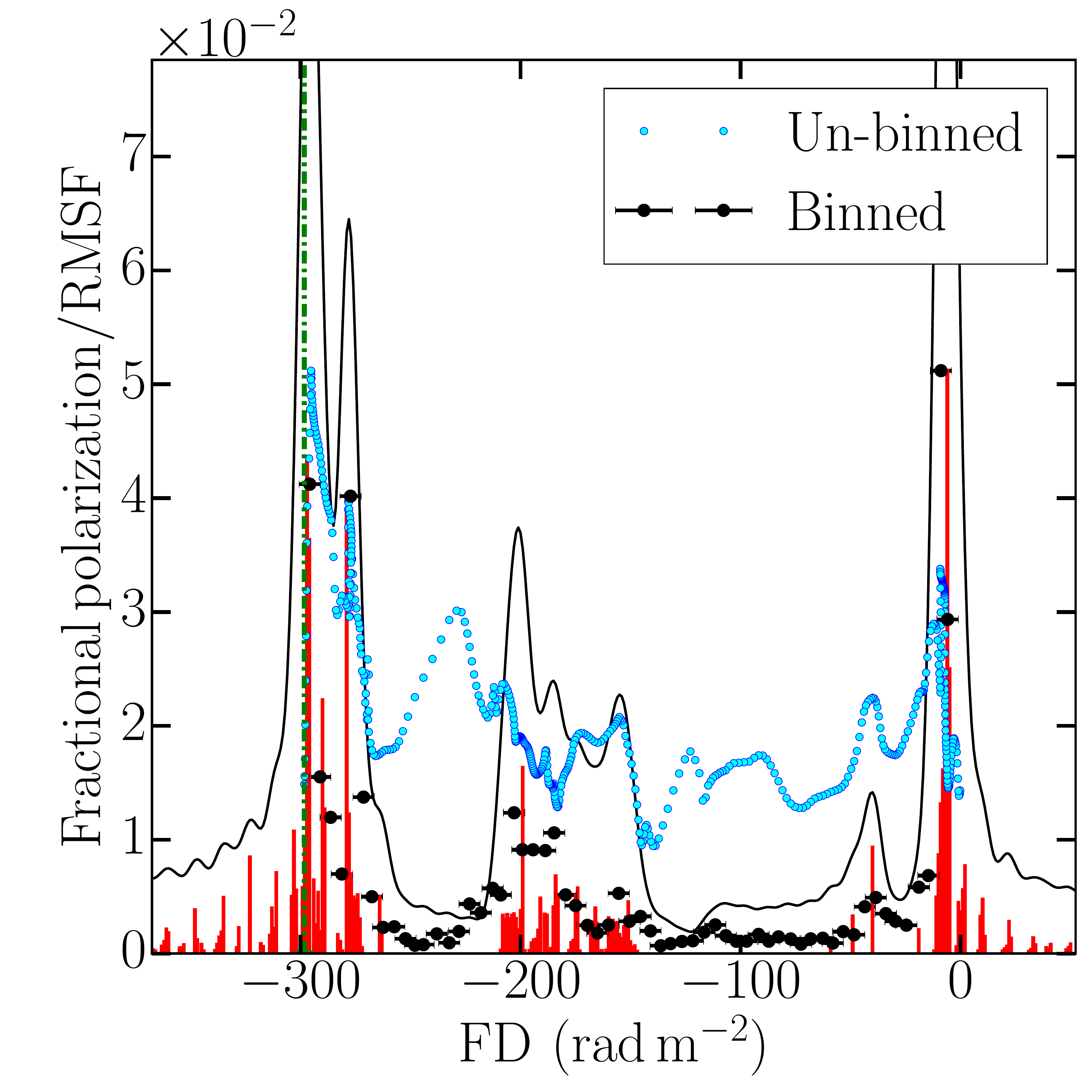}}} \\
{\mbox{\includegraphics[height=4.0cm]{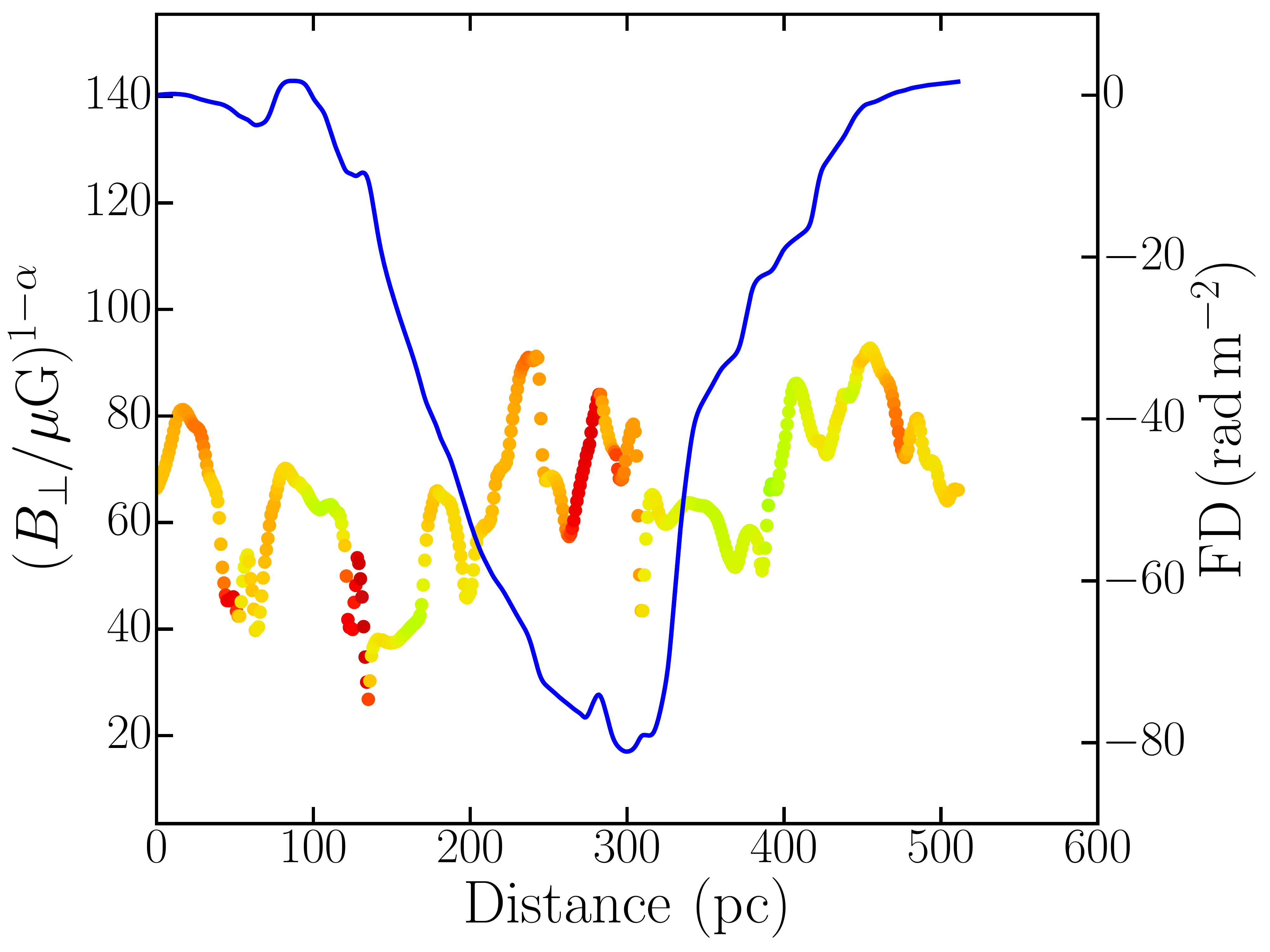}}} &
{\mbox{\includegraphics[height=4.2cm]{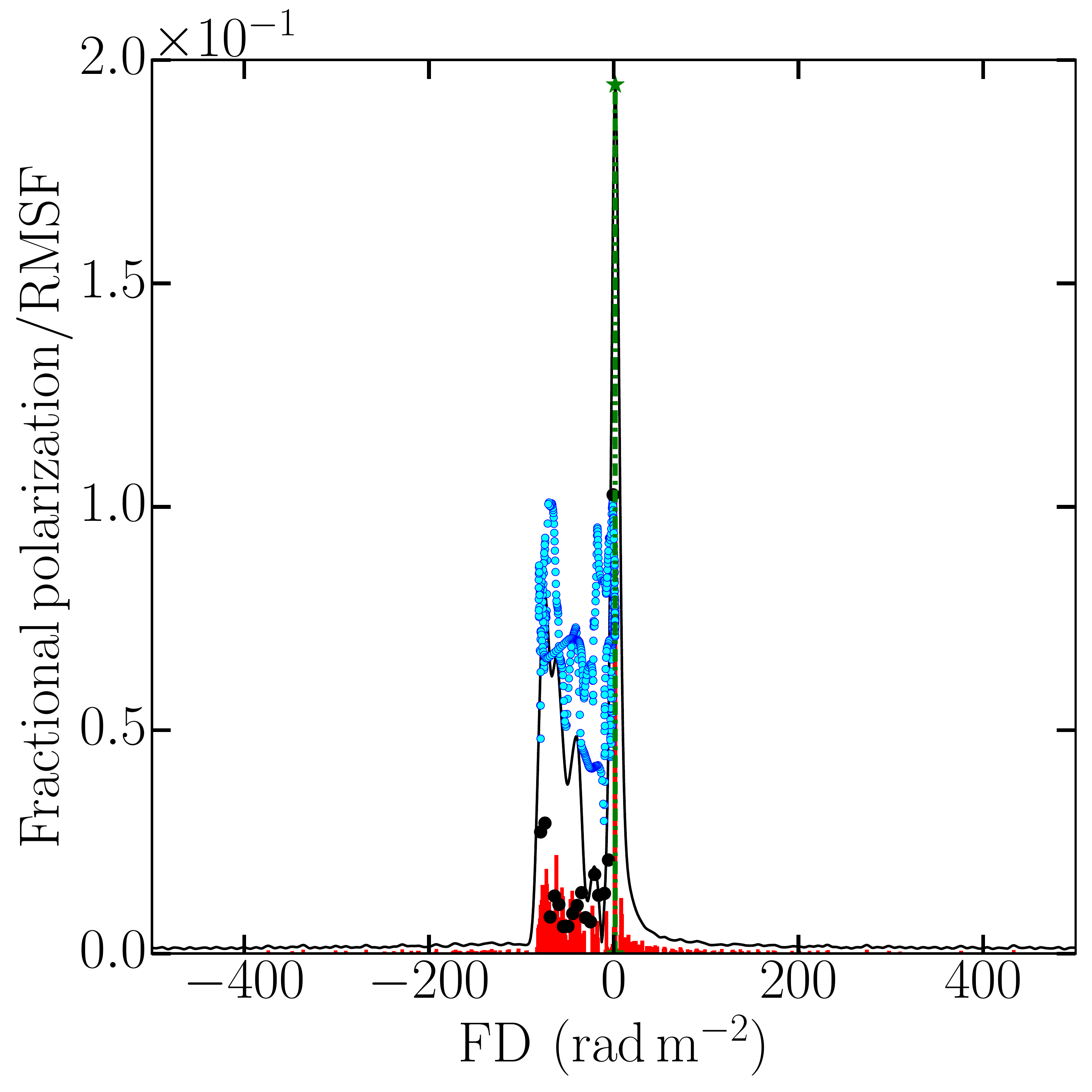}}} &
{\mbox{\includegraphics[height=4.2cm]{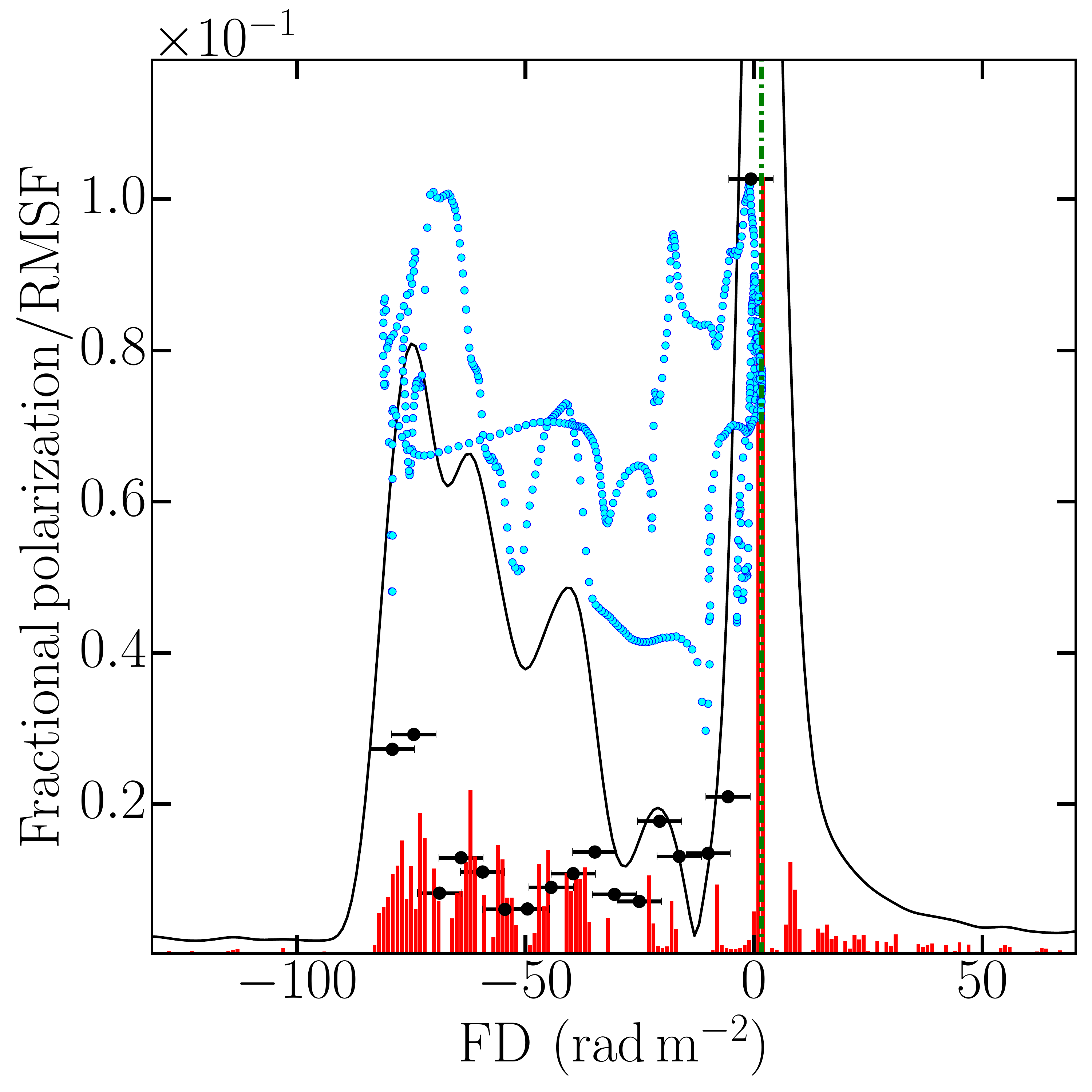}}} \\
{\mbox{\includegraphics[height=4.0cm]{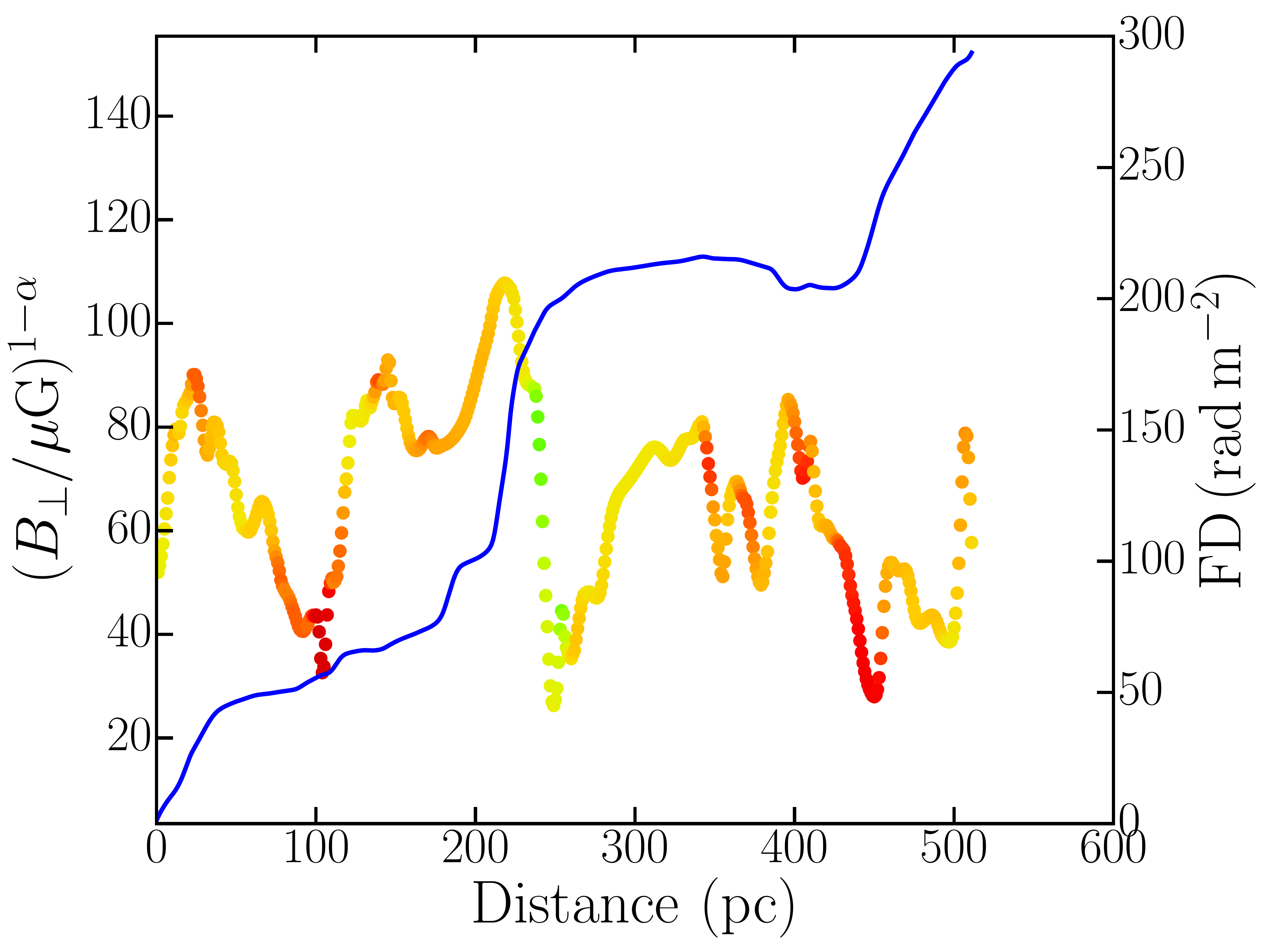}}} &
{\mbox{\includegraphics[height=4.2cm]{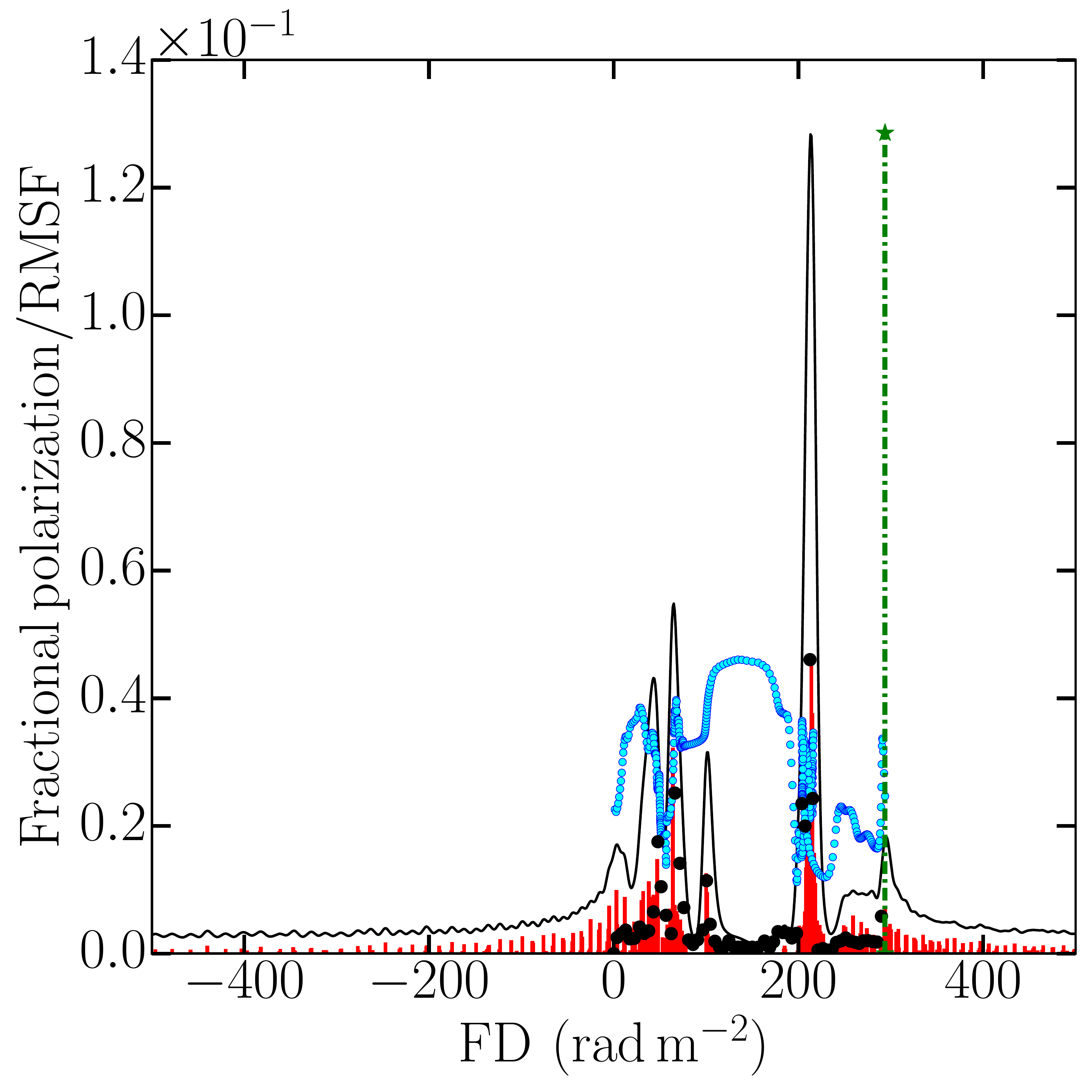}}} &
{\mbox{\includegraphics[height=4.2cm]{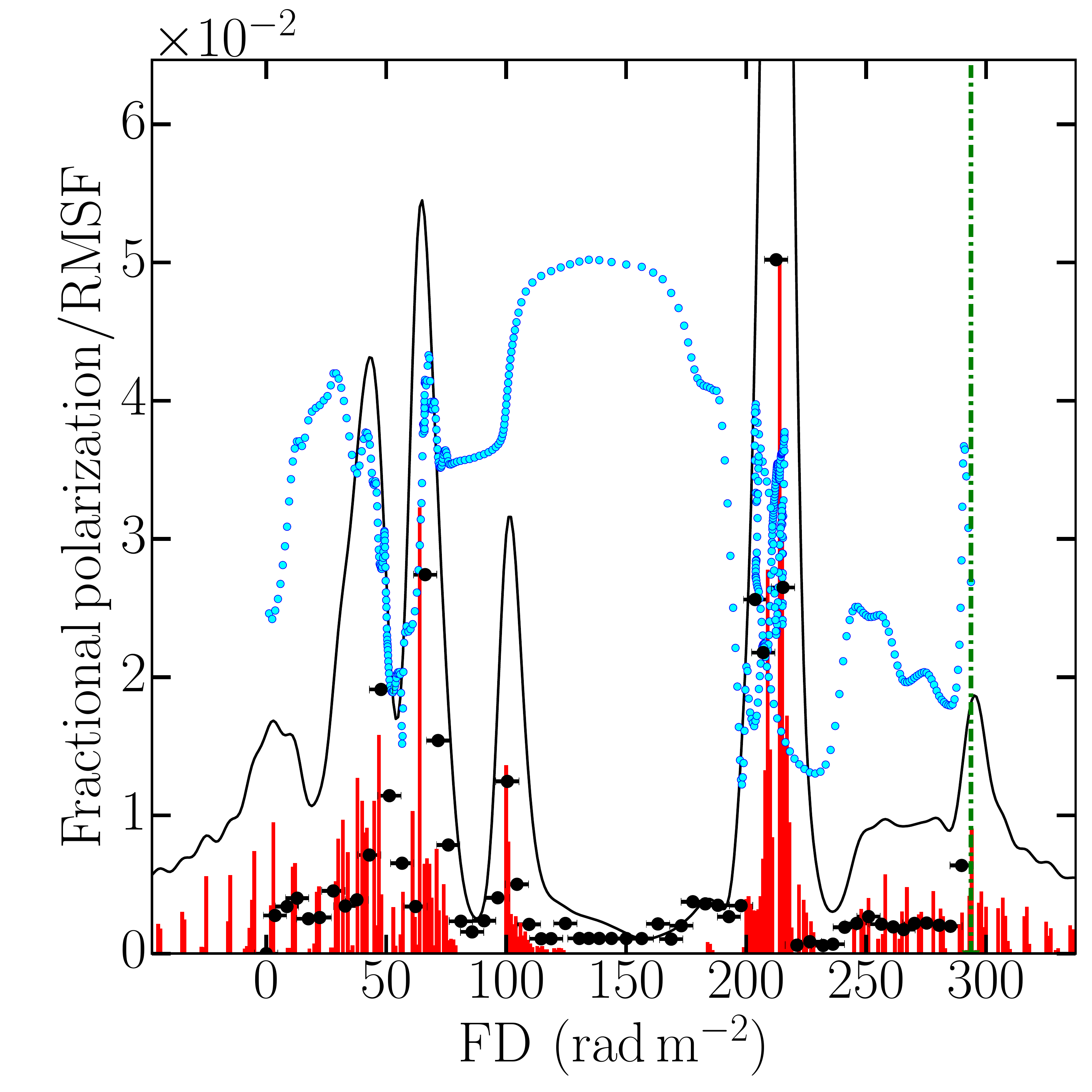}}} \\
\end{tabular}
\caption{{\it Left}: Variation of $B_\perp$ (circles) and cumulative FD (blue
line) with distance along three random LOS each of which is one 
pixel wide. $B_\perp$ is coloured
based on the angle of the polarized synchrotron emission. {\it Middle}: Faraday
depth spectrum for the corresponding LOS. The blue points show the local
polarized emissivity at that FD. The red lines are the clean FD components and
the green dot-dashed line is located at the FD along that LOS. The black points
show the summed linearly polarized synchrotron emissivity per FD bin of size
half the RMSF. The amplitudes of the polarized emissivities are normalized to
the peak FD clean component. {\it Right}: Same as the middle-panel zoomed
around the peak of the Faraday depth spectrum for a clearer visualization.}
\label{fig:rmcomppix}
\end{figure}

Here, we look into the Faraday depth spectrum obtained by performing RM
synthesis on the synthetic observations of MHD turbulence simulations and
compare them to features in the variations of synchrotron emissivity and
Faraday depth along the LOS. We will discuss based on three LOS shown in
Fig.~\ref{fig:rmcomppix}, as they are sufficiently representative of any other
LOS and are therefore general representation of the results we will present. 
The left column displays $B_\perp^{(1 - \alpha)} = B_\perp^{1.8}$ (for $\alpha
= -0.8$), which is directly proportional to the linearly polarized synchrotron
emissivity, at each position along the LOS and also the Faraday depth from that
position to the front of the domain (the side nearest to the notional
observer). The emissivity-proxy fluctuates randomly about the expected mean
value of $\langle B_\perp\rangle^{(1 - \alpha)}={(\langle B_x\rangle^2 +
\langle B_y\rangle^2)}^{(1.8)/2} \approx (10\,\mu {\rm G})^{1.8}$ (see
Section~\ref{sec:mhd}), due to variation in $B_x, B_y$. This will translate
into random structure in the amplitude of the Faraday depth spectrum. In
Fig.~\ref{fig:rmcomppix} (left column), $B_\perp^{(1 - \alpha)}$ is colourized
based on the intrinsic angle of polarization. Because of the strong $B_{\rm
ext}$ along the $x$-direction, the polarization angle do not show strong
fluctuations.


The Faraday depth of each point along the LOS shows a systematic non-random
variation with position. This is because the Faraday depth is cumulative from a
given position to the observer and the fluctuations in the magnetic field
components are not delta-correlated. The correlation scale of $L_{\rm c}\sim
200$~pc is controlled by the forcing used to drive turbulence in the
simulation: in these simulations the forcing scale is about 2.5 times smaller
than the simulation box \citep{burkh09}. The driven turbulence contains
fluctuations on all scales from $L_{\rm c}$ down to the dissipation scale (the
separation between mesh points in this case), but the power-law behaviour of
the magnetic energy means that the field is strongest at $L_{\rm c}$. So, the
Faraday depth builds up systematically over path lengths of roughly 200~pc. An
interesting consequence of this is that, because the domain size is 512~pc, the
Faraday depth variation along the LOS is not necessarily a monotonic function:
the maximum magnitude of Faraday depth along a given LOS may not be that from
the back to the front of the domain. If the domain was half the size, the
Faraday depth variation would tend to be monotonic and if it were much bigger
we would see multiple peaks and troughs in the Faraday depth
variation. For domain sizes $L \gg L_{\rm c}$, FD variations would
eventually appear to be random for $\langle B_\| \rangle = 0~\mu$G. Thus, the
form of the intrinsic Faraday depth variation depends on the number of
turbulent cells along the LOS and does not necessarily look like a random
function for shorter path-lengths.  However, when combined with the variations
in emissivity to produce the intrinsic Faraday depth spectrum the smoothness in
${\rm FD}$ along the LOS is not translated into a smooth spectrum.   

The intrinsic polarized emissivity and Faraday depth combine to give the
intrinsic Faraday depth spectrum shown by the blue points in the middle and
right columns of Fig.~\ref{fig:rmcomppix}. Note that these have been normalized
to the maximum RM clean component, so only the relative variation contains
useful information. Also shown in these panels is the spectrum obtained by RM
synthesis and its clean components. The Faraday depth spectra are highly
complex with multiple peaks and clean components referred to as a ``Faraday
forest'' by \citet{beck12}. These spectra are also clearly different from that
obtained using delta-correlated Gaussian random field as an approximation to
turbulence shown in Fig.~\ref{fig:rm_analytic} (right-hand panel). 

It is clear that the Faraday depth spectra are also very different from the
intrinsic model spectra. RM synthesis tends to select a few strong peaks which
are not always the FD along the entire LOS, nor are they closely
related to features in the intrinsic polarized emission. This illustrates a
fundamental difficulty in interpreting RM synthesis data. It may be
wrong to interpret some of the narrow, high peaks as the action of a Faraday
screen, or a Faraday-thin component: a region in which there is no polarized
emission, but which produces Faraday rotation. However, the corresponding left
panels show that there is emission from all locations along the LOS. 

A further difficulty in interpretation arises because the maximum absolute
value of the Faraday depth along a LOS is not necessarily the Faraday depth
through the entire box. The latter is shown by the green dash-dot line in the
middle and right columns of Fig.~\ref{fig:rmcomppix}. For example, in the first
row of Fig.~\ref{fig:rmcomppix} the spectrum obtained by RM synthesis has
strong peaks at $\rm FD\approx 0$ and $\rm FD\approx -300~\rm rad\,m^{-2}$ and
looking at the left panel we see these correspond respectively to the emission
generated at the front and back of the LOS. However, in the second row, the
emission that produces the strong spectral peak near $\rm FD = 0~\rm
rad\,m^{-2}$ comes from emission from \emph{both} the front and back of the
layer. The Faraday depth through the entire layer is $\rm FD=1.67~\rm
rad\,m^{-2}$ and the maximum along this LOS, $\rm FD\approx -80~\rm
rad\,m^{-2}$ originates from the middle. There is no way yet, that we are aware
of, to reliably recover any of the true properties of the Faraday depth along
the LOS from the Faraday depth spectrum alone. Images produced using the
strongest spectral peak or the maximum absolute FD as proxies for some assumed
property of a source, such as, its intrinsic polarized emissivity or the total
Faraday depth through it, may be misleading. 

Pronounced peaks in the Faraday depth spectrum of a synchrotron emitting medium
can arise in two ways. A strong peak in emissivity from a region where the FD
is changing slowly over a small distance will produce a narrow FD
feature: an example is in the third row of Fig.~\ref{fig:rmcomppix}, where the
strong emission from a distance between 190 and 210~pc, where the FD
profile shows a step-like feature, produces the spectral peak at $\rm
FD\approx 100~\rm rad\,m^{-2}$. Alternatively, when the FD is constant, or
changing slowly over large distances, comparatively weaker
emission builds up at this FD and results in a peak in the spectrum: in the
third row of Fig.~\ref{fig:rmcomppix} we see that $\rm FD\approx
200~rad\,m^{-2}$ between the distance $250$ and $450$~pc along the LOS,
producing a pronounced peak at this FD in the spectrum. Such a
scenario can occur when the angle of the polarized emission is changing slowly
with distance. In neither case does a strong peak in the spectrum result from
a Faraday screen. Whilst the origin of some spectral peaks can be approximately
described by one of these mechanisms, there are also many peaks whose origin is
not so simple and are discussed in the next section.

Multiple polarized components in Faraday depth spectrum is a natural
consequence when the LOS passes through a few coherent scales. We believe, in
the case when the LOS traverse through statistically large (asymptotically
infinite) number of turbulent cells, Faraday depth spectrum similar to that
obtained for delta-correlated random fields could perhaps be obtained. Given
the typical driving scale of turbulence in the ISM, e.g.,
$\sim50\textrm{--}100$~pc when driven by supernova \citep{armst95, elmeg04},
and the typical length of LOS through the ISM of a few kiloparsec, achieving
such a limit is highly unlikely. Unfortunately, the MHD simulations used here
do not allow us to test this scenario. At present we are not aware of any
reliable method which can be used to interpret Faraday depth spectra that
originate in a turbulent, synchrotron emitting region. Since turbulence and
cosmic ray electrons are thought to be present throughout the ISM this problem
requires careful investigation.

\subsection{Comparing clean components to the intrinsic Faraday depth spectrum}
\label{sec:single_pix_fd}

In Fig.~\ref{fig:rmcomppix}, the blue points showing the intrinsic emission at
each Faraday depth are present between the main peaks in the spectra obtained
by RM synthesis and RM clean. This is because, structures in FD are smoothed by
the RMSF. In order to make a better comparison between the observed and
intrinsic spectra it is important to consider the number of emitters in a given
range of Faraday depths that is compatible with the resolution of the RMSF. We
therefore binned synchrotron emissivities within a range of FD, with the
bin-size determined by the RMSF. In our case, we used the bin-size $\Delta
\textrm{FD} = \textrm{RMSF}/2$. 
The choice of this bin-size is motivated by the fact that we wanted to
Nyquist sample the RMSF.
This is equivalent to the sum of polarized
intensities ($\varepsilon_{\rm bin}$) arising in the range $\rm FD \pm \Delta
{\rm FD}/2$ and $\varepsilon_{\rm bin}$ is given by, 
\begin{equation}
\varepsilon_{\rm bin}({\rm FD}) = \sum\limits_{{\rm FD}_i}\,\varepsilon({\rm FD}_i)
\equiv N_{\rm bin}\,\langle \varepsilon_{\rm bin} \rangle,~\textrm{where}~{\rm FD}_i \in [{\rm FD - \Delta FD/2, FD + \Delta FD/2}] .
\label{eq:syncbinFD}
\end{equation}
Here, $N_{\rm bin}$ is the number of synchrotron emitting elements within a FD
bin computed from the simulation and $\langle \varepsilon_{\rm bin}\rangle$ is
the mean polarized synchrotron emissivity of that FD bin. As the intrinsic
polarization angle of the synchrotron emission do not fluctuate strongly (as
can be seen by the smooth colour variation of the angle of the polarized
synchrotron emission in the left column of Fig.~\ref{fig:rmcomppix}), simple
addition of the polarized emissivities will not affect our conclusions.  The
black points in the middle and right columns of Fig.~\ref{fig:rmcomppix} show
$\varepsilon_{\rm bin}$ located at the mean FD of the corresponding bin with
the maximum value of $\varepsilon_{\rm bin}$ normalized to the peak value of
the FD clean components. It is clear from the middle and the right-hand panels
of Fig.~\ref{fig:rmcomppix} that the sum of polarized emission in FD bins
captures the RM clean components, including their relative amplitudes,
remarkably well. 

Note that, strong $\varepsilon_{\rm bin}$, and thus a peak in the Faraday depth
spectrum can originate either due to --- (1) strong emissivity at a location
along the LOS (large $\langle \varepsilon_{\rm bin} \rangle$), or due to (2)
build up from several weaker emissions that have FD within $\pm \rm{RMSF}/4$
(large $N_{\rm bin}$). In fact, both these scenarios can occur along the same
LOS. For example, in the top row of Fig.~\ref{fig:rmcomppix}, the peak near
$-300~\rm rad\, m^{-2}$ originates due to the first case, while the peak near
$0~\rm rad\, m^{-2}$ is because of the second case. In the second row of
Fig.~\ref{fig:rmcomppix}, the peak near $0~\rm rad\, m^{-2}$ is a consequence
of the second case although the polarized emission in that FD bin originate
from both front and back of the sightline. This clearly demonstrates that, for
a turbulent magneto-ionic medium which is simultaneously synchrotron emitting
and Faraday rotating, a peak in the Faraday depth spectrum corresponds to the
polarized emission summed roughly within $\pm \rm RMSF/4$, which
necessarily may not arise from regions along the LOS that are spatially
continuous. Therefore, special care must be taken when interpreting FD maps
constructed through RM synthesis and relating them to the FD of the emitting
volume, especially when the Faraday depth spectrum appears to be well
resolved.

This brings out the important fact that, even if a broad-band observation is
sensitive to extended structures in Faraday depth space and can resolve them,
as our synthetic observations are, the RMSF plays an important role in
determining the amplitudes of the clean components. Moreover, when performing
RM synthesis for diffuse emissions, sharp peaks in the Faraday depth spectrum
that are consistent with the width of the RMSF do not necessarily imply the
presence of a Faraday rotating screen in the foreground of a synchrotron
emitting volume.

Our investigation on Faraday depth spectra is based on a diffuse isothermal
magneto-ionic medium as a representative Galactic ISM that contains transonic
compressible turbulence. It is possible that the conditions of turbulence in
different parts of the Galactic magneto-ionic medium could be of different
type, including the distribution and direction of the regular magnetic fields.
The conclusions regarding the origin of Faraday complexity and various peaks in
the Faraday depth spectrum are expected to be general. However, depending on
the spatial smoothness of the turbulent fields (for example, sub- or
super-sonic turbulence) and strength of the regular fields, the peaks in
Faraday depth spectrum could become smoother or merge together. In other words,
the \textit{clumpyness} of a well resolved Faraday depth spectrum could provide
hints on the nature of turbulence in a magneto-ionic medium. Therefore,
quantifications of the shape of Faraday depth spectrum beyond the recently used
skewness and kurtosis \citep{idegu17} is necessary to be investigated to
distinguish between different types of turbulent medium.

\section{Conclusions} \label{sec:conclusion}

We have investigated in detail various features in Faraday depth spectra
obtained from synthetic broad-band spectro-polarimetric observations in the
frequency range 0.5 to 6 GHz sampled with 500 frequency channels. We have
developed a new software package, {\tt COSMIC}, wherein a user can freely choose
from several possible options to generate synthetic observations. In this work
the synthetic observations were obtained from MHD simulations of an isothermal,
transonic, compressible turbulence in a plasma, similar to that observed in the
Galactic ISM. For comparison, we have also studied the Faraday depth spectrum
for a simple delta-correlated Gaussian random description of turbulent magnetic
fields. We reach the following conclusions:

\begin{enumerate} 

\item For the MHD simulations used, Faraday depth varies smoothly with
distance along the lines of sight due to spatially correlated structures in the
magnetic field. Faraday depth varies on scales of $\sim200$~pc, similar to the
driving scale of turbulence in the simulations used in this work.

\item Strong Faraday depolarization at long wavelengths gives rise to
canal-like small-scale structures in the polarized synchrotron emission. At the
resolution of the MHD simulations used here, the polarized synchrotron emission
below $\sim1$ GHz shows spatial variation on few pixel scales that correspond
to few parsecs.

\item For the choice of our frequency coverage, the synthetic observations are
sensitive to structures extended up to $\sim1250~\rm rad\,m^{-2}$ in Faraday
depth space and resolve them with RMSF of $10~\rm rad\,m^{-2}$. This allowed us
to perform a high resolution investigation of Faraday depth spectra. Faraday
depth spectra for a medium containing Gaussian random magnetic fields that are
delta-correlated are significantly different in structure as compared to 
those obtained from MHD simulations of a medium containing random magnetic
fields that are spatially correlated. The latter is expected to be a closer
representation of the diffuse ISM.

\item Faraday depth spectra of individual sightlines through the MHD cube show
a combination of narrow and broad features, which cannot be described as a
linear combination of simple models that are typically used when polarization
data are analysed. The narrow structures are mostly consistent with the width
of the RMSF containing a single FD clean component, typically considered as a
signature of a medium that is only Faraday rotating, although the entire
simulation volume is emitting synchrotron radiation.

\item We find that, modelling RM clean components of the Faraday depth
spectrum as discrete emitters along a line of sight where, at a physical
distance, the synchrotron emissivity of the emitters is located at the Faraday
depth to that distance does not represent the clean components obtained from
RM clean.

\item The clean components and their relative emissivities obtained from RM
clean are well represented by the sum of polarized synchrotron emissivity in a
Faraday depth bin of bin-size $\rm RMSF/2$.

\item Since the Faraday depth spectrum depends on the interplay of the
emissivity $\varepsilon_{\rm sync}$ and the local Faraday depth, the
complicated structures in Faraday depth can be explained as follows:\\ --
Strong sharp peaks in the spectrum can be produced due to reasons: (i)
strong $\varepsilon_{\rm sync}$ at a single FD, or, (ii) build up of weaker
$\varepsilon_{\rm sync}$ over a range of distance, not necessarily continuous,
that have roughly constant FD, or whose FD lies in the range $\rm FD - RMSF/4$
and $\rm FD + RMSF/4$.\\
-- Broad spectral features are produced by: (iii) a gradient in FD at a
constant $\varepsilon_{\rm sync}$. Deviations from ``constant''
$\varepsilon_{\rm sync}$ or FD produce sub-structure on (ii) and (iii).

\item Our analysis shows that it is highly non-trivial to infer the Faraday
depths, i.e. the integral $0.812\,\int n_{\rm e}\,B_\|\,{\rm d}l$ along the 
entire LOS, for a diffuse medium that emits synchrotron radiation if turbulent
magnetic fields are dominant.

\end{enumerate}

The turbulent magneto-ionic medium in a real astrophysical source is expected
to be further complicated as compared to our simplified assumptions on its
properties, such as, a constant density of CREs having constant power-law
spectral energy distribution, a constant ionization fraction of the neutral gas
and above all, an ideal telescope response. In such scenarios, the generality
of our investigation regarding peaks in Faraday depth spectrum originating due
to accumulation of synchrotron emissivity within a range of FD determined by
the RMSF is expected to hold true. The complication stems from the fact that,
whether the stronger synchrotron emissivity is a consequence of magnetic fields
alone or due to variations in density of CREs or due to variations in their
energy spectrum, will remain degenerate. In the presence of telescope
systematics and noise, complications can be even more difficult to disentangle.
With several upcoming broad-band spectro-polarimetric surveys of the diffuse
Galactic emission and other surveys for extragalactic sources that are
targeting major astrophysical questions, interpreting results obtained from
spectro-polarimetric data analysis techniques in general, and RM synthesis
technique in particular, requires improved statistical methods.


\vspace{6pt} 



\authorcontributions{All authors discussed and conceptualized this work; A.B.
developed the software {\tt COSMIC} based on a basic version by A.F.; B.B.
performed the MHD simulations; A.B. performed all the analyses; and A.B. and
A.F. wrote the paper. All authors discussed the interpretations of the
results.}


\acknowledgments{We thank the two anonymous referees for highly
insightful comments and suggestions. We thank Dominik J. Schwarz, Olaf
Wucknitz and Anirban Lahiri for careful reading and constructive comments on
the paper. A.B. acknowledges financial support by the German Federal Ministry
of Education and Research (BMBF) under grant 05A17PB1 (Verbundprojekt
D-MeerKAT). B.B. acknowledges generous support from the Simons Foundation.
A.F. acknowledges support from the Leverhulme Trust (RPG-2014-427) and the STFC
(ST/N000900/1 Project 2). All RM synthesis calculations were performed using
the {\tt pyrmsynth} package.\footnote{\url{https://github.com/mrbell/pyrmsynth}}
We made use of Astropy, NumPy, matplotlib and joblib in preparing this
manuscript.}

\conflictsofinterest{The authors declare no conflict of interest.} 

\abbreviations{The following abbreviations are used in this manuscript:\\

\noindent 
\begin{tabular}{@{}ll}
RM & Faraday rotation measure\\
FD & Faraday depth\\
MHD & Magnetohydrodynamic\\
CRE & Cosmic ray electron\\
IFD & Internal Faraday dispersion\\
RMSF & Rotation measure spread function\\
FWHM & Full-width at half-maximum\\
\end{tabular}}

\appendixtitles{yes} 
\appendix
\section{Computerized Observations of Simulated MHD Inferred Cubes: {\tt COSMIC}} \label{sec:cosmic_app}
\unskip

Here, we discuss the details of numerical calculations performed by {\tt
COSMIC} to generate synthetic observations. {\tt COSMIC} uses MHD simulation
generated cubes in Cartesian coordinate system and assumes the $z$-axis as the
default line of sight axis. However, any of the perpendicular axis of
the six faces of the cube can be chosen by the user as the line of sight axis.
In the following, we will present our calculations in the default coordinate
system.

\subsection{Synchrotron emissivity} \label{sec:sync_em}

The cubes of $B_x$ and $B_y$ are used to compute the magnetic field strength in
the plane of the sky, $B_\perp = (B_x^2 + B_y^2)^{1/2}$, which is used to
compute the synchrotron emissivity, $\varepsilon_{\rm sync}$, at a frequency
$\nu$ in each cell as,
\begin{equation}
\varepsilon_{\mathrm{sync}, \nu} = N_0\,n_{\rm CRE} \, B_\perp^{1-\alpha}
\, \nu^{\alpha}\, \dfrac{{\rm e}^{-(\nu/\nu_{\rm c})}}{[1 \, + \, (\nu/\nu_{\rm br})^\gamma]}.
\label{eq:sync_em}
\end{equation}
Here, $n_{\rm CRE}$ is the number density of CRE, $N_0$ is an arbitrary
normalization factor, $\alpha$ is the spectral index of the synchrotron
emission, $\gamma$ is the spectral curvature parameter, and $\nu_{\rm br}$ and
$\nu_{\rm c}$ defines the position of a break and cut-off frequency in the
synchrotron spectrum.

Although some MHD simulations do include cosmic rays, here we apply {\tt
COSMIC} to a MHD simulation which does not have information on cosmic ray
electrons, i.e., effects of their propagation and energy losses. In the
scenario when $n_{\rm CRE}$ is unavailable, synchrotron emissivity is computed
using one of the following assumptions:\\
(1) Constant $n_{\rm CRE}$ throughout the 3-D spatial cube.\\
(2) Cosmic rays follow local energy equipartition per mesh point, such that,
$n_{\rm CRE} = B_x^2 + B_y^2 + B_z^2$. We note, however, that such a
distribution of CRE is likely to be unphysical as equipartition conditions are
unlikely to hold on small scales \citep{seta19}.\\
(3) Cosmic rays follows energy equipartition over an user defined volume such
that, $n_{\rm CRE} = \sigma_x^2 + \sigma_y^2 + \sigma_z^2$, where
$\sigma_{x,y,z}$ are the root mean square of the $x$, $y$ and $z$ component of
the magnetic field computed over the defined volume centered at the mesh point
of computation. The typical scale where magnetic fields and CRE are expected to
be in energy equipartition is approximately given by the diffusion length-scale
($l_{\rm diff}$) of CRE \citep[see, e.g.,][]{seta19}. When CRE propagation is
dominated by diffusion, $l_{\rm diff} \approx 1.8\,{\rm kpc}\,\sqrt{D_{\rm
28}\,t_{\rm sync,8}}$. Here, $D_{28}$ is the diffusion coefficient in units of
$10^{28}\,\rm cm^2\,s^{-1}$ and $t_{\rm sync,8}$ is the synchrotron lifetime in
units of $10^8$\,years. If CRE propagation is dominated by streaming
instabilities at the Alfv\'en speed, then $l_{\rm diff} \approx 1\,{\rm kpc} \,
(v_{\rm A,10}\,t_{\rm sync, 8})$.  Here, $v_{\rm A,10}$ is the Alfv\'en speed
in units of $\rm 10\,km\,s^{-1}$.

The MHD simulation box used in the main text has a size of 512~pc. This is
significantly smaller than the typically expected CRE propagation length-scales
mentioned above. We have therefore used the option of constant $n_{\rm CRE}$
and assumed that the CREs have power-law energy distribution so that $\gamma =
0$ and $\nu_{\rm c} \to \infty$.

The normalization $N_0$ is chosen such that the MHD cube produces a certain
integrated flux density $S_0$ (in Jy) at a certain reference frequency $\nu_0$.
Both $S_0$ and $\nu_0$ are user specified. In the main text, we chose $S_0 =
10$~Jy at $\nu_0 = 1$~GHz.

\subsection{Faraday rotation} \label{sec:FaradayRotation}

In the default coordinate system $B_\| = B_z$. The Faraday rotation produced in
each cell is,
\begin{equation}
{\rm FD_{cell}} = 0.812\,n_{\rm e}\,B_\|\, l_{\rm pix}.
\end{equation}
Here, $n_{\rm e}$ is the thermal electron density (in $\rm cm^{-3}$) and
$l_{\rm pix}$ is the physical size of each pixel (in pc). The thermal electron density 
$n_{\rm e}$ is computed from the simulation cube of gas density
using one of the following methods:\\ 
(1) For isothermal simulations, $n_{\rm e} = f_{\rm ion}\, n_{\rm gas}$, where
$f_{\rm ion}$ is a constant ionization fraction throughout the volume.
Here, $n_{\rm gas}$ is the gas number density and is computed from the
simulations. \\
(2) For the case when the gas at temperature $T$ is collisionally ionized,
$n_{\rm e}$ is computed using Saha's ionization formula for hydrogen gas in
thermal equilibrium, 
\begin{equation}
\dfrac{n_{\rm e}^2}{n_{\rm gas} - n_{\rm e}} = \dfrac{2}{\lambda_{\rm B}^3}\,\dfrac{g_1}{g_0}\,\exp\left(\dfrac{-13.6\,{\rm eV}}{k\,T}\right),
\end{equation}
where, $\lambda_{\rm B}$ is the de Broglie wavelength and $k$ is the Boltzmann
constant. In this case, the 3-D spatial distribution of temperature ($T$) is
required as an input.\\ 
(3) In the warm ionized medium, the interstellar gas is mostly ionized by
ultraviolet radiation \citep{haffn09}. In such cases, a simple model for
ionization fraction as a function of temperature can be chosen, with a
continuous arctan-type transition from $f_{\rm ion}=0$ to $f_{\rm ion}=1$ at a
specified temperature.

Since the MHD simulation used in this work is isothermal and we have assumed
that there is no ultraviolet radiation field in the simulated volume, we used
the option of constant $f_{\rm ion}$ in the main text and the choice of its
value is discussed in Section~\ref{sec:mhd}.

In the default coordinate system, the cubes are integrated along the $k$-index
which is the $z$-direction. Thus, the Faraday depth of a pixel at ($i,j,k$) is
computed by summing the ${\rm FD_{cell}}$ of each point along the $k$-index as,
\begin{equation}
{\rm FD}(i,j,k) = \sum_{k^\prime=0}^{k} {\rm FD_{cell}}(i,j,k^\prime).
\label{eq:losFD}
\end{equation}
Here, $i, j$ and $k$ are index along the $x$-, $y$- and $z$-axes, respectively.

\subsection{Linear polarization parameters} \label{sec:polInt}

The maximum fractional polarization of the synchrotron emission ($p_{\rm max}$) 
in each cell of the 3-D spatial cube depends on the spectral index $\alpha$ as, 
\begin{equation}
p_{\rm max} =  \dfrac{1 - \alpha}{5/3 - \alpha},
\label{eq:pmax}
\end{equation}
and, the Stokes $Q$ and $U$ emissivities ($\varepsilon_{Q, \nu}$ and
$\varepsilon_{U,\nu}$, respectively) at each cell at a frequency $\nu$ are
computed as; 
\begin{equation}
\varepsilon_{Q, \nu} = p_{\rm max} \, \varepsilon_{\mathrm{syn}c, \nu} \, \cos \left(2\,\theta_0\right).
\end{equation}
\begin{equation}
\varepsilon_{U, \nu} = p_{\rm max} \, \varepsilon_{\mathrm{sync}, \nu} \, \sin \left(2\,\theta_0\right).
\end{equation}
Here, $\theta_0$ is the intrinsic angle of the linearly polarized emission,
\begin{equation}
\theta_0 = \dfrac{\pi}{2} + \arctan\left(\dfrac{B_y}{B_x}\right).
\end{equation}

In the presence of Faraday rotation, the 2-D-projected Stokes $Q$ and $U$
parameters at a frequency $\nu$ are,
\begin{equation}
Q_\nu(i,j) = \sum_k p_{\rm max} \, \varepsilon_{\mathrm{sync}, \nu}(i,j,k)\, \cos \left[ 2\,\left\{\theta_0(i,j,k) + {\rm FD^\prime}(i,j,k)\,c^2/\nu^2\right\}\right],
\label{eq:compQ}
\end{equation}
\begin{equation}
U_\nu(i,j) = \sum_k p_{\rm max} \, \varepsilon_{\mathrm{sync}, \nu}(i,j,k)\, \sin \left[ 2\,\left\{\theta_0(i,j,k) + {\rm FD^\prime}(i,j,k)\,c^2/\nu^2\right\}\right].
\label{eq:compU}
\end{equation}
Here, ${\rm FD^\prime}(i,j,k) = {\rm FD}(i,j,k) - {\rm
FD_{cell}}(i,j,k)/2$. Because each of the cell in the simulation is also
emitting polarized synchrotron emission, the amount of Faraday rotation it
undergoes within itself is ${\rm FD_{cell}}/2$ instead of ${\rm FD_{cell}}$
\citep{sokol98}. It is important to note, in addition to the Faraday rotation
experienced by a cell, the emission also undergoes Faraday depolarization due
to LOS component of magnetic field within its own cell. However,
the above expressions assumes that the Faraday depolarization within each cell
to be negligible. Depending on the typical values of $\rm FD_{cell}$ in a MHD
simulation, this assumption can break down towards low frequencies and thereby
limits application of {\tt COSMIC} below certain frequencies. For the
simulations used in this work, {\tt COSMIC} can be used to generate synthetic
observations above $\sim0.5$~GHz (see Section~\ref{sec:mhd}).

The linearly polarized intensity map is then computed as,
\begin{equation}
PI_\nu(i,j) = \sqrt{Q_\nu^2(i,j) + U_\nu^2(i,j)}
\label{eq:compPI}
\end{equation}

\subsection{Current limitations of {\tt COSMIC}} \label{sec:limitations}

Here, we list the computational approximations used in {\tt COSMIC} and the
limitations of applying {\tt COSMIC} for obtaining the observables.

\begin{enumerate} 

\item All spatial averaging of intensities over a desired sky area are
calculated as simple arithmetic averages instead of the Gaussian averaging of a
telescope beam.

\item All observable intensities --- total and polarized synchrotron emission,
and Stokes $Q$ and $U$ parameters --- are computed for monochromatic
frequencies. In other words the frequency channels are assumed to be
sufficiently narrow such that bandwidth depolarization is negligible.

\item We assume Faraday depolarization effects within each 3-D mesh point of
the simulation to be negligible. 

\item We do not sample images using the $u-v$ coverage of an interferometer
array. Therefore emission from all spatial scales are included in the images
which is equivalent to observations performed using a single-dish radio
telescope. Note that, a mis-match in $u-v$ coverage can manifest itself as
complicated structures in the total and polarized synchrotron intensity maps.

\item We assume that the total synchrotron emission is optically thin and the
frequency spectrum follows either a simple power-law for the entire frequency
range or a curved spectrum parametrized by $\nu_{\rm br}$ and/or $\nu_{\rm c}$
in Eq.~\eqref{eq:sync_em}. The diffuse synchrotron emission is expected to be
optically thick at frequencies below $\sim50$~MHz. Such a low frequency is not
investigated here and will require rigorous radiative transfer equations to be
incorporated into {\tt COSMIC}.

\item {\tt COSMIC} integrates through sightlines parallel to the axis of the
input MHD cubes and thus is only applicable for emission from a region far away
from the observer. We plan to implement diverging lines of sight in a later
version.

\item Currently, {\tt COSMIC} only computes the different polarization
parameters that describe the synchrotron emission and not other emission
mechanisms, like the thermal free--free emission. Estimation of the free--free
emission is important to study the broad-band emission properties in galaxies.

\end{enumerate}



\reftitle{References}


\externalbibliography{yes}
\bibliography{abasu_etal_mdpi.bbl} 



\end{document}